\documentclass[useAMS,usenatbib]{mn2e}

\usepackage{url,times,graphicx,amsmath,amsfonts,amssymb,aas_macros,epsfig,epstopdf}
\usepackage[dvipsnames,svgnames,hyperref]{xcolor}
\usepackage[
    pdftex,
    a4paper=true,
    breaklinks=true,
    bookmarks=true,
    bookmarksopen=false,
    bookmarksopenlevel=2
    bookmarksnumbered=true,
    bookmarkstype=toc,
    colorlinks=true,
    citecolor=Teal,
    linkcolor=blue,
    menucolor=green,
    urlcolor=magenta,
]{hyperref}
\usepackage{comment}

\newcommand{\ahf}{\textsc{AHF}}
\newcommand{\bdm}{\textsc{BDM}}
\newcommand{\subfind}{\textsc{SUBFIND}}
\newcommand{\asohf}{\textsc{ASOHF}}
\newcommand{\voboz}{\textsc{VOBOZ}}
\newcommand{\mendieta}{\textsc{Mendieta}}
\newcommand{\sdfof}{\textsc{6DFOF}}
\newcommand{\adaptahop}{\textsc{AdaptaHOP}}

\newcommand{\hot}{\textsc{HOT}}
\newcommand{\hottd}{\textsc{HOT3D}}
\newcommand{\hotsd}{\textsc{HOT6D}}

\newcommand{\rockstar}{\textsc{Rockstar}}
\newcommand{\hbt}{\textsc{HBT}}
\newcommand{\hsf}{\textsc{HSF}}
\newcommand{\stf}{\textsc{STF}}
\newcommand{\pso}{\textsc{pSO}}
\newcommand{\phop}{\textsc{pHOP}}
\newcommand{\denmax}{\textsc{DENMAX}}
\newcommand{\hop}{\textsc{HOP}}
\newcommand{\mhf}{\textsc{MHF}}
\newcommand{\mht}{\textsc{MHT}}
\newcommand{\origami}{\textsc{ORIGAMI}}
\newcommand{\skid}{\textsc{SKID}}
\newcommand{\grasshopper}{\textsc{GRASSHOPPER}}
\newcommand{\psb}{\textsc{PSB}}
\newcommand{\pfof}{\textsc{pFOF}}
\newcommand{\ntropyfof}{\textsc{ntropyFOF}}
\newcommand{\jumpd}{\textsc{JUMP-D}}
\newcommand{\lanl}{\textsc{LANL}}
\newcommand{\fof}{\textsc{FOF}}
\newcommand{\surv}{\textsc{SURV}}

\newcommand{\dd}{{\rm d}}
\newcommand{\deriv}[2]{\frac{\dd#1}{\dd#2}}
\newcommand{\hMpc}{{\ifmmode{h^{-1}{\rm Mpc}}\else{$h^{-1}$Mpc}\fi}}
\newcommand{\hkpc}{{\ifmmode{h^{-1}{\rm kpc}}\else{$h^{-1}$kpc}\fi}}
\newcommand{\hMsun}{{\ifmmode{h^{-1}{\rm {M_{\odot}}}}\else{$h^{-1}{\rm{M_{\odot}}}$}\fi}}
\newcommand{\ltsima}{$\; \buildrel < \over \sim \;$}
\newcommand{\gtsima}{$\; \buildrel > \over \sim \;$}
\newcommand{\lsim}{\lower.5ex\hbox{\ltsima}}
\newcommand{\gsim}{\lower.5ex\hbox{\gtsima}}
\def\LCDM{$\Lambda$CDM}
\def\nbody{$N$-body}
\def\lesssim{\mathrel{\hbox{\rlap{\hbox{\lower4pt\hbox{$\sim$}}}\hbox{$<$}}}}
\def\gtrsim{\mathrel{\hbox{\rlap{\hbox{\lower4pt\hbox{$\sim$}}}\hbox{$>$}}}}

\newcommand{\Tab}[1]{Table~\ref{#1}}
\newcommand{\Sec}[1]{Section~\ref{#1}}
\newcommand{\App}[1]{Appendix~\ref{#1}}
\newcommand{\Eq}[1]{Eq.~(\ref{#1})}
\newcommand{\Fig}[1]{Fig.~\ref{#1}}
\newcommand{\beq}{\begin{equation}}
\newcommand{\eeq}{\end{equation}}
\def\beqa{\begin{eqnarray}}
\def\eeqa{\end{eqnarray}}
\def\hMpc{$h^{-1}\,{\rm Mpc}$}
\def\hkpc{$h^{-1}\,{\rm kpc}$}
\def\LCDM{\ensuremath{\Lambda}CDM}

\def\Vmax{$V_{\rm max}$}
\def\Rmax{$R_{\rm max}$}

\title[Structure Finding in Cosmological Simulations]
{Structure Finding in Cosmological Simulations: The State of Affairs}
\author[Knebe et. al]
       {Alexander Knebe,$^{1}$\thanks{E-mail: alexander.knebe@uam.es}
  Frazer~R.~Pearce,$^{2}$ 
  Hanni~Lux,$^{2,30}$
  Yago Ascasibar,$^{1}$
  Peter~Behroozi,$^{4,5,6}$
  \newauthor
  Javier Casado,$^{1}$
  Christine~Corbett~Moran,$^{7}$
  Juerg Diemand,$^{7}$
  Klaus Dolag,$^{8,9}$
  \newauthor
  Rosa Dominguez-Tenreiro,$^{1}$
  Pascal~Elahi,$^{2,10}$
  Bridget Falck,$^{18}$
  Stefan Gottl\"ober,$^{11}$
  \newauthor
  Jiaxin~Han,$^{10,12,13}$
  Anatoly~Klypin,$^{14}$
  Zarija Luki\'c,$^{15,34}$
  Michal Maciejewski,$^{8}$
  \newauthor
  Cameron K. McBride,$^{16,33}$
  Manuel~E.~Merch\'{a}n,$^{17}$
  Stuart~I.~Muldrew,$^{2}$ 
  Mark~Neyrinck,$^{3}$
  \newauthor
  Julian~Onions,$^{2}$
  Susana~Planelles,$^{31,32}$
  Doug~Potter,$^{7}$
  Vicent~Quilis,$^{19}$
  Yann Rasera,$^{20}$
  \newauthor
  Paul M. Ricker,$^{21,22}$
  Fabrice Roy,$^{20}$
  Andr\'{e}s N. Ruiz,$^{17}$ 
  Mario A. Sgr\'{o},$^{17}$ 
  Volker Springel,$^{23,24}$ 
  \newauthor
  Joachim Stadel,$^{7}$
  P. M. Sutter,$^{21,27,28,29}$
  Dylan Tweed,$^{25}$
  Marcel Zemp$^{26}$
\\
\\
  $^{1}$Departamento de F\'isica Te\'{o}rica, M\'{o}dulo 15, Facultad de Ciencias, 
  Universidad Aut\'{o}noma de Madrid, 28049 Madrid, Spain\\
  $^{2}$School of Physics \& Astronomy, University of Nottingham, Nottingham, NG7 2RD, UK\\
 $^{3}$Department of Physics and Astronomy, Johns Hopkins University, 3701 San Martin Drive, Baltimore, MD 21218, USA\\
  $^{4}$Kavli Institute for Particle Astrophysics and Cosmology, Stanford, CA 94309, USA\\
  $^{5}$Physics Department, Stanford University, Stanford, CA 94305, USA\\
  $^{6}$SLAC National Accelerator Laboratory, Menlo Park, CA 94025, USA\\
 $^{7}$University of Zurich, Institute for Theoretical Physics, Winterthurerstrasse 190, CH-8057 Zurich, Switzerland\\
 $^{8}$Max-Planck Institut f\"{u}r Astrophysik, Karl-Schwarzschild Str. 1, D-85741 Garching, Germany\\
 $^{9}$University Observatory M\"unchen, Scheinerstr. 1, 81679, M\"unchen, Germany\\
  $^{10}$Key Laboratory for Research in Galaxies and Cosmology, Shanghai Astronomical Observatory, Shanghai 200030, China\\
 $^{11}$Leibniz-Institut f\"ur Astrophysik Potsdam (AIP), An der Sternwarte 16, 14482 Potsdam, Germany\\
  $^{12}$Graduate School of the Chinese Academy of Sciences, 19A, Yuquan Road, Beijing, China \\
  $^{13}$Institute for Computational Cosmology, Department of Physics, Durham University, South Road, Durham DH1 3LE, UK\\
 $^{14}$Department of Astronomy, New Mexico State University, Las Cruces, NM 88003-0001, USA\\
 $^{15}$Computational Cosmology Center, Lawrence Berkeley Lab, Berkeley CA 94610, USA\\
 $^{16}$Vanderbilt University, Department of Physics \&  Astronomy, 6301 Stevenson Center, Nashville, TN 37235, USA\\
  $^{17}$Instituto de Astronom\'ia Te\'{o}rica y Experimental (CCT C\'{o}rdoba, CONICET, UNC), Laprida 922, X5000BGT, C\'{o}rdoba, Argentina\\
  $^{18}$Institute of Cosmology and Gravitation, University of Portsmouth, Dennis Sciama Building, Portsmouth, PO1 3FX, UK\\
  $^{19}$Departament   d'Astronomia   i   Astrof\'{\i}sica,   Universitat de Val\`encia,   46100  -   Burjassot  (Valencia),   Spain\\
 $^{20}$CNRS, Laboratoire Univers et Th\'eories (LUTh), UMR 8102 CNRS, Observatoire de Paris, Universit\'e Paris Diderot; 5 Place  Jules Janssen, 92190 Meudon, France\\
 $^{21}$Department of Physics, University of Illinois at Urbana-Champaign, Urbana, IL 61801-3080, USA\\
 $^{22}$National Center for Supercomputing Applications, University of Illinois at Urbana-Champaign, Urbana, IL 61801, USA\\
 $^{23}$Heidelberg Institute for Theoretical Studies, Schloss-Wolfsbrunnenweg 35, 69118 Heidelberg, Germany\\ 
 $^{24}$Zentrum f\"ur Astronomie der Universit\"at Heidelberg, ARI, M\"{o}nchhofstr. 12-14, 69120 Heidelberg, Germany \\
  $^{25}$Racah Institute of Physics, The Hebrew University, Jerusalem 91904, Israel \\
 $^{26}$Kavli Institute for Astronomy and Astrophysics, Peking University, Yi He Yuan Lu 5, Hai Dian Qu, Beijing 100871, P. R. China\\
 $^{27}$ UPMC Univ Paris 06, UMR7095, Institut d'Astrophysique de Paris,  F-75014, Paris, France \\
 $^{28}$ CNRS, UMR7095, Institut d'Astrophysique de Paris, F-75014, Paris, France\\
 $^{29}$ Center for Cosmology and Astro-Particle Physics, Ohio State University, Columbus, OH 43210\\
 $^{30}$ Department of Physics, University of Oxford, Denys Wilkinson Building, Keble Road, Oxford, OX1 3RH, UK\\
 $^{31}$ Astronomy Unit, Department of Physics, University of Trieste, via Tiepolo 11, I-34131 Trieste, Italy\\
 $^{32}$ INAF, Osservatorio Astronomico di Trieste, via Tiepolo 11, I-34131 Trieste, Italy\\  
 $^{33}$ Harvard-Smithsonian Center for Astrophysics, 60 Garden Street, Cambridge, MA 02138, USA\\  
 $^{34}$T-2, Theoretical Division, Los Alamos National Laboratory, P.O. Box 1663, Los Alamos, NM 87544, USA\\
 }

\begin{document}

\date{Accepted XXXX . Received XXXX; in original form XXXX}

\pagerange{\pageref{firstpage}--\pageref{lastpage}} \pubyear{2010}

\maketitle

\label{firstpage}

\clearpage

\begin{abstract}

The ever increasing size and complexity of data coming from simulations of cosmic structure formation demands equally sophisticated tools for their analysis. During the past decade, the art of object finding in these simulations has hence developed into an important discipline itself.  A multitude of codes based upon a huge variety of methods and techniques have been spawned yet the question remained as to whether or not they will provide the same (physical) information about the structures of interest. Here we summarize and extent previous work of the "halo finder comparison project":  we investigate in detail the (possible) origin of any deviations across finders. To this extent we decipher and discuss differences in halo finding methods, clearly separating them from the disparity in definitions of halo properties. We observe that different codes not only find different numbers of objects leading to a scatter of up to $20$~per cent in the halo mass and \Vmax\ function, but also that the particulars of those objects that are identified by all finders differ. The strength of the variation, however, depends on the property studied, e.g. the scatter in position, bulk velocity, mass, and the peak value of the rotation curve is practically below a few per cent, whereas derived quantities such as spin and shape show larger deviations.  Our study indicates that the prime contribution to differences in halo properties across codes stems from the distinct particle collection methods and -- to a minor extent -- the particular aspects of how the procedure for removing unbound particles is implemented. We close with a discussion of the relevance and implications of the scatter across different codes for other fields such as semi-analytical galaxy formation models, gravitational lensing, and observables in general. 

\end{abstract}
\noindent
\begin{keywords}
  methods: $N$-body simulations -- galaxies: haloes -- galaxies:
  evolution -- cosmology: theory -- dark matter
\end{keywords}

\section{Introduction} \label{sec:introduction}
Over the last 30 years great progress has been made in the development of simulation codes that model the distribution of the dissipationless dark matter that makes up most of the Universe's dynamical mass. Some codes also simultaneously follow the substantially more complex baryonic physics of the visible and hence directly observable Universe. Nowadays we have a great variety of highly reliable, cost effective (and in some cases publicly available) codes designed for the simulation of cosmic structure formation \citep[e.g.][]{Couchman95, Pen95, Kravtsov97, Bode00, Springel01, Knebe01, Teyssier02, OShea04, Quilis04, Dubinski04, Merz05, Springel05, Springel10, Doumler10}.  However, producing the data is only the first step in the process; the ensembles of billions of tracers generated still require interpreting so that their distribution may be somehow compared to the real Universe. This necessitates access to analysis tools to map the phase-space which is being sampled by the tracers onto `real' 
objects in the Universe. Therefore, to take advantage of sophisticated \nbody\ codes and to optimise their predictive power one needs equally sophisticated structure finders.

Halo finders mine \nbody\ data to find locally over-dense (either in configuration or phase-space) gravitationally bound systems, i.e.~ the dark matter haloes we currently believe surround galaxies. This type of analysis has led to critical insights into our understanding of the origin and evolution of cosmic structure and galaxies. Theoretically, the properties of the simulated objects are often reduced to readily usable functional forms, e.g. the dark matter halo density profile, \citep{Navarro96, Navarro97, Moore99}, concentration-mass relation \citep{Bullock01, Wechsler02}, mass accretion histories \citep{Wechsler02,DeLucia04, Diemand07b, DeLucia07,Behroozi12c}, shape distributions \citep{Dubinski91, Kauffmann99, Bullock01b}, clustering properties \citep{Mo96, Smith03}, environmental effects \citep{Baugh96, Moore96b}, merger rates \citep{Lacey94, Bower06,Fakhouri08,Fakhouri10,Behroozi12c}, disruption timescales \citep{Ghigna98, Zentner05}. All these properties and interconnections have been derived from 
simulations, encoded using analytical formulae, and subsequently been used as input in, for instance, semi-analytical models \citep[e.g.][]{Cole94, Cole00, Croton06}, gravitational lensing calculations \citep[e.g.][]{Kaiser93, Bartelmann98, Diemand08}, or directly in comparison to observations \citep[e.g.][]{Davis85, Klypin99s, Springel05b, Komatsu11}. And one of the questions we will address here is whether some or all of these relations depend sensitively upon the choice of the applied halo finder? 

\begin{table}
  \caption{Chronological list of halo finders and methods since the dawn of computational cosmology.}
\label{tab:finders}
\begin{center}
\begin{tabular}{lll}
\hline
year 		& method/code 				& reference\\
\hline
1974		&	SO						&	\citet{Press74}\\
1985		&	FOF						&	\citet{Davis85}\\
1991		&	DENMAX					& 	\citet{Bertschinger91}\\
1994		&	SO						&	\citet{Lacey94}\\
1995		&	adaptive FOF				&	\citet{vanKampen95}\\
1996		&	IsoDen					&	\citet{Pfitzner96}\\
1997		&	\bdm						&	\citet{Klypin97}\\
1998		&	\hop						&	\citet{Eisenstein98hop}\\
1999		& 	hierarchical \fof				&	\citet{Gottloeber99}\\
2001		&	\skid						&	\citet{Stadel01}\\
2001		& 	enhanced \bdm			&	\citet{Bullock01}\\
2001		&	\subfind					&	\citet{Springel01subfind}\\
2004		&	\mhf\ \& \mht				&	\citet{Gill04a}\\
2004		&	\adaptahop				&	\citet{Aubert04}\\
2004		&	\denmax$^2$				&	\citet{Neyrinck04}\\
2004		&	\surv						&	\citet{Tormen04}\\
2005		&	improved \denmax			&	\citet{Weller05}\\
2005		&	\voboz					&	\citet{Neyrinck05}\\
2006		&	\psb						&	\citet{Kim06}\\
2006		&	\sdfof					&	\citet{Diemand06}\\
2007		&	further improved \denmax		&	\citet{Shaw07}\\
2007     	&	\ntropyfof					&      \citet{Gardner07a}\\
2009		&	\hsf						&	\citet{Maciejewski09}\\
2009		&	\lanl\ finder				&	\citet{Habib09}\\
2009		&	\ahf						&	\citet{Knollmann09}\\
2010		&	\phop					&	\citet{Skory10}\\
2010		&	\asohf					&	\citet{Planelles10}\\
2010		&	\pso						&	\citet{Sutter10}\\
2010		&	\pfof						&	\citet{Rasera10}\\
2010		&	\origami					&	\citet{Falck12}\\
2010		&	\hot						&	Ascasibar, in prep.\\
2010		&	\rockstar					&	\citet{Behroozi12}\\
2010		&	\mendieta					&	\citet{Sgro10}\\
2010		&	enhanced \surv				&	\citet{Giocoli10}\\
2011		&	\hbt						&	\citet{Han12}\\
2011		&	\stf						&	\citet{Elahi11}\\
2012		&	\grasshopper				&	Stadel et al., in prep.\\
2012		&	\jumpd					&	Casado \& Dominguez-Tenreiro, in prep.\\
\hline
\end{tabular}
\end{center}
\end{table}

\subsection{History of Halo Finding} \label{sec:history}
While for decades the focus was on getting the simulations themselves under control, it is now obvious that halo finding is equally important and, unfortunately, not that well understood as yet. Or to put it another way, producing the raw simulation data is only the first step in the process; the model requires reduction before it can be compared to the observed Universe we inhabit.  In recent years, this field has also seen great development in the number and variety of object finders as shown in \Tab{tab:finders}, where we chronologically list the emergence of codes or methods. We can clearly see the increasing pace of development in the past decade reflecting the necessity for state-of-the-art codes: in the last ten years the number of existing halo finding codes has practically tripled. While for a long time the spherical overdensity method first mentioned by \citet[SO, ][]{Press74} as well as the friend-of-friends algorithm introduced\footnote{This is strictly speaking only true for astrophysics as the 
friends-of-friends method is widely used in molecular dynamics since the 1960's to find bound clusters (e.g. liquid droplets in a gas). In that field it is well known as the ``Stillinger method'' \citep{Stillinger63}.} by \citet[FOF, ][]{Davis85} remained the standard techniques, the situation changed in the 90's when new methods were developed \citep{Gelb92, Lacey94, vanKampen95, Pfitzner96, Klypin97, Eisenstein98hop, Gottloeber99}.

While the first generation of halo finders primarily focused on identifying isolated field haloes the situation dramatically changed once it became clear that there was no such thing as `overmerging': the premature destruction of haloes orbiting inside larger host haloes \citep{Klypin99} was a numerical artifact rather than a real physical process. Now post-processing tools face the challenge of finding both haloes embedded within the (more or less uniform) background density of the Universe as well as subhaloes orbiting within the density gradient of a larger host halo. The past decade has seen a substantial number of codes and techniques introduced in an attempt to cope with this problem (see \Tab{tab:finders}). One approach was to make use of the additional information available in a simulation where all six phase-space 
variables are typically known. Additionally, some modern finders make use of the time co-ordinate too, as large structures are not expected to suddenly appear out of nothing. The use of such extra information makes possible the investigation of structures beyond the traditional bound objects. For instance disrupted objects can be studied either by tracking the debris from a once known object that has been disrupted or identifying such an object as a distinct entity in six-dimensional phase-space (see Elahi et~al., submitted). Streams of stars are of course a highly topical example of work relevant to near-field cosmology  \citep[e.g.][]{Belokurov06}. 

Further, as simulations became much larger this also led to a trend towards parallel analysis tools. The simulation data had become too large to be analysed on single CPU architectures and hence halo finders had to be adjusted to cope with this.  The recent profusion of new codes is also a reflection of the drive to build halo finders and associated analysis tools into the simulation codes themselves. Such an approach obviates the need to frequently save the raw simulation output, instead only requiring the storing of much smaller reduced catalogues of the interesting structures and their properties \citep{Angulo12}. This approach of course founders if the analysis applied is either not robust or incomplete in some way, as there is no longer any ability to return to the raw data and reprocess it without rerunning the entire simulation. 

For the upcoming generation of trillion particle production simulations that represent the forefront of numerical cosmology the approach of storing only the reduced halo catalogues therefore appears to be essential unless a dramatic storage breakthrough is made. With such a clear direction it is essential that any post-processing scheme adopted is both robust and well understood. This is one of the key drivers of this entire project.

\subsection{What is a Halo?} \label{sec:whatisahalo}
Although the question of what is a halo appears straightforward a direct answer is not immediately obvious and has been the subject of some previous studies already \citep[e.g.][]{Maccio03,Prada06,Cuesta08,Anderhalden11,Diemer12}. While one can argue that a halo is a `gravitationally bound object' \citep[cf.][]{Knebe11}, this still leaves the definition of the outer edge unresolved. While we go on to discuss this topic in more detail in the following Sections,  we nevertheless consider it sufficiently important to attempt to address it here in the Introduction, too.

Assuming for a moment that we agree upon the aforementioned definition of boundness, we already face the problem that haloes may contain substructure: will the mass of the subhaloes (which certainly are also bound to the host itself) be considered part of the host or should they be excised from it? While the answer to this uncertainty depends on the scientific problem under investigation (e.g. gravitational lensing studies would require the full mass, including substructure, whereas mass profile investigations would likely prefer to live without these extra peaks), one needs to be aware that some halo finders return halo masses including `sub-masses' (e.g. \ahf) while others do not (e.g. \subfind).

Another -- directly related and actually affected -- question is that of the edge of the subhalo. Note that we liberally talk about the mass and edge at the same time as they are most commonly defined simultaneously via one equation (cf. \Eq{eq:virialradius} in \Sec{sec:methodsmass} below), i.e. the mass enclosed within the radius of a halo has to be some multiple of a reference density (usually either the background or the critical density of the Universe) times the (spherical) volume defined by that radius. But as we have just asked, should sub-masses be included as well? It will certainly change the enclosed mass and hence radius of the object. Irrespective of this `sub-mass issue', the edge of a halo is not a well-defined quantity. Even though the most commonly used working definition assumes spherical symmetry and adopts some theory-driven pre-factor for the reference density based upon a spherical top-hat collapse, this pre-factor is nevertheless a loosely defined parameter for which different halo finders use slightly 
different (and possibly cosmology and redshift dependent) values. Additionally, it is not obvious that dark matter (sub-)haloes should be characterised by a spherical radius, especially when they are subject to severe tidal distortion. Friends-of-Friends based finders bypass this problem by merely stating as the halo mass the sum of all the particle masses linked together by their favourite choice of linking length; for a more elaborate discussion about the relation between SO and FOF mass please refer to \citet{Lukic09} and \citet{More11} as well as the pioneering comparison found in \citet{Lacey94} and \citet{Cole96}. Thus the edge of FOF haloes are by definition non-spherical. And there are examples for halo finders that circumvent the conventional edge definition by linking the halo's boundary to the dynamics of the particles \citep[\origami][cf. \Sec{sec:origami}]{Falck12}.

Is either of these approaches a reasonable or suitable strategy? As we will explore in more detail later, one could also think of rather different definitions for the halo edge (and hence its mass) inspired by, for instance, a desire to truncate subhaloes at the saddle point of the density field or the tidal radius. 

An alternative approach to quantify the `size' of a halo which avoids this problem is to use a related quantity rather than the mass: for instance, the peak of the rotation curve as characterised by \Vmax\ or the radial location of this peak by \Rmax. These quantities do indeed provide a physically-motivated scale \citep[e.g.][]{Ascasibar08}. While the physical properties derived from particles at distances beyond \Rmax\  might exhibit scatter and systematic trends arising from differ definitions of a halo's edge, the quantities derived from the inner regions such as  \Rmax\ and \Vmax\ prove to be far stabler against such numerical uncertainties.  Another advantage of using \Vmax\ is that it is more closely related to certain observable properties (such as galaxy rotation curves) than the halo mass. However, the peak of the rotation curve is reached quite close to the centre of the halo, and its measurement is sensitive to numerical resolution. Being set by the central particles, it is less sensitive to tidal stripping than mass, which may be seen as 
either an advantage or a disadvantage depending on the scientific question under study.

In summary, this brief discussion should serve to alert users of halo catalogues that there are choices that need to be made before a halo catalogue can be produced. Different code authors naturally make difference choices, as often there is no `correct' method and more often than not the definition adopted depends upon the problem being addressed. In what follows we will discuss the range in derived properties that arises due to these different choices as well as addressing the question of whether or not the halo finders agree when applied to the same data set with a common set of assumptions.

\subsection{The Workshops} \label{sec:workshops}

We initiated the halo finder comparison project that has brought together practically every expert/code developer in the field at a series of bi-annual workshops focusing on the comparison of their respective codes.

\subsubsection{Haloes going MAD 2010}
The start-up gathering and first comparison with respect to mock and
field haloes: during the last week of May 2010 we held the workshop
``Haloes going MAD'' in Miraflores de la Sierra close to Madrid dedicated to the issues surrounding identifying haloes in cosmological simulations. Amongst other participants 15 halo finder representatives were present. The aim of this workshop was to define (and use!) a unique set of test scenarios for verifying the credibility and reliability of such programs. We applied each and every halo finder to our newly established suite of test cases and cross-compared the results.

To date most halo finders were introduced (if at all) in their respective code papers which presented their underlying principles and generally subjected them to tests within a full cosmological environment, primarily matching (sub-)halo mass functions to theoretical models and fitting functions. Hence no general benchmarks such as the ones designed at this workshop existed prior to this meeting. Our newly devised suite of test cases is designed to be simple yet challenging enough to assist in establishing and gauging the credibility and functionality of all commonly employed halo finders.  These tests include mock haloes with well defined properties as well as a state-of-the-art cosmological simulation. They involve the identification of individual objects, various levels of substructure, and dynamically evolving systems. The cosmological simulation has been provided at various resolution levels with the best resolved containing a sufficient number of particles (1024$^3$) that it can only presently be 
analysed in parallel.

All the test cases and their analysis are publicly available from \url{http://popia.ft.uam.es/HaloesGoingMAD} under the tab ``The Data''.

\subsubsection{Subhaloes going Notts 2012}
While ``Haloes going MAD'' primarily dealt with either mock halo set-ups containing well behaved substructure or field haloes, the next natural question was how halo finders perform and compare when it comes to subhaloes as found in high-resolution simulations. Within the hierarchical structure formation scenario \citep{Davis85} the quantification of the amount of substructure (both observationally and in simulations of structure formation) is an essential step towards what is nowadays referred to as ``Near-Field Cosmology'' \citep{Freeman02}. We therefore utilized the data for one of the haloes from the Aquarius project \citep[][courtesy VIRGO consortium\footnote{\url{http://www.virgo.dur.ac.uk/}}]{Springel08} that consists of multiple dark matter only re-simulations of a Milky Way like halo at a variety of resolutions performed using \textsc{GADGET3} \citep[based upon \textsc{GADGET2},][]{Springel05}.

The follow-up meeting ``Subhaloes going Notts'' then took place during the second week of May 2012 in Dovedale, probably one of the most remote locations in England. The focus of this meeting was to better understand the differences in (sub-)halo properties that emerged during the analysis of the two comparison papers \citet{Knebe11} and \citet{Onions12}.\footnote{Note that the ``Subhaloes going Notts'' paper had been published prior to the workshop.}  Furthermore, we also took the collaboration of the 9 code representatives present at that meeting and all other participants actively interested in halo finding one step further: having at our disposal the analysis and expertise for various codes in the field we intended to address scientific questions (as opposed to academic comparisons) using the `code scatter' as error bars on the results. To this extent we focused on the development of a common post-processing pipeline. Further, a lot of effort during this meeting went into an improved understanding of 
where the differences between the codes came from.

Again, all the test cases and their analysis are available from \url{http://popia.ft.uam.es/SubhaloesGoingNotts} following the instructions given under the tab ``Data''; access to the data (also including the Aquarius simulations) requires registration which will certainly be granted to everyone scientifically interested in the data.

\subsection{Intention of this Work} \label{sec:intention}
The aim of this paper is firstly to acquaint the reader with the
general concepts commonly applied to the problem of finding objects in
simulations of cosmic structure formation. These assumptions and
choices underpin the production of halo catalogues that are then often
used by other fields, for instance, semi-analytical galaxy formation,
or -- most importantly -- direct observational comparisons. We address
such questions as ``what can I expect from halo finders?'' as well as ``to what accuracy can I trust these catalogues?''. The latter is obviously of great relevance to anyone employing halo catalogues, especially as we have entered the era of precision cosmology \citep{Smoot03,Primack05,Coles05,Primack07}.

In order to address such questions we first have to plunge into the details of halo finding. What are the various methods applied by the community and how do these different approaches drive any scatter in the derived halo properties? To this extent, parts of this article serve as a reference for anyone interested in the technical details which are discussed in Sections~\ref{sec:methods} and~\ref{sec:precisioncosmology} where we describe the (sub-)halo finding methods and the technical issues on the way to high precision (sub-)halo finding, respectively; Sections~\ref{sec:definition}, \ref{sec:recovery}, and \ref{sec:relationandapplication}) are of particular interest to users of halo catalogues and address the definition of halo properties, the uncertainties in their recovery and their applications in other fields, respectively. A summary of the content in the respective Sections is given here to better guide the reader and allow quicker access to the information.

\begin{itemize}
 \item [\textit{\ref{sec:methods}}] \textbf{\textit{Halo Finding Methods:}} We
 	separate the actual working methodology of a halo finder from
 	the subsequently applied definitions for halo properties
 	discussed in the following Section. This section therefore is
 	of a technical nature with likely little interest to the
 	end-user of halo catalogues. It allows for greater insight
 	into the possible origin of any (dis-)similarities between the
 	finders. But note that we are not discussing or presenting
	individual codes here, we rather talk about the methods in general.
	 
 \item [\textit{\ref{sec:definition}}] \textbf{\textit{Definition of Halo
 	Properties:}} Given an identical set of particles belonging to
 	a halo, there are still various possibilities for how to define
 	(and hence calculate) its properties. In this Section we
 	present the most commonly adopted working definitions which
 	are -- in principle -- independent of the applied halo finder.
 
 \item [\textit{\ref{sec:recovery}}] \textbf{\textit{Recovery of Halo
 	Properties:}} This Section compares the results from different
 	halo finders applied to various identical data sets. While it
 	is in part a summary of the work presented in \citet{Knebe11}
 	and \citet{Onions12} it extends these works by digging deeper
 	and quantifying the errors.

 \item [\textit{\ref{sec:precisioncosmology}}] \textbf{\textit{Precision
 	Cosmology:}} After presenting hard numbers for the differences
 	between codes the questions remain about the origin of the
 	scatter, possible ways to improve agreement, and the impact
 	for the era of precision cosmology. These topics shall be
 	discussed in this Section.
	
 \item [\textit{\ref{sec:relationandapplication}}] \textbf{\textit{Relation
 	and Application to other Fields:}} While all previous Sections
 	primarily dealt with cosmological simulations and academic
 	test cases, here we talk about the relevance of halo finding
 	for other fields such as semi-analytical galaxy formation
 	models, gravitational lensing, and observables in general. The
 	focal point will be to gauge the significance of differences
 	in halo finders (and property definitions) for the respective
 	fields.
\end{itemize}

Even though we clearly separated the finding methods in \Sec{sec:methods} from the property definitions in \Sec{sec:definition}, we emphasize that there is a great interplay between those two parts, especially when it comes to the centre and velocity of the halo: both these quantities are essential for the procedure of removing gravitationally unbound particles (forming part of the methods) and hence one needs to bear in mind that the division is not entirely straightforward.

The data used and results presented throughout this work are based upon the two earlier comparison projects ``Haloes going MAD'' and ``Subhaloes going Notts''. However, there are subtle differences to these sets as the former allowed code representatives to return their values for halo properties as derived from their respective codes, whereas the latter project based the comparison on catalogues obtained via a common post-processing pipeline applied to the provided particle ID lists. Here we also go one step further using at times only those objects found by all finders and directly comparing the same haloes across codes.

\section{Halo Finding Methods} \label{sec:methods}
Here we present a summary of all the steps commonly employed in the process of generating halo catalogues starting from the raw output of a cosmological simulation. For the general reader this may be a rather technical section and we therefore encourage everyone not interested in `flowcharts for halo finding' to skip to the next \Sec{sec:definition} where working definitions for halo properties will be given. But a lot of the points discussed here will actually be relevant and of importance when it comes to understanding the origin of differences between halo catalogues obtained with different finders for identical data sets: the steps outlined here and realised in practice actually define a halo finder and distinguish it from others.

In any case, the first two halo finders mentioned in the literature, i.e. the spherical overdensity (SO) method \citep{Press74} and the friends-of-friends (FOF) algorithm \citep{Davis85} remain the foundation of nearly every code: they often involve at least one phase where either particles are linked together or (spherical) shells are grown to collect particles. While we do not wish to invent stereotypes or a classification scheme for halo finders there are unarguably two distinct groups of codes:

\begin{itemize}
 \item density peak locator (+ subsequent particle collection)
 \item direct particle collector
\end{itemize}

\noindent
The density peak locators -- such as the classical SO method -- aim at identifying by whatever means peaks in the matter density field. About these centres (spherical) shells are grown out to the point where the density profile drops below a certain pre-defined value normally derived from a spherical top-hat collapse. Most of the methods utilising this approach merely differ in the way they locate density peaks. The direct particle collector codes -- above all the FOF method -- connect and link particles together that are close to each other (either in a 3D configuration or in 6D phase-space). They afterwards determine the centre of this mass aggregation. Please note that there is a subtle difference between codes utilizing a hybrid approach, i.e. a sole SO finder will be different from a finder that first applies a FOF method and then crops the halo by means of SO.

For a brief technical presentation of all the finders participating in the comparison project in one way or the other, we refer the reader to \App{app:codedescriptions} where their mode of operation is presented.

\subsection{Candidate Identification} \label{sec:methodscandidateidentification}
The first step for nearly all non-FOF based halo finders is to generate a list of potential halo centres. In most cases, and in particular for SO based finders, this is achieved by locating peaks in the density field or troughs in the gravitational potential field. Some techniques such as phase-space (\rockstar) or velocity based finders (\stf) may have their very own approach, but almost any finder comes up with such an initial list which is then processed further.

A related issue, which is often the last step of any algorithm, would be the prescription followed to decide which of the objects found are indeed real and which ones are spurious. This may be as simple as a threshold on the particle number, but more elaborate statistical criteria, often specifically tuned for each particular technique, are also implemented by several codes. This `catalogue cleaning' process is in practice a problem area as it may be difficult to implement in parallel for the largest datasets. The issue is that particles from rejected haloes may need to be re-added to either another halo or the background pool in a self-consistent way. Thus whether or not the cleaning process is carried out `on-the-fly' as the halo catalogue is built up or as a separate post-processing step can make a difference to the finally generated catalogue. We would advocate that the final catalogue generated after any halo cleaning algorithm has been applied should be independent of the location of the halo 
cleaning in the analysis chain. Unfortunately this is presently not always the case.

\subsection{Particle Collection} \label{sec:methodsparticlecollection}
Once a candidate list has been generated, one needs to gather those particles that likely belong to each and every object. In practice, there is again a lot of room for variety in how to achieve this, and we will see later on that it may have an influence on the actual halo properties. For instance, when dealing with simulations containing substructure (like the Aquarius data used during the ``Subhaloes going Notts'' workshop), one question is whether or not the particles belonging to a subhalo should also be affiliated to the host halo. This essentially boils down to a decision on whether or not any single particle can be in more than one object at the same time. Further, haloes will also be affected by either collecting particles in spherical regions, from arbitrary geometries, or in phase/velocity-space. All these issues will leave their imprint on the final halo catalogue.

\subsection{Halo Centre \& Bulk Velocity Determination} \label{sec:methodscentre}
Once a candidate particle list for each halo has been obtained it is important to locate its centre as this defines the physical location of the object as well as being used within many subsequent analyses. There are a variety of possibilities and implementations imaginable for the identification of the centre (see \Sec{sec:definitionscentre}). For instance, some codes simply stick to the location of the density peak or gravitational potential minimum used during the candidate identification (e.g. \ahf) whereas others use the centre of mass of some fraction of the particles. And in the case of extended or stream like structures the halo centre can be quite ill defined. In that regards, iterative refinement techniques for a robust centre may need to be implemented also impacting upon the particle collection discussed before.

Similar issues arise when calculating the velocity of the object, which could be calculated from all the particles or some central subset or by simply taking the velocity of the most bound particle for example. An added complication is any unbinding procedure, which may lead to a need to iteratively recalculate the centre and bulk velocity as unbound particles are removed. Some codes determine the position of the centre first, and then use that information to determine the velocity, while others find both at the same time. All this will subtly affect the decision of whether a particle is considered bound or not as it is the particle velocity relative to the bulk velocity that matters for unbinding.

\subsection{Unbinding Procedure} \label{sec:methodsunbinding}
We have just seen that the centre and bulk velocity determination may actually form part of any unbinding procedure during which gravitationally unbound particles are iteratively removed. These two steps are therefore not necessarily separate tasks. Furthermore, the scheme for collecting particles touched upon in \Sec{sec:methodsparticlecollection} is certainly not fully disconnected from the unbinding process either: while some codes prefer to adhere to a conservative initial particle collection others rely on the fact that a stringent unbinding procedure will remove any incorrectly collected particles again. Adherents to this approach point out that if a particle is not included in the initial candidate list it can never be added back in later.

Other obvious differences come from the calculation of the potential $\phi$ entering the calculation of the escape velocity $v_{\rm esc}=\sqrt{2\phi}$ (against which each particle's velocity is compared), the order which particles are removed, and the termination criterion for the unbinding process itself. Most of the codes discussed in \citet{Onions12} calculate $\phi$ using a tree, whereas others make simplifying assumptions such as spherical symmetry (\ahf) or detailed surmises about the radial density profile (\hottd\ \& \hotsd). Some codes remove one particle at a time, always using the least bound one (\grasshopper), whereas others remove every particle considered unbound in one go before re-iterating, and yet others restart the iteration when a certain fraction of the particles have been removed. Finally there are various termination criteria for the iterations: no more particles are removed, only a negligible fraction of the particles have been removed, etc.

It should be noted that presently some codes \textit{do not} feature an unbinding procedure. The necessity for unbinding is tightly linked to the particle collection method. Configuration-space finders will always include some dynamically unrelated particles with high relative velocities \citep[see, for instance][]{Onions13}. Consequently, we argue that all configuration-space based finders, regardless of how conservative the initial particle collection is, require an unbinding step to remove false positives -- unless the scientific question(s) to be addressed are based upon all gravitating matter within the objects, e.g. lensing studies, Sachs-Wolfe effect, Sunyaev-Zeldovich effect, X-ray properties, etc. In practice, the addition of an unbinding process essentially converts any configuration-space based finder into a mock phase-space finder, as unbound particles are dynamically unrelated and will be some distance away from the object in phase-space. The nature of unbound particles means that  phase-space finders can be more reliable when it comes to picking bound particles in the first place. However, we caution that unbinding is not actually a physically motivated method of pruning the particle list \citep{Behroozi12d}. Particles can become marginally unbound momentarily, for instance when a subhalo passes through dense regions of its host halo, and later become bound. Pruning a particle list based on a particle's instantaneous binding energy will not return the mass that is dynamically associated with an object, regardless of whether the particles have been collected in configuration or phase-space. 

The need for an unbinding procedure depends not only upon the algorithm but the problem being addressed. If the parameter being quantified is not sensitive to a small fraction of interlopers (such as the total mass for field haloes) then the errors are likely to be small. However, as we show, some properties (such as halo spin) are highly sensitive to the presence of unbound particles and for such measures an unbinding procedure is essential. Additionally, halo catalogues produced by configuration-space based finders without an unbinding procedure suffer from a significant amount of contamination from spurious small objects. Consequently, the number of particles required within an object before it can be trusted is correspondingly much higher than for similar finders with an unbinding process. For this reason alone it makes sense to always utilize an unbinding procedure unless the inclusion of unbound material is specifically desired, for instance if studying diffuse streams, tidal relics, or gravitational lensing. 

Finally we would like to state that there are also similarities in the unbinding procedures adopted by all codes: we all consider the object in isolation (i.e. not embedded within an inhomogeneous background, be that a host halo or the surrounding universe itself) and we all agree that considering the Hubble flow has little if any impact (and hence the Hubble flow is not taken into account in some codes) .

\subsection{Mass \& Edge Determination} \label{sec:methodsmass}
Once a set of (gravitationally bound) particles has been found for each halo, one of the most nebulous steps arises: how to find the halo edge, a quantity which will in turn also determine its mass \citep[see, for instance, ][]{Maccio03,Prada06,Cuesta08,Anderhalden11,Diemer12}. This is an important procedure because for many purposes we require a rank ordering of our objects, whether this be by mass or size, and then subsequently attempt to find conversion relations between one property and another based on this ranking.

This topic is closely related to the aforementioned question ``what is a halo?'' (cf.~\Sec{sec:introduction}). The answer is not as straightforward as one might hope  and depends on the halo finder and the scientific questions in mind. For instance, in studies of gravitational lensing, one certainly needs to include the substructure in the host halo's mass; for (stellar) stream investigations one is actually more interested in the unbound rather than the bound particles; for shape distributions and correlations with environment spherically cut haloes are not the best choice, etc. To add to confusion already created with all the ambiguity arising from the steps presented before, let us quote here several statements from the lively discussion about this subject at the ``Subhaloes going Notts" workshop:

\begin{itemize}
 \item the halo edge is the distance to the farthest bound particle
 \item the halo edge is defined via the spherical top-hat collapse model
 \item the halo edge is the `zero-velocity' radius
 \item the halo edge is defined by the outer 3D caustic in the transformation from Lagrangian to Eulerian coordinates
 \item as a large region of the universe is bound to every object, we should simply use the first isodensity contour that goes through a saddle point
 \item an object should be defined dynamically: whatever particles stay with the object over several dynamical times are part of it
 \item do not try to define an edge, provide best-fit parameters to some function describing the density profile of each object
 \item do not try to define an edge, just provide the (bound) particle lists to the user
\end{itemize}

It will be up to the user and the actual scientific problem at hand to decide which definition serves best. But note that most finders adhere to some form of \Eq{eq:virialradius} (see \Sec{sec:definitionsmass} below). One noteworthy exception to this rule though is \origami\ \citep{Falck12} that uses the outer caustic approach to both collect particles and assign an edge to them (cf. \Sec{sec:origami}). 

\subsection{Tracking of Haloes} \label{sec:methodstracking}
Finding objects in simulations is not necessarily a task that is only limited to a single time snapshot. On the contrary, in most cases we are actually interested in the temporal evolution of our haloes. While this could be achieved by tracking objects between multiple halo catalogues separated in time, one may also think of basing the halo finder upon this approach, as has recently been done for \hbt\ \citep{Han12} or for \mht\ \citep{Gill04a} and \surv\ \citep{Tormen04,Giocoli08,Giocoli10}. A sophisticated tracking algorithm may in fact improve the accuracy and credibility of halo finders: one can use the tracking results to adjust the halo catalogues and remove spurious identification and/or recover missing objects \citep[][]{Springel01subfind,Gill04a,Tormen04,Giocoli08,Tweed09,Giocoli10,Behroozi12b,Benson12b}.

Note that most of the aforementioned papers are concerned with the proper tracking of subhaloes; they all, with the exception of \citet{Behroozi12b},  devised methods to follow subhaloes after infall into their host. \citet{Behroozi12b}, however, extended this  idea to halo catalogues in general: large, established haloes should not be expected to suddenly appear or vanish and the location of the halo centre and bulk velocity should not change by unphysical amounts between any two outputs.

\subsection{Treatment of Baryons} \label{sec:methodsbaryons}
The treatment of baryons is something that is becoming more and more important given the fact that the simulations are routinely including them these days. It is well established that baryonic physics alters the particulars of dark matter haloes and subhalo populations
\citep[e.g.][]{Blumenthal86,Tissera98,Libeskind10,Romano-Diaz10,Schewtschenko11,DiCintio11,Zemp12},
there remains the question of how this will influence the performance of an object finder. Additionally the gas particles themselves carry not only kinetic but also thermal energy giving rise to thermal pressure in the medium (specified by the adopted equation of state). At present, halo finders deal with these subtleties differently, with some accounting for the gas' thermal energy $u=\frac{3}{2} \frac{k_B}{m} T$ and others ignoring it. Finders like \ahf\ or \subfind\ add $u$ to the total specific energy $e_{\rm gas}$ (required to be negative for 
bound particles):

\begin{equation}
 e_{\rm gas} = \phi + \frac{1}{2} v^2 + u
\end{equation}

\noindent
where $\phi$ the gravitational potential and $v$ the gas particle's velocity. A different approach used by some codes (e.g. \jumpd) is to remove all the hot gas from a two-phase medium by making use of its bi-modal energy distribution.

In a recent study \citep{Knebe13} we compared a set of finders when applied to a simulation that not only models gravity, but simultaneously follows the evolution of the baryonic material by incorporating a self-consistent solution to the hydrodynamics; the simulation further included a model for star formation and stellar feedback. We found that the diffuse gas content of the haloes shows great disparity, especially for low-mass satellite galaxies. We nevertheless acknowledged that the handling of gas in halo finders is something that needs to be dealt with carefully, and the precise treatment may depend sensitively upon the scientific problem being studied. We therefore refrain from any in-depth discussions of this subject here and only will present a key plot in \Sec{sec:recoverygalaxies}; the details can be found in \citet{Knebe13}.

\subsection{Summary} \label{sec:methdossummary}
We have seen that halo finding is not as simple as passing the raw simulation data through some filter (may that be velocity or position filtering). It involves several steps starting from initially generating a putative list of halo candidates to eventually locating an edge for an object possibly defined by them. Presenting the detailed implementation of each of these steps in every single halo finder is beyond the scope of this article. We nevertheless provide in \App{app:codedescriptions} a brief descriptions of all the codes that participated in one way or the other in the comparison project; please refer to the  references therein for more details. But we would also like to highlight that there is no unique implementation: each code applies its own way of realising the necessary steps for going from the raw simulation data to the final halo catalogue. In fact, one cannot come up with a unique candidate identification or particle collection method as these parts clearly define and characterize a halo 
finder. For instance, phase-space finders usually base their particle collection on an intrinsically different algorithm than configuration-space finders (but see e.g. \hot). It should be noted that FOF based finders combine several of the steps outlined here due to their intrinsic simplicity: their candidate identification and particle selection is in practice just one step; they further may also not apply an edge definition other than the one given by the isodensity contour defined via the applied linking length and they do not intrinsically involve any unbinding step. And one should not forget that most of the methods outlined here are linked to each other. For instance, adhering to a certain edge definition method will lead to a code that upfront collects its particles in a way tailored to suite that definition. For example, a code aiming at collecting out to the zero-velocity radius certainly collects particles differently than a code using the first shell crossing approach. But the particle collection will influence the 
unbinding procedure as we will have different centres and bulk velocities to start with.

We will see below that this `freedom of realisation' will lead to unavoidable scatter when recovering halo properties with different finders. Although there are of course many possible variations, the steps outlined here underlie the architecture of almost every halo finder, and each of them will introduce some scatter in the physical properties of the haloes returned by the different algorithms. For end users, it is important to know the magnitude of the scatter associated with the most important properties of the haloes (i.e. position, velocity, and mass), to be quantified in \Sec{sec:recovery}. Advanced users -- and, above all, developers -- will also be interested in the amount of scatter due to the particular implementation of the different steps, which will be investigated in \Sec{sec:precisioncosmology} using the mass function of dark matter subhaloes as a reference test case.

It is important to note, though, that this scatter should not be confused with the discrepancies arising from the different definitions of the same quantity that may be adopted by any given algorithm. These are discussed in more detail in the following Section. However, as we will also see in \Sec{sec:recovery} and \Sec{sec:precisioncosmology} there are ways to unify the post-processing once an initial set of particles has been gathered and added to the list of putative halo centres.

\section{Definition of Halo Properties} \label{sec:definition}
Even if one had a perfect, uncontaminated set of particles associated with an object of interest representing the dark matter halo of a galaxy or a galaxy cluster, or maybe a stream of tidal debris material, it would still remain unclear how to define its physical properties.  The most relevant and fundamental are probably the position, mass, radius, and bulk velocity of the object.  All other properties (e.g. \Vmax, spin parameter, shape, velocity dispersion, concentration, etc.) are actually derived properties that mostly require the determination of the position and bulk velocity in order to place the associated particles into the rest frame of the halo.

This is an important Section even for end users of halo finders, as not every code uses the same definitions for extracting halo properties. This leads to different, yet still internally correct, results.  The general user should be aware that the adopted definitions will have a significant impact on the final halo catalogues.


\subsection{Centre Position \& Bulk Velocity} \label{sec:definitionscentre}
Most finders use the peak of the local (phase-space) density field to define the centre of a halo.  Its spatial location and bulk velocity are determined by either the (weighted) average over all the (bound) particles in the object or only a certain `central' fraction of them. The chosen fraction, as well as the criterion to define `central' (e.g. spatial and/or velocity distance, binding energy, etc) are specific to each halo finder.  Usually -- but not always -- the same prescription is applied to field haloes and subhaloes.

The position and velocity of the centre play an important role in several of the steps described in \Sec{sec:methods}, as well as on some of the other properties of the halo (see \Sec{sec:recovery}):  differences of the order of ten per cent in the centre and bulk velocity are expected when using different reference frames \citep[][cf. also \Sec{sec:precisioncosmologycm}]{Ascasibar08,Knebe11,Han12}.

\subsection{Mass \& Edge definition} \label{sec:definitionsmass}
In \Sec{sec:methodsmass} we have raised the more general problem of determining an object's extent once the problem of associating particles to it has been accomplished. Different codes might use different approaches of which some had been sketched in the bullet-item list in that Sub-Section. Here we simply like to go one step further highlighting that even adhering to one of these methods, e.g. the commonly used virial edge definition via the spherical top-hat collapse, might lead to degeneracies in halo mass.

While most FOF-based finders simply report the cumulative mass of all linked particles and derive the radius of a spherical region equivalent to the total extent of the halo, other finders use the following working definition for both halo mass and radius:
\begin{equation} \label{eq:virialradius}
  \frac{M_{\rm ref}(<R_{\rm ref})}{\displaystyle \frac{4\pi}{3} R_{\rm ref}^3} = \Delta_{\rm ref} \times \rho_{\rm ref} \ ,
\end{equation}
where $\Delta_{\rm ref}$ is a parameter (usually determined from the spherical top-hat collapse of an overdense region in an expanding universe, and hence a function of the cosmological parameters $\Omega_m$ and $\Omega_{\Lambda}$, as well as the redshift $z$; see, for instance, \citealp{Courtin11}) and $\rho_{\rm ref}$ is a reference density (normally either the critical density $\rho_{\rm crit}=3H^2/8\pi G$ or the background density $\rho_b = \Omega_m \times \rho_{\rm crit}$).  The `freedom' in choosing these two parameters already hints at the possible ambiguities in the location of the halo edge (and therefore mass).  It must be stressed, though, that there is no right or wrong way; users of halo finder catalogues just need to be aware that several alternative definitions exist and which one of these has been used. For a relation between SO and FOF masses as defined above, the reader is referred to \citet{Lukic09} and \citet{More11}; and for a thorough discussion of different choices for $\Delta_{\rm ref}
$ and $\rho_{\rm ref}$ please refer to, for instance, \citet[][]{Maughan06} and the Appendix in \citet{Sembolini12}, respectively. But also note that we are not entering the discussion here about the applicability of such a definition and its implications for the redshift evolution of halo mass. We just like to state that \Eq{eq:virialradius} may not be the appropriate choice in the end as it leads to spurious (and unphysical) evolution as, for instance, shown and discussed by \citet{Diemand07b}, \citet{Diemer12}, and \citet{Kravtsov12}. Those authors have shown that even though the physical density profile remains constant over time, the evolution of the reference density with redshift causes changes in the mass of the object. To avoid such ambiguities it might therefore be more meaningful to use intrinsic scales to characterize and quantify the mass and size of haloes such as, for instance, \Vmax\ and \Rmax\ \citep[][]{Ascasibar08, Knebe11} -- as already advocated before in \Sec{sec:whatisahalo}.

Analogously, not all finders consider the mass of a subhalo to be part of the mass of the host halo. Which definition is to be used depends on the scientific problem at hand, but the end user needs to be aware of what the code returns.  The mass of the subhaloes themselves depends on how their particles are collected.  Again, some codes identify a spherical `tidal radius', whereas others use isodensity contours (or other prescriptions) to define the subhalo edge.

\subsection{Derived Properties} \label{sec:definitionsderivedproperties}
The position, velocity, mass, and radius of a halo are the basic properties to be returned by any halo finder.  However, most codes provide additional information (e.g. \Vmax, spin parameter, concentration, shape, etc.). We will discuss here some of the quantities that we consider especially relevant for a large number of users of halo catalogues.  The relation to several particular fields is explored in more detail in \Sec{sec:relationandapplication} below. Note that, since most of these properties depend on the reference frame of the halo, their actual values may be subtly dependent on the adopted prescriptions.

\subsubsection{Rotation Curve}
As already mentioned in \Sec{sec:whatisahalo}, the peak of the circular rotation curve
\begin{equation}
  V_{\rm max} = \max_{r}  \sqrt{\frac{GM(<r)}{r}}  
\end{equation}
may be a more physically meaningful and stable measure of halo mass than any value based upon an ambiguous edge definition (especially for subhaloes).  Moreover, this quantity is much closer to the observational data, as it is possible to measure rotation curves (and hence \Vmax), whereas all the ambiguities related to the outer edge of a galaxy apply to observations as well. 

On the other hand, measuring \Vmax\ requires sufficient resolution to determine the circular velocity accurately enough:  the peak position \Rmax\ will always be reached relatively close to the centre of the object.  In addition, it should be mentioned that the actual procedure for the determination of \Vmax\ could be considered to be part of the `methods' of the halo finder: some codes directly use the list of particles sorted in distance from the halo centre, with or without smoothing, and locate the peak via some form of interpolation; other codes prefer to bin $M(<r)$ prior to the peak determination. These choices again introduce subtle differences in the derived quantity. Further, while \Vmax\ might be easily determined, \Rmax\ is more ambiguous as the velocity profile can be quite flat. And as \Vmax\ itself is measured closer to the halo's centre than an edge-based mass, it likely is more affected by the details of gas physics, star formation, and feedback than the (virial) mass, so the differences in this quantity between $N$-body vs. hydrodynamic simulations could be significant \citep[e.g.][]{DiCintio11}.

The main message is that, due to the noise inherent to the mass profile, the \emph{practical} definition of \Vmax\ and \Rmax\ implemented in each halo finder is not as simple as `the maximum of the rotation curve'.  The scatter due to the different definitions/methods will be discussed below in \Sec{sec:recovery}.

\subsubsection{Spin}
There are two commonly used definitions for the spin parameter of a halo 

\begin{equation} \label{eq:spinparameter}
\begin{array}{lcl}
\lambda_P & = & \displaystyle  \frac{L \ \sqrt{|E|}}{G M^{5/2}}\\
\\
\lambda_B & = & \displaystyle  \frac{L}{\sqrt{2} M V R}\\
\end{array}
\end{equation}

\noindent
where 

\begin{equation}
{\bf L} = \sum_{i=1}^N m_i {\bf r}_i \times {\bf v}_i
\end{equation}
is the angular momentum vector of all $N$ particles in the halo, $M$ is the virial mass enclosed at a virial radius $R$, $V=\sqrt{GM/R}$ the circular velocity at $R$, and $E$ its total energy. The former is the classical definition originally introduced by \citet[][subscript $P$]{Peebles69} and the latter a simplification of it first introduced by \citet[][subscript $B$]{Bullock01b} (which reduces to the standard $\lambda_P$ when measured at the virial radius of a truncated singular isothermal halo).

The spin parameter can be seen to be a measure of the amount of coherent rotation in a system compared to random motions. For a spherical object, it is approximately the ratio of its own angular velocity to the angular velocity needed for it to be supported against gravity solely by rotation \citep[see e.g.][]{Padmanabhan93}. A detailed account of the merits and drawbacks of the two alternative definitions of the spin parameter is provided in \citep{Hetznecker06}.

\subsubsection{Shape}
Having identified the set of particles belonging to (and defining the shape of) an object, it is common practice to compute their moment of inertia tensor $I_{jk}$.  For a distribution of discrete point masses, $I_{jk}$ is expressed as
\begin{equation}
\label{eq:Inertia} 
I_{jk} = \sum_{i=1}^N m_i (r^2_i \delta_{jk} - x_{ij}x_{ik}) \quad \ 
	{\rm with}\ j,k = \{1;2;3\} \quad, 
\end{equation} 
\noindent 
where $m_i$ is the mass of particle $i$, $N$ the number of particles and $r_i = \sqrt{x_{i1}^2+x_{i2}^2+x_{i3}^2}$ is the distance of particle $i$ from the centre of mass of the particles. The eigenvalues and eigenvectors of $I_{jk}$ are related to the axis ratios and orientation, respectively, of the ellipsoid that fits best
the particle distribution. A similar ellipsoid can be obtained from the tensor
\begin{equation}
  \label{eq:commoninertiatensor}
	{\cal M}_{jk} = \sum_{i=1}^N m_i x_{ij}x_{ik}\ ,
\end{equation} 
which has been widely used in previous studies.  Both forms provide axis ratios and orientations that are identical (though the individual eigenvalues are different).

Note that the simplest method determines the shape and orientation of a halo using all particles within a spherical volume or shell at a given radius \cite[e.g.][]{Frenk88,Kasun05,Hopkins05, Bailin05}. While this method robustly recovers the orientation of the halo, the resulting axis ratios tend to be biased towards larger values (i.e. haloes are predicted to be rounder).

An alternative, iterative approach to the problem consists in using all the particles within a spherical volume or shell, but the initial surface is deformed along the principal axes of the best-fitting ellipsoid, and the process is repeated until convergence is reached
\cite[e.g.][]{Dubinski91,Warren92,Allgood06,Vera-Ciro11}.
Both \cite{Jing02} and \cite{Bailin05} have noted that iterative methods have difficulty in achieving convergence in simulations in which haloes are very well resolved and contain a population of satellites. Satellites tend to lead to a distortion of the shape, and this is most pronounced in the outermost parts of host haloes, where recently accreted satellites are most likely to be found.

The impact of substructure can be reduced by working with the reduced inertia tensor
\begin{equation}
\label{eq:reducedinertiatensor}
\hat{{\cal M}}_{jk} = \sum_{i=1}^N \frac{m_i x_{ij}x_{ik}}{r_i^2}\ .
\end{equation}
where each particle is weighted by the inverse square of its distance to the centre of the halo. While this recovers accurately the orientation of the ellipsoid, the axis ratios are systematically overestimated, and thus haloes are predicted to be more spherical than they actually are \cite[e.g.][]{Bailin04}.  For a more elaborate discussion of all these possibilities we refer the reader to a recent study by \citet{Zemp11}; here we would only like to reiterate that there is more than one definition of halo shape.

\subsection{Summary} \label{sec:summarydefinitions}
Not only will halo finders vary in the method used to determine certain properties, such as position, mass, and bulk velocity (as covered in \Sec{sec:methods}); they may also use different definitions as discussed here. While the precise way to gather the particles belonging to an object is indeed the essence of the halo finder, the exact definition of its physical properties could in principle be passed on to the end user by supplying just the associated particle lists and/or physically-motivated fits to the particle distribution.  Arguably, this would impose an unnecessary burden on the user, and it is normally considered much more convenient that the halo finder returns actual numeric values for halo properties, according to any particular definition of its choice (that should hopefully be explicitly stated in the halo finder documentation).  It is the responsibility of the user to understand those definitions and use them consistently when comparing to other numerical, observational, or analytical work. 
Conversion factors or more elaborate recipes may need to be applied to switch from one definition to another and have been the subject of various investigations in the literature.

\section{Recovery of Halo Properties} \label{sec:recovery}
In this section we address the following question: do halo finders (dis-)agree when applied to identical data sets? More precisely, we would like to discuss the scatter in the fundamental properties returned by \emph{any} halo finder (i.e. position, velocity, and mass of each object) as well as in some of the most popular derived quantities, such as the maximum of the circular velocity, halo shapes, spin parameter, and halo number counts as a function of mass or circular velocity. Much in the spirit of previous comparison papers \citep{Knebe11, Onions12}, from which several analyses will be borrowed and extended, we will use the scatter of the values recovered by different codes as a first attempt to quantify the uncertainties that are nowadays associated to the process of halo finding. The origin of such scatter, and possible ways to reduce it, will be discussed in \Sec{sec:precisioncosmology}.

For this project we have utilised pre-existing datasets from a range of sources. Note that all comparisons will be done using redshift $z=0$ data only. \Tab{tab:recoverydatasets} summarises which datasets are used in each of the following subsections.

\begin{table*}
  \caption{A recap of the data used for the distinct Sub-Sections. $B$ is the side lenght of the computational box, $\Omega_m$ the total matter density parameter, $\Omega_\Lambda$ the vacuum energy density parameter, $\sigma_8$ the normalisation of the input power spectrum of density perturbations at redshift $z=0$, $m_p$ is the particle mass, $\epsilon$ is the Plummer-equivalent gravitational softening length.}
\label{tab:recoverydatasets}
\begin{center}
\begin{tabular}{ll}
\hline
\hline
  Sub-Section & data set and its particulars\\
\hline
\hline
  \ref{sec:recoveryfieldhaloes} -- Field Haloes		& $-$ MareNostrum (large-scale structure) simulation at $z=0$ \citep{Gottloeber06, Knebe11}:\\
   & \ \ \ \ \ $B=500$\hMpc, $\Omega_m=0.3, \Omega_{\Lambda}=0.7, \sigma_8=0.9, m_p=9.8\times 10^{9}$\hMsun$, \epsilon=15$\hkpc \\
    & $-$ each halo finder returned its own analysis\\
\hline
\ref{sec:recoverysubhaloes} -- Sub-Haloes		& $-$ Aquarius A-4 (zoom simulation of Milky Way type halo) at $z=0$ \citep{Springel08,Onions12}:\\
   & \ \ \ \ \ $B=100$\hMpc, $\Omega_m=0.25, \Omega_{\Lambda}=0.75, \sigma_8=0.9, m_p=2.9\times 10^{5}$\hMsun$, \epsilon=0.25$\hkpc \\
    & $-$ common post-processing of supplied particle ID lists\\
\hline
~\ref{sec:recoveryposvel} -- \ref{sec:recoveryshapespin} -- Error Quantification	& $-$ Aquarius A-4 at $z=0$ \citep{Springel08,Onions12}: see above for details\\
    & $-$ common post-processing of supplied particle ID lists\\
    & $-$ sub-set of halo finders featuring reliable unbinding\\
    & $-$ sub-set of subhaloes commonly found by all finders\\
\hline
\hline
\end{tabular}
\end{center}
\end{table*}

\subsection{Field Haloes}\label{sec:recoveryfieldhaloes}
 \begin{figure*}
   \includegraphics[width=\columnwidth]{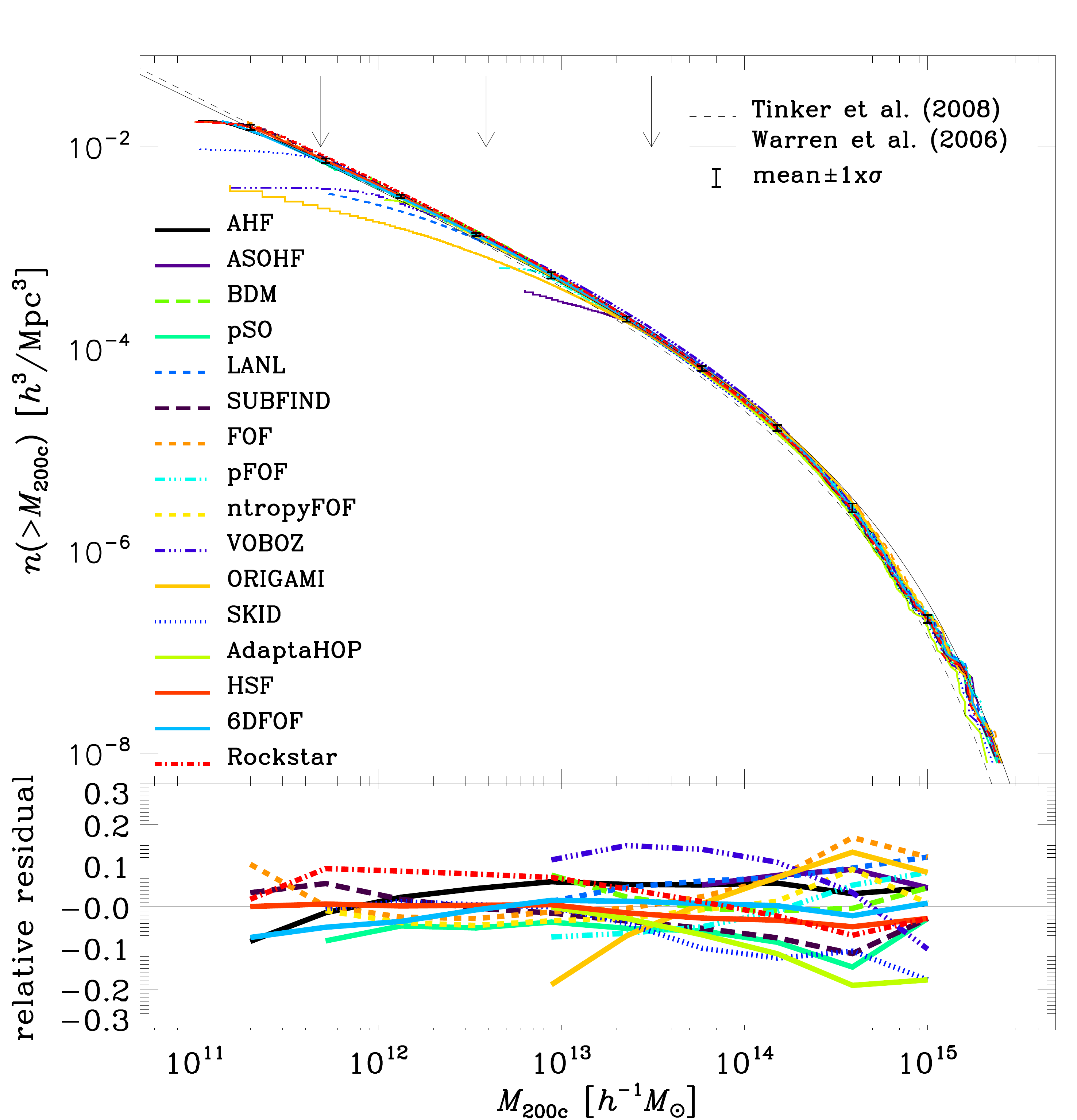}
   \includegraphics[width=\columnwidth]{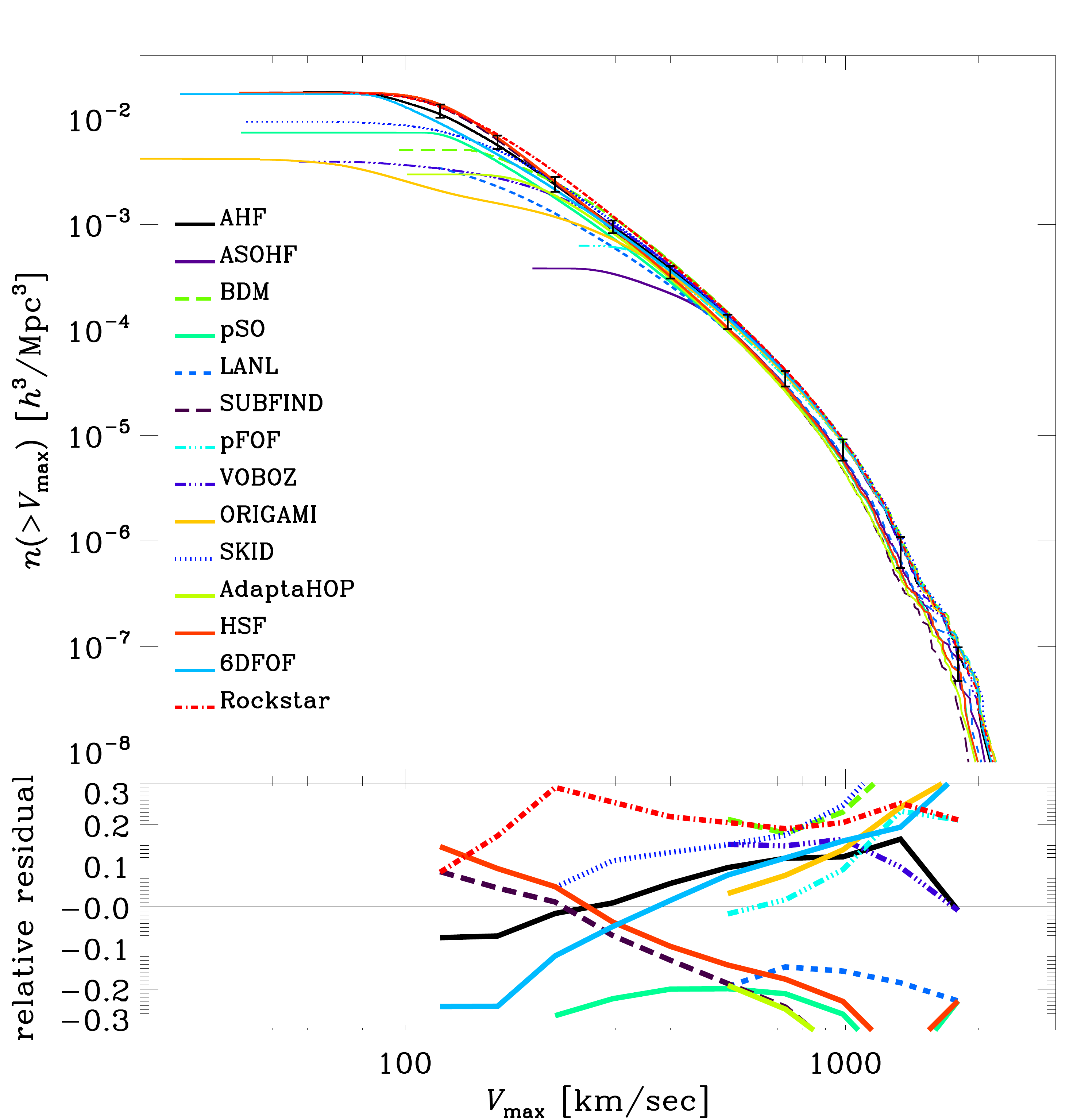}
   \caption{\textbf{Comparison of field haloes.} Upper left panel: the cumulative mass ($M_{200c}$)
   function. The arrows indicate the 50 particle limit for the
   1024$^3$ (left), 512$^3$ (middle), and 256$^3$ (right) simulation
   data. The thin black lines crossing the whole plot corresponds to
   the mass function as determined by \citet[][(solid)]{Warren06} and
   \citet[][(dashed)]{Tinker08}. The error bars represent the mean
   mass function of the codes ($\pm 1 \sigma$).  Lower left panel: the
   fractional difference of the mean and code halo mass
   functions. Upper right panel:  Cumulative number count of haloes above the indicated
   \Vmax\ value. Lower right panel: the
   relative offset from the mean of the cumulative count. The pair of
   solid lines in each of the residual plots simply indicates the 10~per cent
   error bars. Note that both properties (i.e. mass and \Vmax) have
   been determined individually by each code.}
 \label{fig:fieldhaloes}
 \end{figure*}

We begin by discussing field haloes extracted by the finders from the MareNostrum simulation \citep{Gottloeber06} at a range of resolutions and previously discussed by \citet{Knebe11}. This is perhaps the easiest scenario for catalogue generation as at this numerical resolution there is effectively little substructure and the vast majority of the haloes found are isolated. In this case the choices made and discussed above are not as crucial as we shall see later and the different finders generally agree well even if no common post-processing pipeline is employed and we just take the mass and velocity values returned directly by each group. Even the lack of any unbinding procedure in some codes has little impact as for a general halo this removes very few particles as the haloes themselves {\it are} the background.

\Fig{fig:fieldhaloes} shows in the upper panels the cumulative mass\footnote{Defined by using $\Delta_{\rm ref}=200$ and $\rho_{\rm ref}=\rho_{\rm crit}$ for \Eq{eq:virialradius}} $M_{200c}$ (left) and \Vmax\ (right) functions alongside the mean and 1-$\sigma$ standard variation for a selection of mass/\Vmax\ points as error bars; different finders are encoded using a combination of colour and linestyle. The lower panels show the scatter of each halo finder about those mean values. Note that these plots are showing results at various mass resolution levels \citep[i.e. 1024$^3$, 512$^3$, and 256$^3$ particles, see][for more details]{Knebe11} as not all finders have the capability to analyse the largest data set; the vertical arrows in the mass function indicate the 50 particle limit for the respective resolution. The two thin lines in the upper left mass function panel represent two analytical mass functions based upon fits to the numerical mass functions found in cosmological simulations: \citet{Warren06}, who use a FOF-based finder for their best-fit model, and \citet{Tinker08}, who applied an SO-finder. The difference between these "semi-theoretical" functions stems from the fact they are originally based on fits to numerical mass functions derived using different halo finders. Therefore it only appears natural that they span the scatter seen in \Fig{fig:fieldhaloes}, i.e. a non-unified post-processing of the halo catalogues in a cosmological box.\footnote{We encourage the interested reader to confirm this by using the online mass function calculator \url{http://hmf.icrar.org} where every analytical mass function from the literature can be calculated and plotted against each other \citep[see also][]{Murray13}.}

We find that the scatter in mass is at the $10$~per cent level and actually within the limits given by the two analytical functions. This scatter is driven by a variety of sources and is due to the different choices made by the finders. In particular the finders differ on whether or not they include the mass of any substructures in the halo mass, whether or not they do unbinding and the precise definition both the outer edge and halo centre. We note that the differences in \Vmax\ are -- in the case that each code uses its own method to determine \Rmax\ and \Vmax\ -- substantially larger than we shall see later and of order $20-30$~per cent. 

This level of accuracy may be perfectly acceptable for some measurements and indeed within any particular code where the assumptions employed are conserved convergence would be expected at a much higher level. However, these errors should be indicative of the level of accuracy at which we can {\it absolutely} measure the cumulative halo mass function within a cosmological model given that, as we stressed earlier, all of the range of assumptions adopted here are perfectly physically acceptable and it is hard to argue one set is any better than another.

\subsection{Sub-Haloes} \label{sec:recoverysubhaloes}
 \begin{figure*}
   \includegraphics[width=\columnwidth]{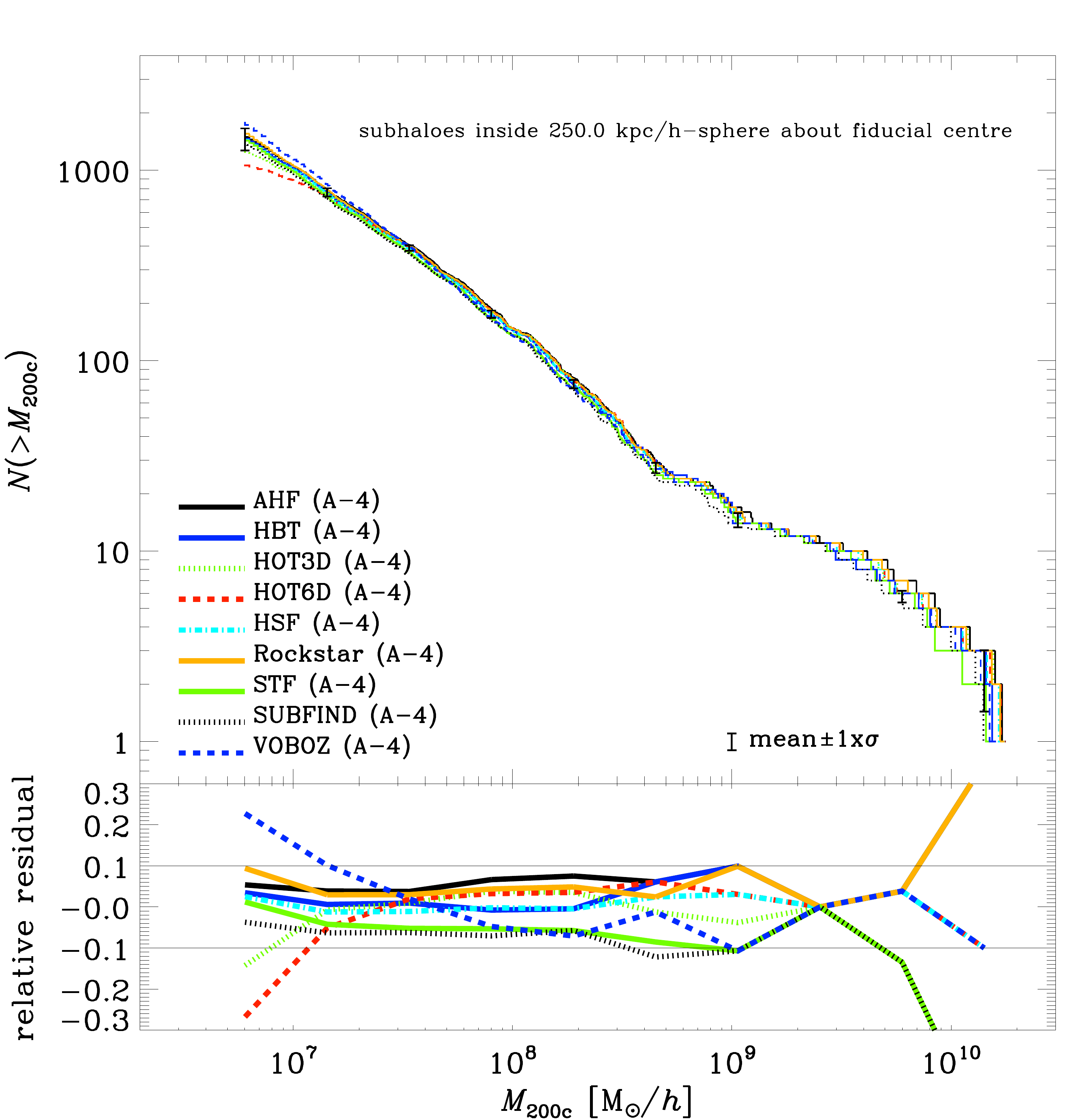}
   \includegraphics[width=\columnwidth]{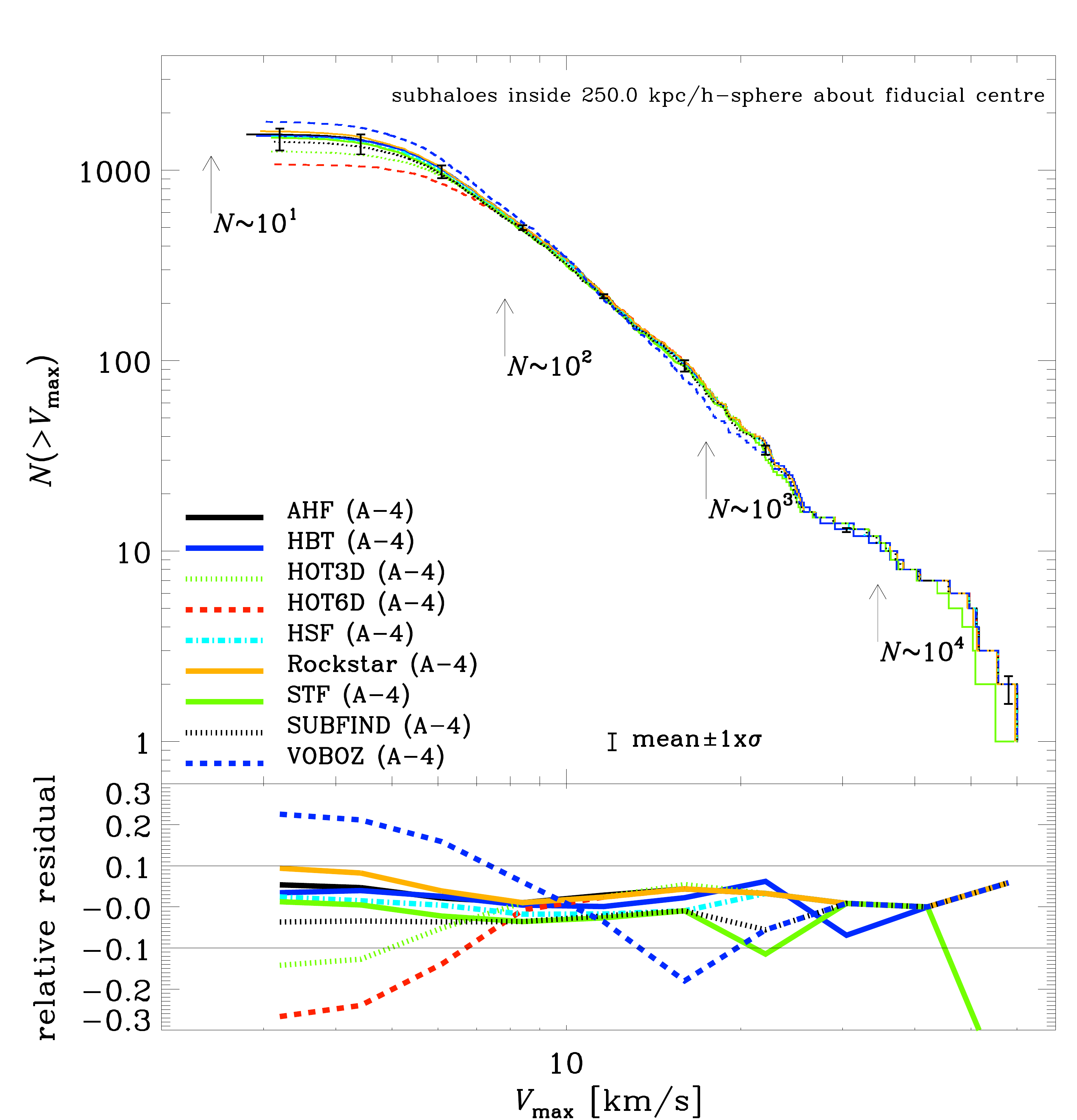}
   \caption{\textbf{Comparison of subhaloes.} Cumulative mass $M_{200c}$ (upper left) and \Vmax\ (upper right)
   functions for the subhaloes on Level 4 of the Aquarius host A
   \citep{Springel08}. The arrows in the \Vmax\ function indicate the
   number of particles interior to \Rmax, the position of the peak of
   the rotation curve.  The error bars represent the mean mass and
   \Vmax\ function, respectively, of the codes ($\pm 1 \sigma$).
   Lower left panel: the fractional difference of the mean and code
   subhalo mass functions.  Lower right panel: the relative offset
   from the mean of the cumulative count.  The pair of solid lines in
   each of the residual plots simply indicates the 10~per cent error
   bars. Note that both properties (i.e. mass and \Vmax) have been
   determined by a common post-processing pipeline.}
 \label{fig:subhaloes}
 \end{figure*}

For the rest of this section we will employ a more challenging and realistic dataset to answer the question: how well could we expect to do if we force the finders to use a common set of definitions? The Aquarius simulations \citep{Springel08} are a set of Milky Way sized haloes studied at a range of resolutions. We have processed the A-4 dataset using a wide range of substructure finders and compared the results in \citet{Onions12}. This is a more difficult problem as now a single host halo contains several thousand subhaloes and it is the properties of these we wish to compare. Using the same ordering of panels as for \Fig{fig:fieldhaloes}, we show in \Fig{fig:subhaloes} the subhaloes' mass $M_{200c}$ and \Vmax\ functions. In contrast to the field halo comparison, the mass and \Vmax\ were calculated using a common post-processing pipeline, i.e. \textit{only} the candidate identification, particle collection, and unbinding procedure were different (cf. \Sec{sec:methods}). For more details about this pipeline and the way to calculate $M_{200c}$ and \Vmax\ we refer the reader to Section 4.1 of \citet{Onions12} where also the choice of using $M_{500c}$ is discussed.

We find that for this more complex problem a successful implementation of unbinding is essential in order to obtain reliable number counts anywhere near the resolution threshold. Note that the two finders \adaptahop\ and \mendieta\ do not feature a (reliable) unbinding procedure: \adaptahop\ (without any unbinding) finds far too many small objects; \mendieta\  does not contain a reliable unbinding procedure and hence finds too few objects across a large range in mass. If we were to use a common unbinding scheme for both their pre-unbinding datasets their results then agree with the majority of the finders. For these reasons we drop both \adaptahop\ and \mendieta\ results from the rest of this discussion. 

Neglecting the results from these two finders we see in \Fig{fig:subhaloes} that the scatter for the cumulative mass function is roughly similar to the field halo case studied in \Fig{fig:fieldhaloes} despite the fact we are now using a common post-processing routine. However, the scatter in the \Vmax\ function is considerably less than in \Fig{fig:fieldhaloes} (for subhaloes composed of more than 100 particles). Both of these results are to be expected: the mass of a subhalo is sensitive to both the particle collection scheme and the unbinding procedure whereas the maximum circular velocity is less sensitive to these assumptions as this quantity only depends on a small fraction of the most central particles.

However, one may raise the issue that differences in the scatter can be due to either moving from field to sub-haloes or the fact that the processing of the particles has been outsourced. To shed some more light into this we also calculated the subhalo mass function (i.e. left panel of \Fig{fig:subhaloes}) for the values directly returned by the respective code (not shown here though): we found that the differences are minuscule and hence the scatter -- at least in mass -- is not driven by the implementation to calculate it. As \Vmax\ values have not been returned by the finders themselves we are unfortunately unable to draw any conclusions about the reduction in scatter seen in the right panels of \Fig{fig:fieldhaloes} and \Fig{fig:subhaloes} due to our common post-processing.

Further, for subhaloes, which can be significantly tidally stripped, it is not guaranteed that the mass profile reaches the peak of the velocity curve. That is, the subhalo's radius  can be smaller than \Rmax. Can we hence trust the \Vmax\ values presented here? First, this will affect all subhaloes for all finders equally due to our common post-processing. Second, we actually checked the ratio $R_{\rm max}/R$ and found  it to never exceed 0.5, i.e.  \Rmax\ is always substantially smaller than the subhalo's radius. However, we also acknowledge that our \Rmax\ (and \Vmax) values are based upon a single snapshot analysis at redshift $z=0$, i.e. after a subhalo entered the influence of its host and experienced tidal stripping; therefore, the values reported here are solely based upon the present mass profile of the subhalo and do not necessarily reflect the original ones prior to infall.

We would like to close with a cautionary remark about the relative residual curves presented in the lower panels of \Fig{fig:fieldhaloes} and \Fig{fig:subhaloes}: these ratios are actually measuring the difference in the number of objects found by a certain halo finder above a given mass or \Vmax\ threshold, respectively. They do \emph{not} directly measure the differences in mass or \Vmax. In that regard, there are two errors entering into these residuals: variations in the number of identified haloes (above a threshold) and differences in the recovered mass of the same object between finders. The following Sub-Section will now focus on the latter effect, quantifying the scatter across finders for \emph{the same} object.

\subsubsection{A Common Set of Objects} \label{sec:recoveryset}
\begin{table}
  \caption{Total number of subhaloes and the ones found in excess of
  the common set of 823 objects of the Aquarius A-4 data set. All subhaloes are requested to contain 20 or more particles and have their centres
  within a sphere of radius 250\hkpc\ from the fiducial centre. \Fig{fig:superplot-mass-level4-nbins4} indicates that the majority of these missing objects are small, containing less than 200 particles. 
}
\label{tab:excessobjects}
\begin{center}
\begin{tabular}{lcc}
\hline
code 		& total number of objects & `excess' objects\\
\hline
\ahf			& 1599	& 776\\
\hbt			& 1544	& 721\\
\hottd		& 1265	& 442\\
\hotsd		& 1075	& 252\\
\hsf			& 1544	& 721\\
\rockstar		& 1707	& 884\\
\stf			& 1521	& 698\\
\subfind		& 1433	& 610\\
\voboz		& 1863	& 1040\\

\hline
\end{tabular}
\end{center}
\end{table}

Before quantifying the actual deviations between various properties, we want to define a set of objects that could be used for this purpose. Note that distribution functions will not only suffer from differences in individual halo properties but also encode the fact that some finders may have identified different numbers of objects. To circumvent this we aim at directly comparing quantities on a halo-to-halo basis and move on from general distribution functions and their variations as discussed above. 

To cross-identify objects we use a halo matching technique that correlates all haloes found by a given halo finder to the catalogue of another finder by examining the particle ID lists and maximizing the merit function $C=N_{\rm shared}^2/(N_{1}N_{2})$, where $N_{\rm shared}$ is the number of particles shared by two objects, and $N_{1}$ and $N_{2}$ are the number of particles in each object, respectively \citep[see e.g.][for more details,  as well as \App{app:crosscorrelations} for different merit functions]{Klimentowski10, Libeskind10}. By restricting ourselves to the set of objects found by \textit{every} halo finder we are able to directly compare the properties of \textit{the same object} across all finders. We will discuss the excess objects in Section~\ref{sec:precisioncosmologycandidates} and caution reader that this common set can be dictated by one finder not finding a sufficient number of haloes in the first place. 

Please note that for this analysis we also only used the ``Subhaloes going Notts" data set; this project featured a common post-processing pipeline based upon individual particle ID lists.  These lists make subhalo cross-matching as outlined above feasible. In order to avoid any possible ambiguities with the exact definition of position, bulk velocity, mass, and \Vmax\ calculation implemented by every algorithm, participants were asked to return only the lists of those particles that they consider bound/belonging to each object; the centre, bulk velocity, edge/mass, as well as various derived quantities were then calculated by a common post-processing pipeline, i.e. positions are iteratively determined centre-of-masses using the innermost 50~per cent of particles, the bulk velocity is the mean velocity of all particles, the mass corresponds to $M_{200c}$ (as defined by \Eq{eq:virialradius} when applying $\Delta_{\rm ref}=200$ and $\rho_{\rm ref}=\rho_{\rm crit}$), \Vmax\ is the peak value of the rotation curve, the 
shape is the ratio between the smallest and largest eigenvalue of the moment of inertia tensor $\mathcal{M}_{jk}$ (cf. \Eq{eq:commoninertiatensor}), and the spin parameter $\lambda_B$ as given in \Eq{eq:spinparameter}. This approach and the use of a common data set, which might be biased towards rather clean subhaloes that are easier to detect, means that the scatter reported here should be considered lower limits.

The plots in the following Sub-Sections~\ref{sec:recoveryposvel} through to \ref{sec:recoveryshapespin} now all follow the scheme: the $x$-axis shows the median of the subhalo mass ${\rm med}(M)$ whereas the $y$-axis gives the normalized difference between the lower and upper percentiles equivalent to the 3rd and 7th ranked of the distribution across all nine (sub-)halo finders. We deliberately chose to use medians and percentiles as the distribution of properties across finders is highly non-Gaussian and at times biased by just one or two outliers. The plots further show medians in four mass bins as histograms to highlight any possible dependence on mass. And those points for which the difference between the 3rd and 7th percentile is zero are shown at the bottom of the $y$-axis. The number of cross-matched subhaloes is 823 and should be compared against the total number of objects found by each individual (sub-)halo finder given in Table~2 of \citet[][Aq-A-4 row]{Onions12}; however, for convenience we list here in \Tab{tab:excessobjects} the number of subhaloes found by each code in excess of the common 823 objects. \Fig{fig:superplot-mass-level4-nbins4} indicates that the majority of these missing objects are small, containing less than 200 particles. 

\subsubsection{Position \& Bulk Velocity} \label{sec:recoveryposvel}
\begin{figure}
  \includegraphics[width=\columnwidth]{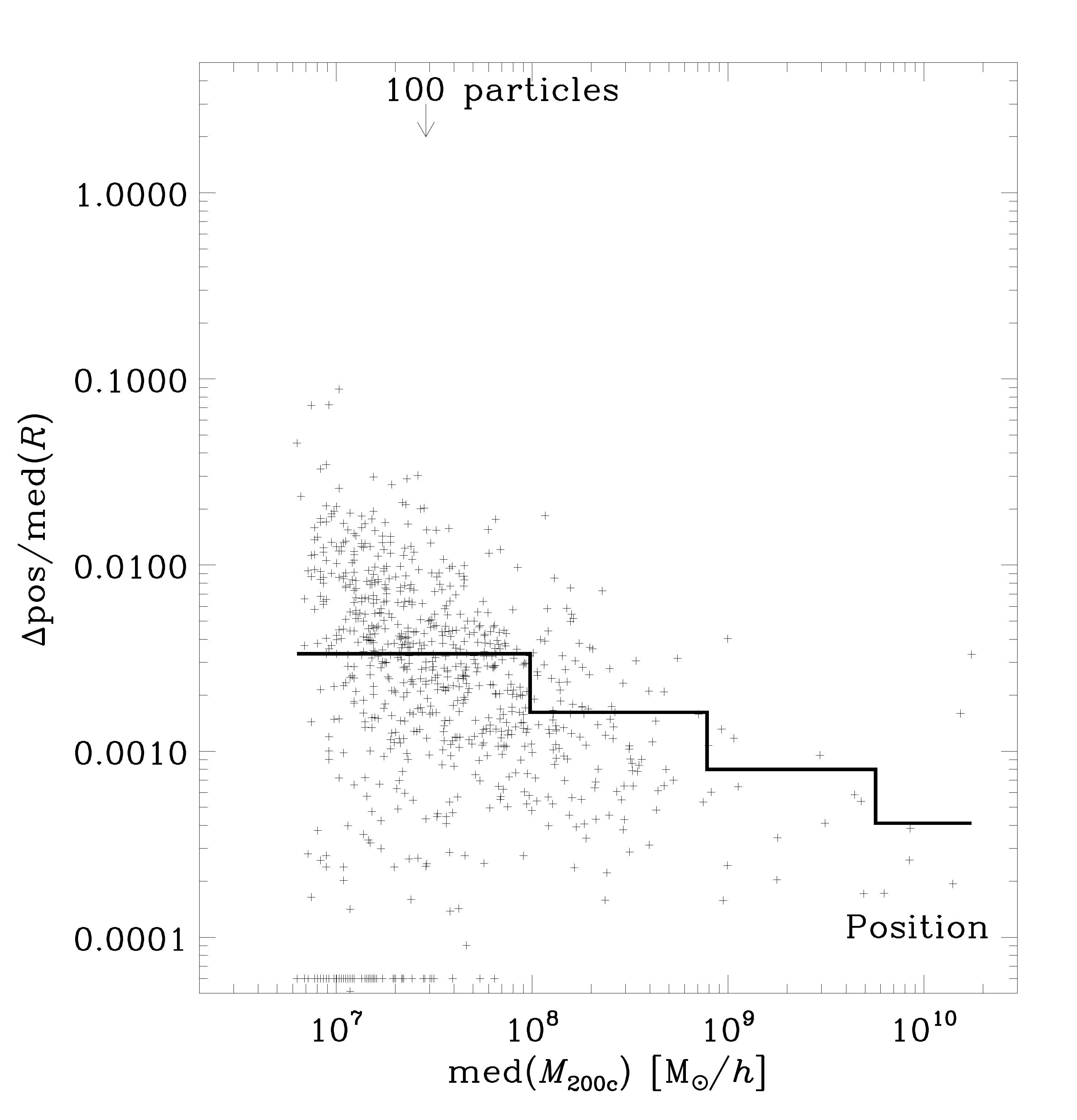}
  \includegraphics[width=\columnwidth]{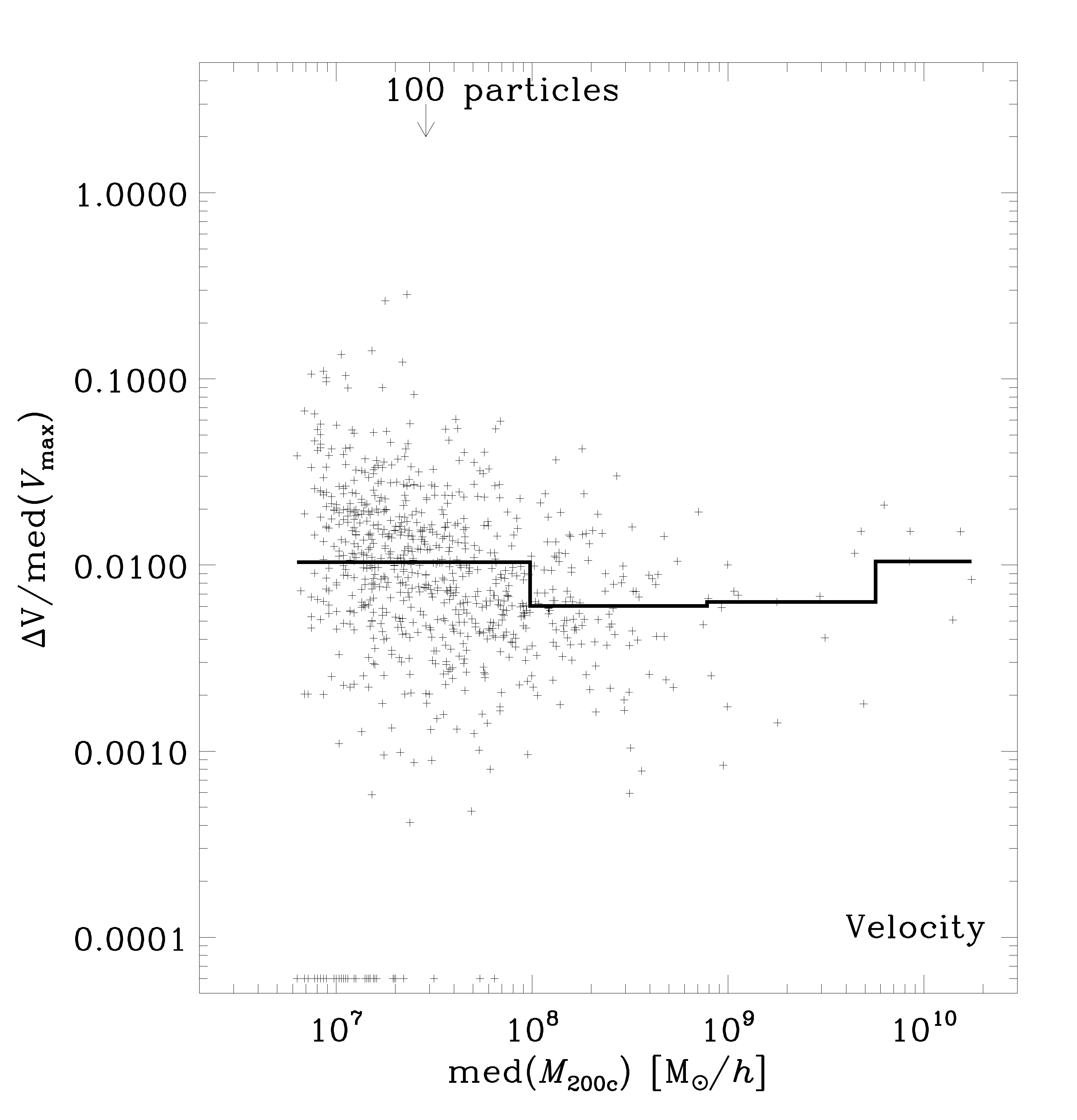}
  \caption{Relative errors in the recovered positions (top) and
  velocities (bottom) of the subhaloes found by all finders. The errors
  are scaled by the median size of the object and the median \Vmax, respectively.}
\label{fig:errorPosVel}
\end{figure}
                                                                                                                                                                                                                                                                                                                                                                                                                                                                                                                                                                                                                                                                                                                                                               
In \Fig{fig:errorPosVel} we start by inspecting the errors in position and velocity with the former deviation $\Delta_{\rm pos}$ normalized by the median radius for the object, ${\rm med}(R)$ and the latter $\Delta_{\rm vel}$ by the median of the peak of the object's rotation curve, ${\rm med}$(\Vmax). We can see a trend for both variations to decrease for more massive objects (especially for the position), but the errors are rarely larger than a few percent with the overall median error being 0.2~per cent and 0.8~per cent for position and velocity, respectively. The trend with mass reflects the resolution dependence  of the accuracy of both the position and bulk velocity of the (sub-)haloes. This scatter is consistent to that observed in \citet{Knebe11} for mock haloes.

\subsubsection{Mass} \label{sec:recoverymass}
\begin{figure}
  \includegraphics[width=\columnwidth]{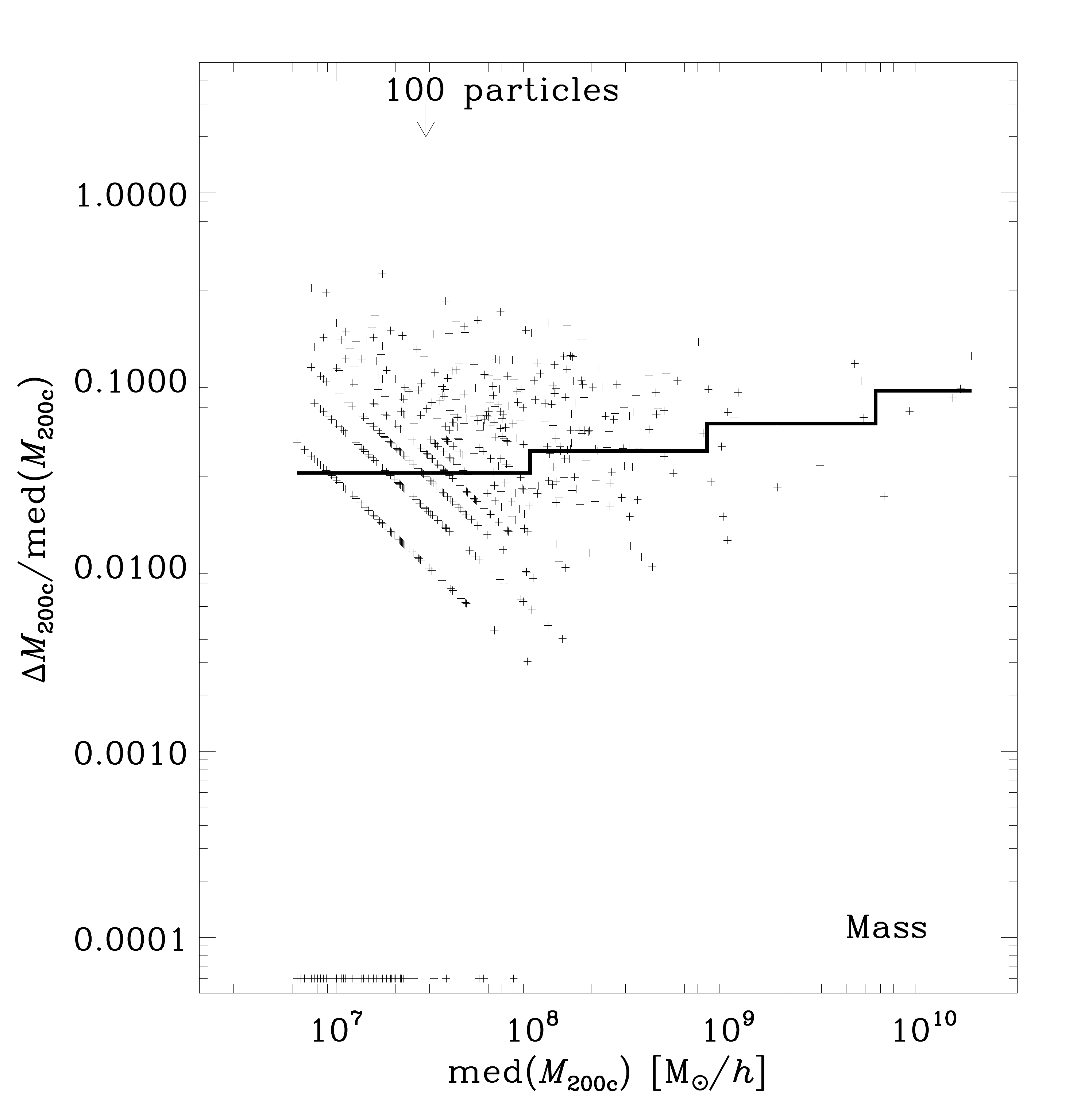}
  \caption{Relative errors in the recovered mass for the subhaloes
  found by all the finders. The errors for each object are scaled by
  the median mass found for the object.}
\label{fig:recoverymass}
\end{figure}

The recovery of subhalo mass is presented in \Fig{fig:recoverymass}. We no longer see a prominent trend with mass anymore. The discrete nature of the particle masses is evident; the difference $\Delta M$ (again normalized by the median of the mass itself) can only be a multiple of the actual particle mass which gives rise to the diagonal stripes visible in the plot for lower-mass objects.  The overall median of the scatter is found to be 3~per cent. 

\subsubsection{\Vmax\ \& \Rmax} \label{sec:recoveryrmaxvmax}
\begin{figure}
  \includegraphics[width=\columnwidth]{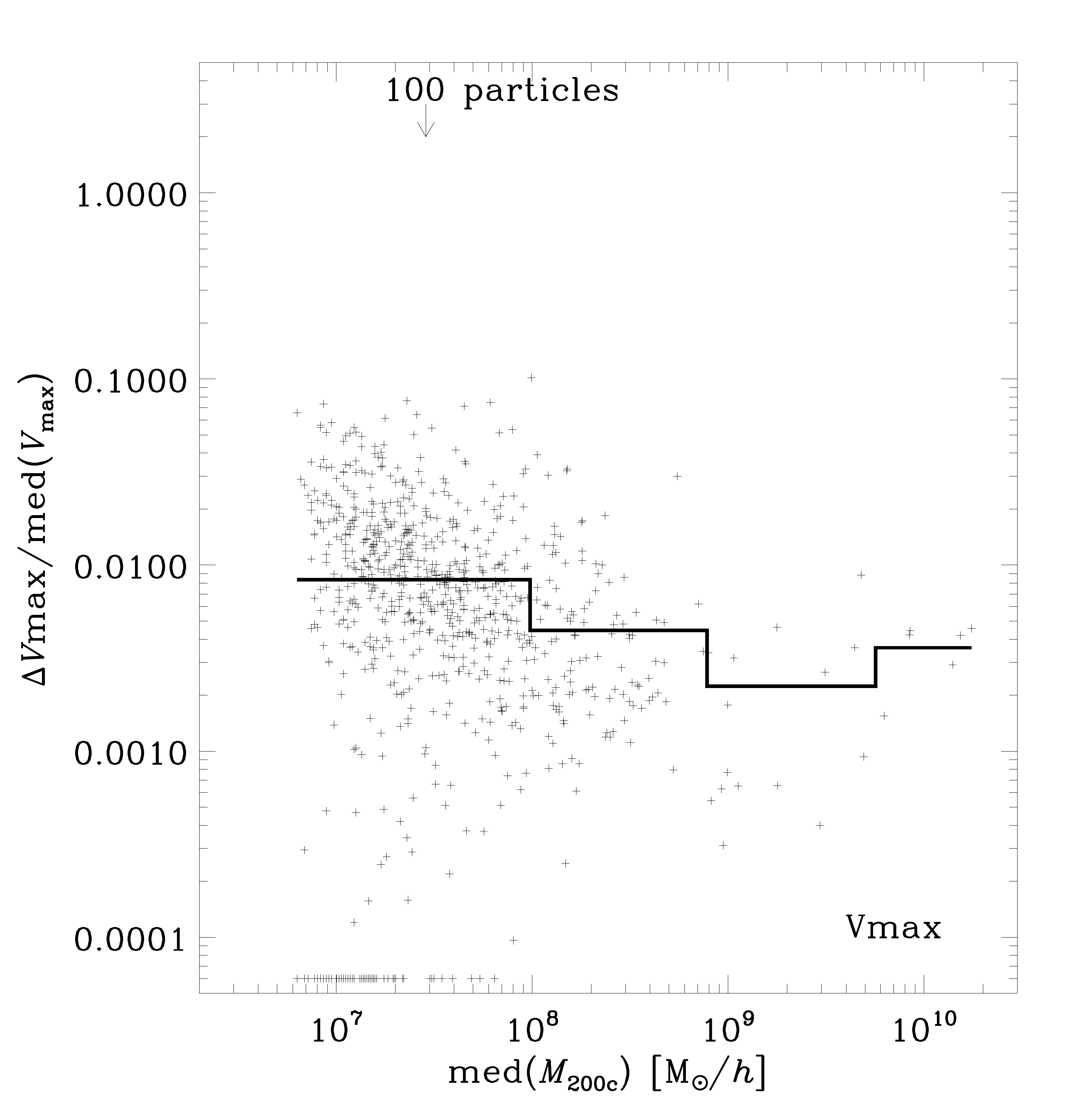}
  \includegraphics[width=\columnwidth]{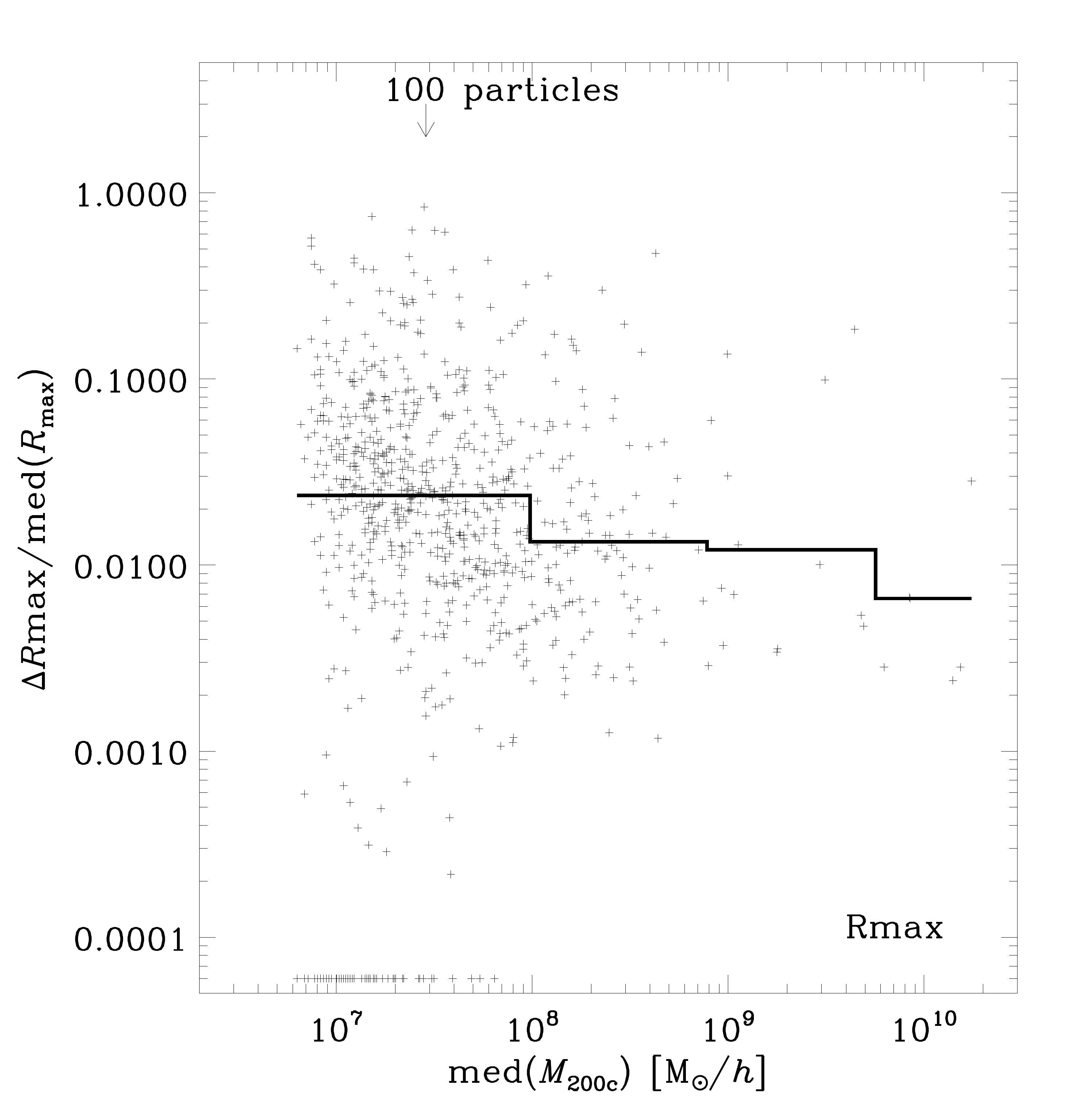}
  \caption{Relative error in \Vmax\ (top) and \Rmax\ (bottom).}
\label{fig:recoveryvmaxrmax}
\end{figure}

Let us consider next the magnitude and radial location of the peak in the rotation curve of the halo, characterised by the values of \Rmax\ and \Vmax, respectively. It has been claimed that these quantities provide a good proxy for the mass and spatial scale of the object \citep[see e.g.][]{Ascasibar08, Muldrew11}, and our previous comparisons \citep{Knebe11,Onions12} show that this may indeed be the case, especially for the maximum circular velocity. We can indeed see in \Fig{fig:recoveryvmaxrmax} that the scatter in the \Vmax\ value (normalized to \Vmax\ itself) is lower than for the mass having a median of a mere 0.6~per cent. However, the variations in \Rmax\ are naturally larger due to the uncertainty in the determination of the peak position: the rotation curves show a flat behaviour about \Rmax\ leading to a median error of 2~per cent.

\subsubsection{Shape \& Spin} \label{sec:recoveryshapespin}
\begin{figure}
  \includegraphics[width=\columnwidth]{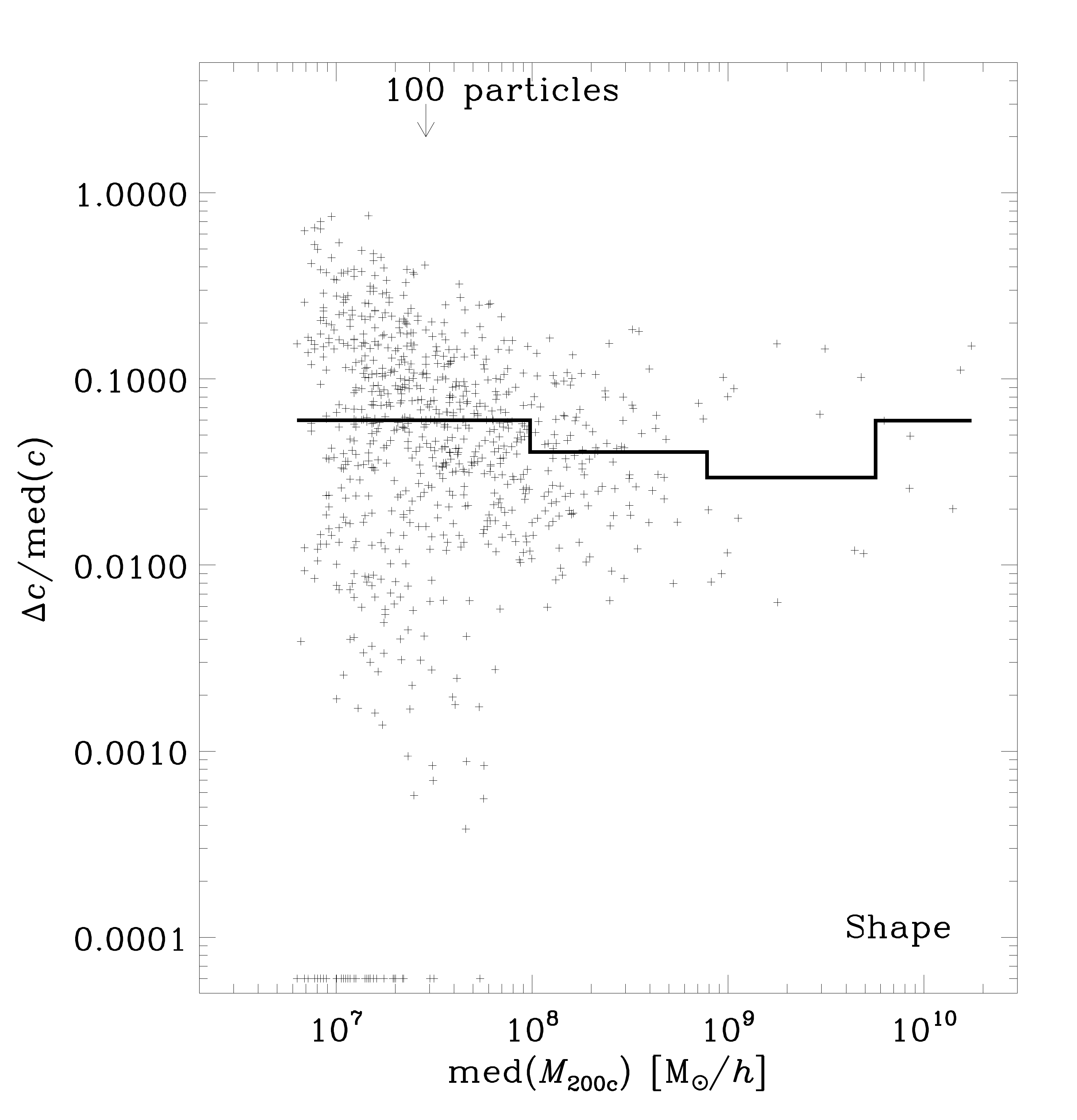}
  \includegraphics[width=\columnwidth]{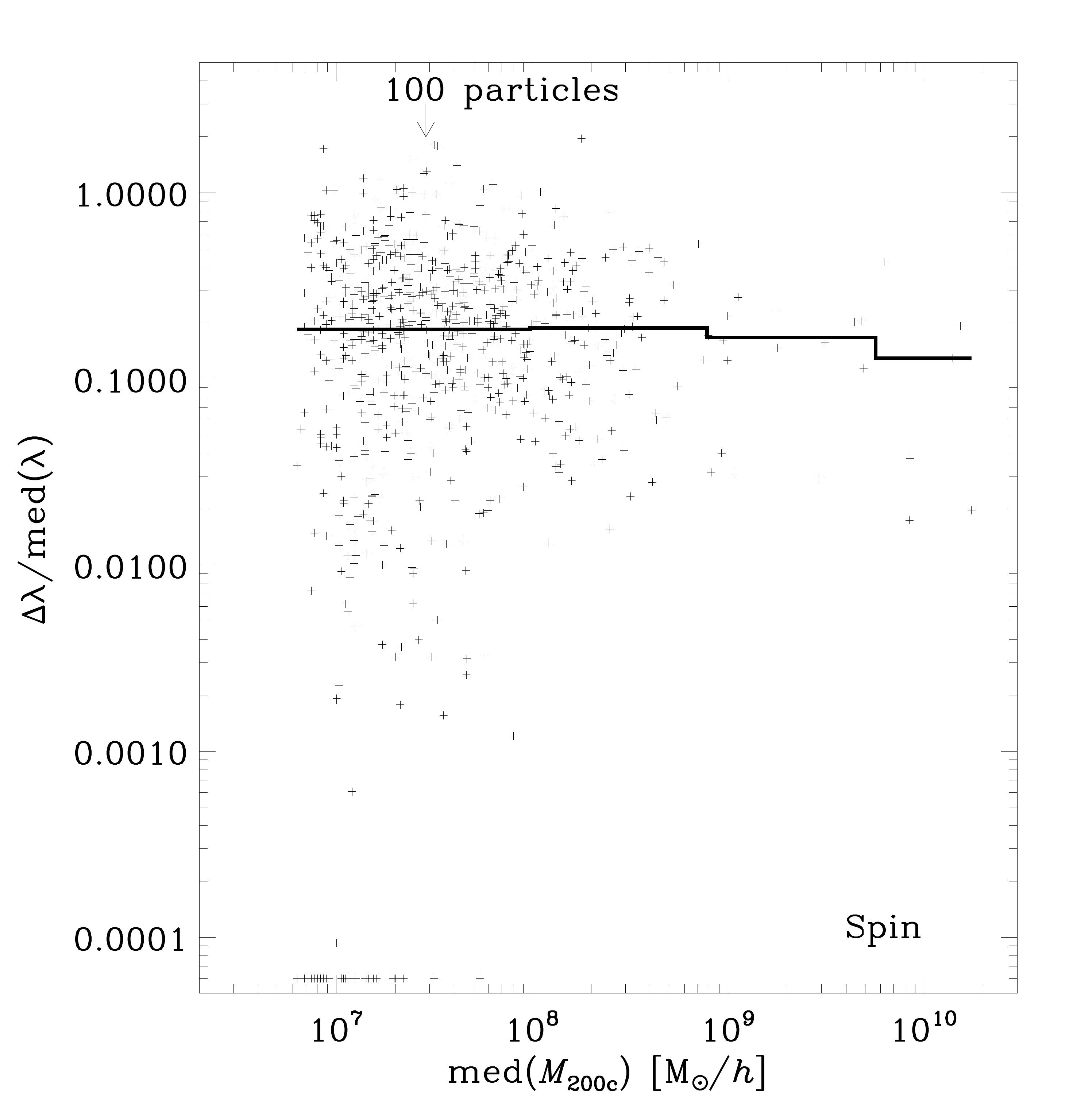}
  \caption{Relative errors in shape (top) defined as the ratio between the smallest and largest eigenvalue of the moment of inertia tensor defined by \Eq{eq:commoninertiatensor} and spin parameter $\lambda_B$ (bottom).}
\label{fig:recoveryshapespin}
\end{figure}
In \Fig{fig:recoveryshapespin} we will now turn our attention towards the shape and spin of the objects identified by the halo finders. The calculation of these quantities is more involved and hence we expect the scatter to be larger, e.g. several of the errors already reported here will propagate in a non-linear fashion to these properties. While the shape (defined here as sphericity, i.e.  the ratio between the smallest and largest eigenvalue of the moment of inertia tensor defined by \Eq{eq:commoninertiatensor}) appears to be determined to approximately the same order of magnitude as the previous quantities giving a median of 5~per cent, the spin is less precisely determined with a median of 18~per cent. For a more elaborate discussion of the spin using the same data and finders as presented here we like to refer the reader to \citet{Onions13}.

 \begin{figure}
   \includegraphics[width=\columnwidth]{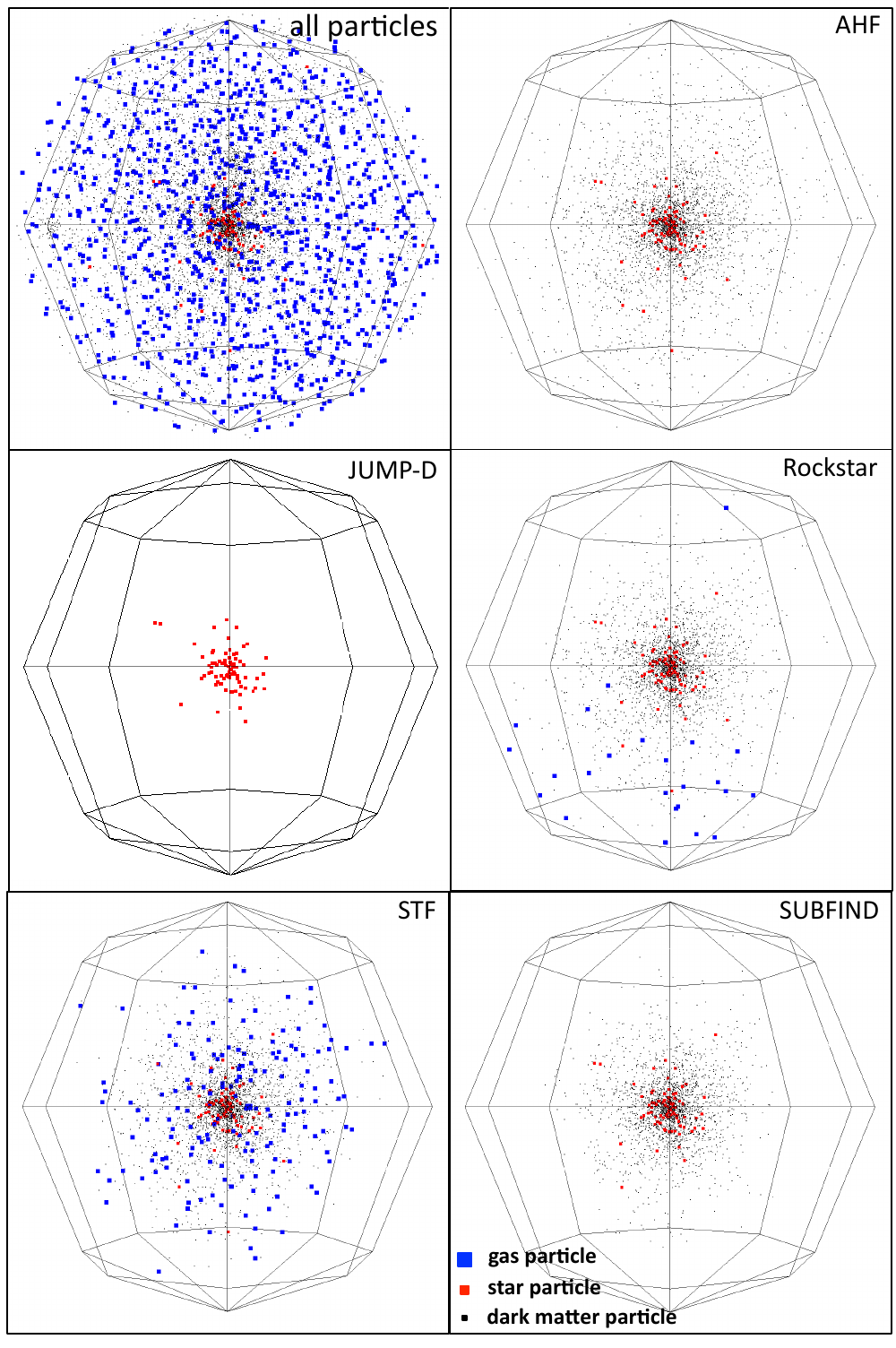}
   \caption{Visualization of a subhalo showing all particles inside a spherical region about the identified centre (upper-left panel) versus the actually identified particles of the individual halo finder showing all types of particles: gas (blue), stars (red) and dark matter (black).}
 \label{fig:galaxies}
 \end{figure}

\subsection{Galaxies} \label{sec:recoverygalaxies}
In \citet{Knebe13} we presented a comparison of codes as applied to the Constrained Local UniversE Simulation (CLUES) of the formation of the Local Group which incorporates much of the physics relevant for galaxy formation. We compared both the properties of the three main galaxies in the simulation (representing the Milky Way, Andromeda, and M33) as well as their satellite populations for a variety of halo finders ranging from phase-space to velocity-space to spherical overdensity based codes, including also a new mere baryonic object finder. We obtain agreement amongst codes comparable to our previous comparisons -- at least for the total, dark, and stellar components of the objects. However, the diffuse gas content of the haloes shows great disparity, especially for low-mass satellite galaxies. This is primarily due to differences in the treatment of the thermal energy during the unbinding procedure. We acknowledge that the handling of gas in halo finders is something that needs to be dealt with carefully,
 and the precise treatment may depend sensitively upon the scientific problem being studied.

To give an impression of the differences found we extracted all the particles from the simulation data in a spherical region about the centre of a certain subhalo. The results can be viewed in \Fig{fig:galaxies}. We can clearly see that the region about the object's centre contains a substantial number of gas particles (shown in the upper left panel).  All codes featuring a treatment of the gas thermal energy either during or prior to the unbinding (i.e. \ahf\ and \subfind) remove essentially all gas from the subhalo; whereas \rockstar, which does not include the thermal energy during the unbinding, and \stf, which does not process the gas and stars through an unbinding routine, are left with a residual amount of gas. Note that \jumpd\ is designed to find galaxies and ignores the dark matter. 

When using \ahf\ in a mode where the gas thermal energy has been ignored, \ahf\ basically considers \emph{all} gas particles seen in the left panel to be part of the subhalo. In contrast, due to their phase-/velocity-space nature, both \rockstar\ and \stf\ consider the majority of the gas particles to belong to the background host and keep only a small amount of them. For the object considered here the effective thermal velocity of each gas particle is always larger than its kinetic velocity (not shown here) and hence the grouping in phase- or velocity-space will naturally remove (hot) gas whenever a gas particle is considered not belonging to it based upon kinetic velocity only. Or put differently, the gas component forming part of the background halo is prone to be removed by such finders as they inherently use velocity information when grouping and collecting the initial set of particles, whereas configuration space finders only deal with velocities (either kinetic or thermal) in a (post-processing) 
unbinding procedure. On a side note, a visual inspection of a larger region about this particular sample satellite galaxy indicates that it has passed extremely close to its host already and been subjected to severe tidal forces; this might also explain why \rockstar\ associates gas to one side of the galaxy.

\subsection{Summary} \label{sec:recoverysummary}
There are several sources of uncertainty in the properties computed by halo finders. While \Sec{sec:definition} was concerned with the ambiguities arising from the definition of each quantity, here we have also quantified the scatter due to the different procedures followed by each halo finder in order to compute \emph{the same quantities} from \emph{the same data}. Our results are succinctly summarized in \Tab{tab:summaryrecovery}. While the differences are below 1~per cent for position, bulk velocity and \Vmax\ (and still marginal for masses), they can rise to over $15$~per cent for certain derived quantities such as spin parameter. Without any further investigations, these numbers could in fact indicate reasonable error bars to be attached to \textit{any} study based upon the results derived from a single halo finder. However, the values listed in \Tab{tab:summaryrecovery} should be considered lower limits as we have restricted the comparison to objects found by all halo finders and used a common post-processing 
pipeline. 
\begin{table}
\caption{Scatter in the main properties computed by the halo finders. Note
that this error is only a lower limit, especially for the numbers
based upon the set of common objects.}
\label{tab:summaryrecovery}
\begin{center}
\begin{tabular}{lc}
\hline
Quantity		& Scatter\\
\hline
\underline{set of common objects:}
\\
Position		& $< 1$ \% $R_{200}$\\
Bulk velocity	& $< 1$ \% \Vmax\\
$M_{200c}$	& $3$ \%\\
\Vmax		& $< 1$ \%\\
\Rmax		& $2$ \%\\
Shape		& $5$ \%\\
Spin			& $18$ \%\\
\\
\underline{full catalogues:}
\\
Mass function	& $10$ \%\\
\Vmax\ function	& $20-30$ \%\\
\hline
\end{tabular}
\end{center}
\end{table}

However, this is a rather academic situation as in reality neither are the objects found by each halo finder restricted to some common set nor will the properties be calculated applying the same method or definition or even common post-processing pipeline. To this extent we also presented the scatter in the general mass and \Vmax\ functions noting that it is in fact larger than the lower limits derived before. But what is responsible for this scatter? Are there ways to understand its origin and hence possible post-correct for it? The sources of the scatter are certainly two-fold, i.e. the individual particle collection and unbinding procedures of each finder. Different finders do not necessarily find the same set of objects and different finders retrieve variations in the properties of the same object. We explore these possibilities in more detail in the next section and discuss the relevance of this variation for scientific applications in \Sec{sec:relationandapplication}.


\section{Precision Cosmology?} \label{sec:precisioncosmology}
One of the most pressing questions arising from the plots presented in the previous Section is: ``Why is there a residual scatter between halo finders of up to 10~per cent?''  The finders have been applied to the same data, subjected to a common post-processing pipeline, and, in some cases, even compared on a set of objects identified in common.  As mentioned before, unless we can be certain which halo-finding technique is the best (if such a statement can be made at all), the observed scatter indicates the accuracy to which we can determine these properties in cosmological simulations. In this section, we will try to pinpoint the origin of the observed scatter by examining the contribution of the different steps outlined in \Sec{sec:methods}. Our final goal is to see if it is possible to bring the errors incurred by halo finding down to the one per cent level demanded by precision cosmology \citep[e.g.][]{Tinker08, Komatsu11}.

\subsection{Origin of the Scatter} \label{sec:precisioncosmologyorigin}
What are possible sources for the observed scatter? We have already seen that there are two fundamental steps involved in obtaining halo catalogues, i.e. 
\begin{itemize}
 \item halo finder methodology (\Sec{sec:methods})
 \item definition of halo properties (\Sec{sec:definition})
\end{itemize}

\noindent
In that regard, the differences seen in \Fig{fig:fieldhaloes} combine uncertainties in both these steps whereas \Fig{fig:subhaloes} is only subjected to the first point thanks to the common post-processing pipeline. For the time being we will actually leave the investigation of differences due to the second point aside and only focus on the mode of operation of halo finders. But one also has to go one step further and ask whether one is interested in the actual halo-to-halo scatter or systematic errors affecting the whole halo ensemble. This is an important point as we have seen that the variations in, for instance, the halo mass for cross-identified objects is of order 3~per cent (cf. \Tab{tab:summaryrecovery}) whereas the ensemble mass function of the same data set seen in \Fig{fig:subhaloes} clearly shows larger scatter: while the number of uniquely found haloes is 823, each finder nevertheless found approximately the same number of objects not part of the common pool (cf. \Tab{tab:excessobjects}). This suggests 
that once the halo finders do find the same set of objects, the errors across them should decrease approaching the first set of values listed in \Tab{tab:summaryrecovery}. As we have shown, most of the missing objects are small. Hence one way to improve the `purity' of a catalogue is to only rely on objects containing more than 300 particles. This shouldn't be too surprising as this is the same limit found elsewhere \citep[e.g.][]{Onions13} to be required if stable derived halo properties (such as the spin parameter) are desired. 

Focusing on the halo finder methodology and following the scheme of \Sec{sec:methods}, all the differences between one halo finder and another must fall into one of the following categories, representing the basic steps required to end up with a halo catalogue: 
\begin{itemize}
  \item candidate identification
  \item halo centre and bulk velocity determination
  \item particle collection and unbinding procedure
  \item mass and edge determination
\end{itemize}

We will now follow a divide-and-conquer strategy, in an attempt to isolate those steps that are responsible for most of the scatter, again using the Aquarius A, Level 4 data set and restricting ourselves to the commonly post-processed particle ID lists of some of the finders featuring a credible unbinding procedure. While we include mass and edge determination in this list we do not discuss it below as with a common post-processing pipeline this step is the same for all finders and introduces no scatter. This highlights the issue mentioned above, that in actuality a common post-processing routine is not used by everybody and significant scatter can be introduced at this stage unless great care is taken. 

\subsubsection{Candidate Identification \& "Excess" Objects}\label{sec:precisioncosmologycandidates}
\begin{figure}
  \includegraphics[width=\columnwidth]{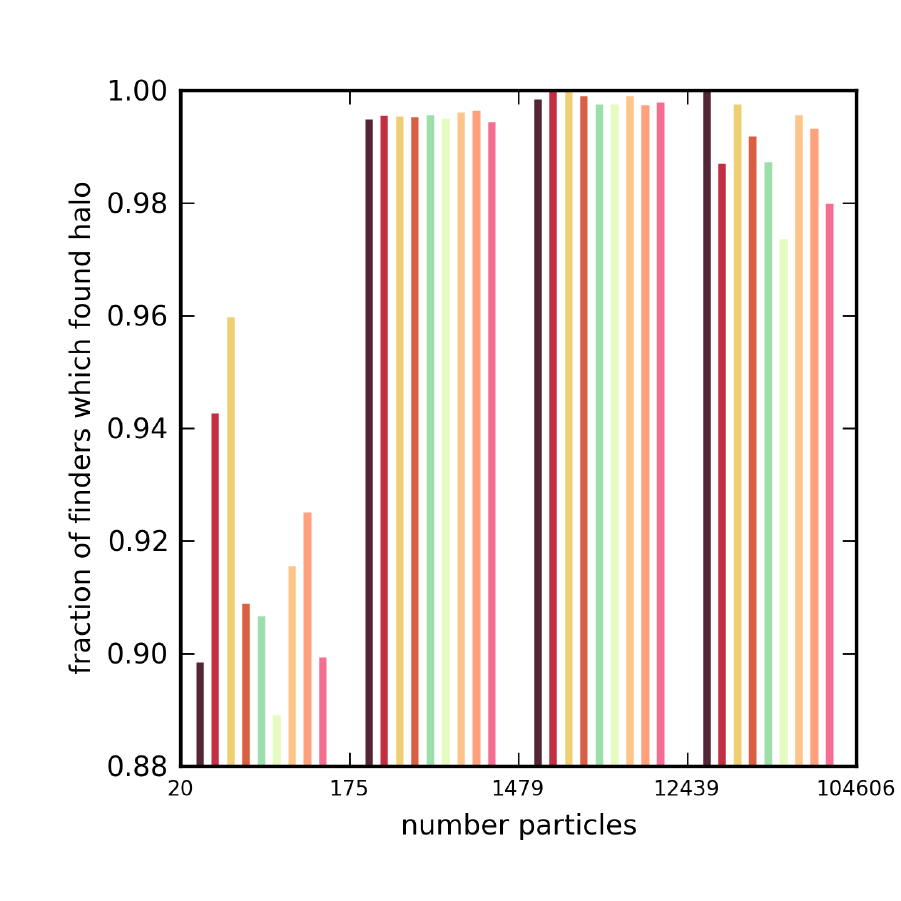}
  \includegraphics[width=\columnwidth]{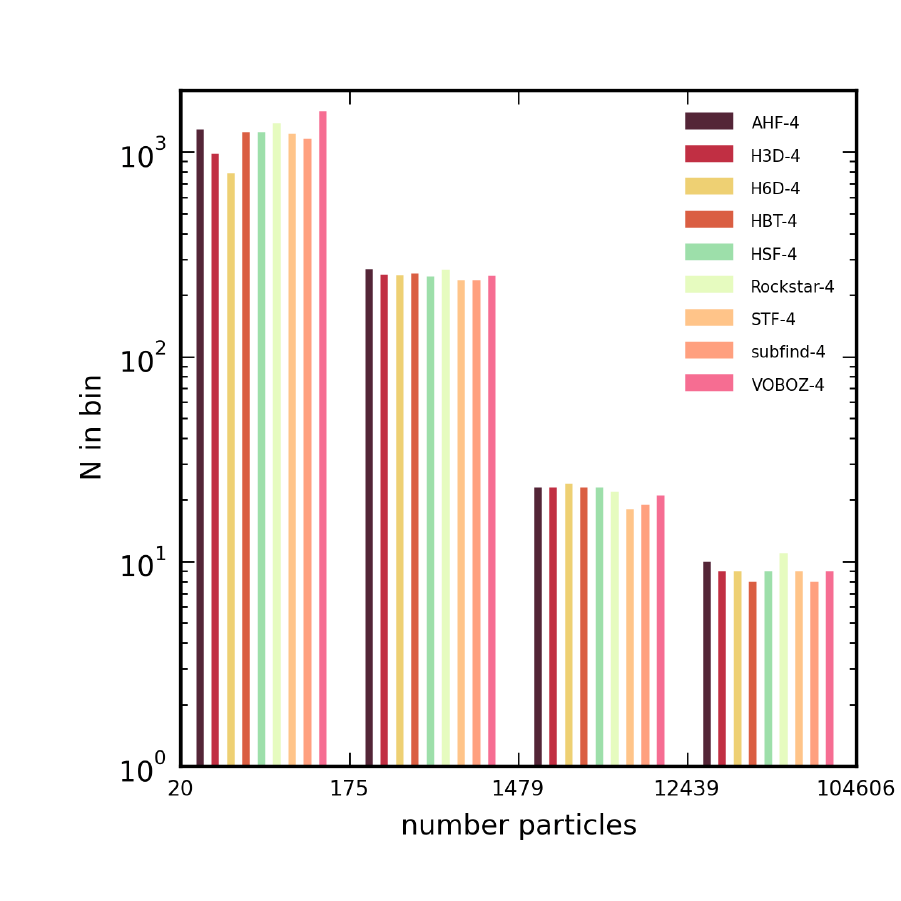}
  \caption{The fraction of codes finding the same object as the
  reference code given in the legend as a function of the binned
  object's number of particles in that reference code (upper panel) as well as the
  actual number of objects entering the comparison in the respective
  number of particle bin (lower panel).}
\label{fig:superplot-mass-level4-nbins4}
\end{figure}

\begin{figure}
  \includegraphics[width=0.95\columnwidth]{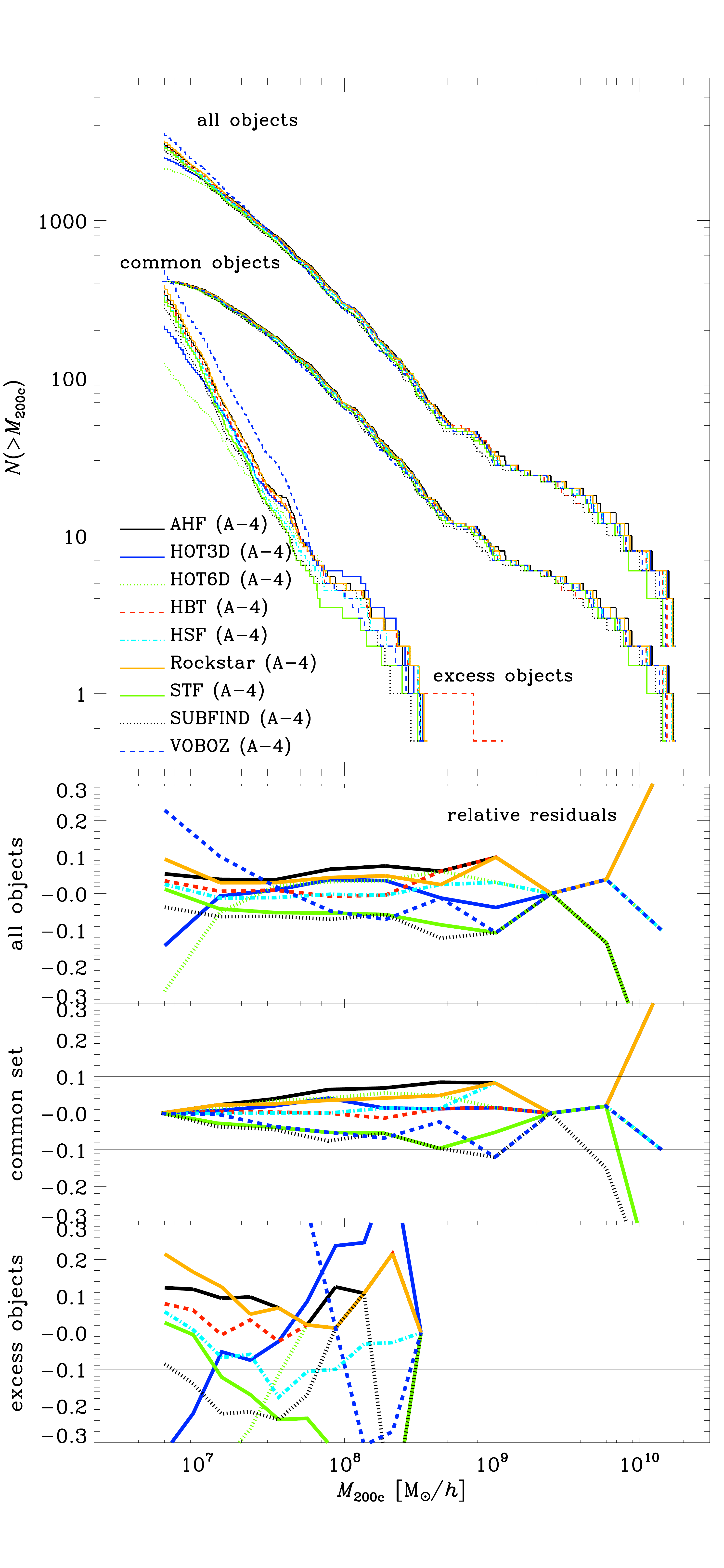}
  \caption{Subhalo mass $M_{200c}$ functions for all
  identified objects (upper set of lines), the common set of
  objects (middle lines), and the excess objects (lower lines). The lines for all objects have been shifted upwards by a factor of 5 whereas the other two sets have been shifted downwards by a factor of 5 for clarity. The lower three panels show the relative residuals with respects to the mean (as in the plots before), and the pair of solid  lines in each of them indicates the 10~per cent error  bars.}
\label{fig:submassfuncCroCo}
\end{figure}

\begin{figure}
  \includegraphics[width=0.95\columnwidth]{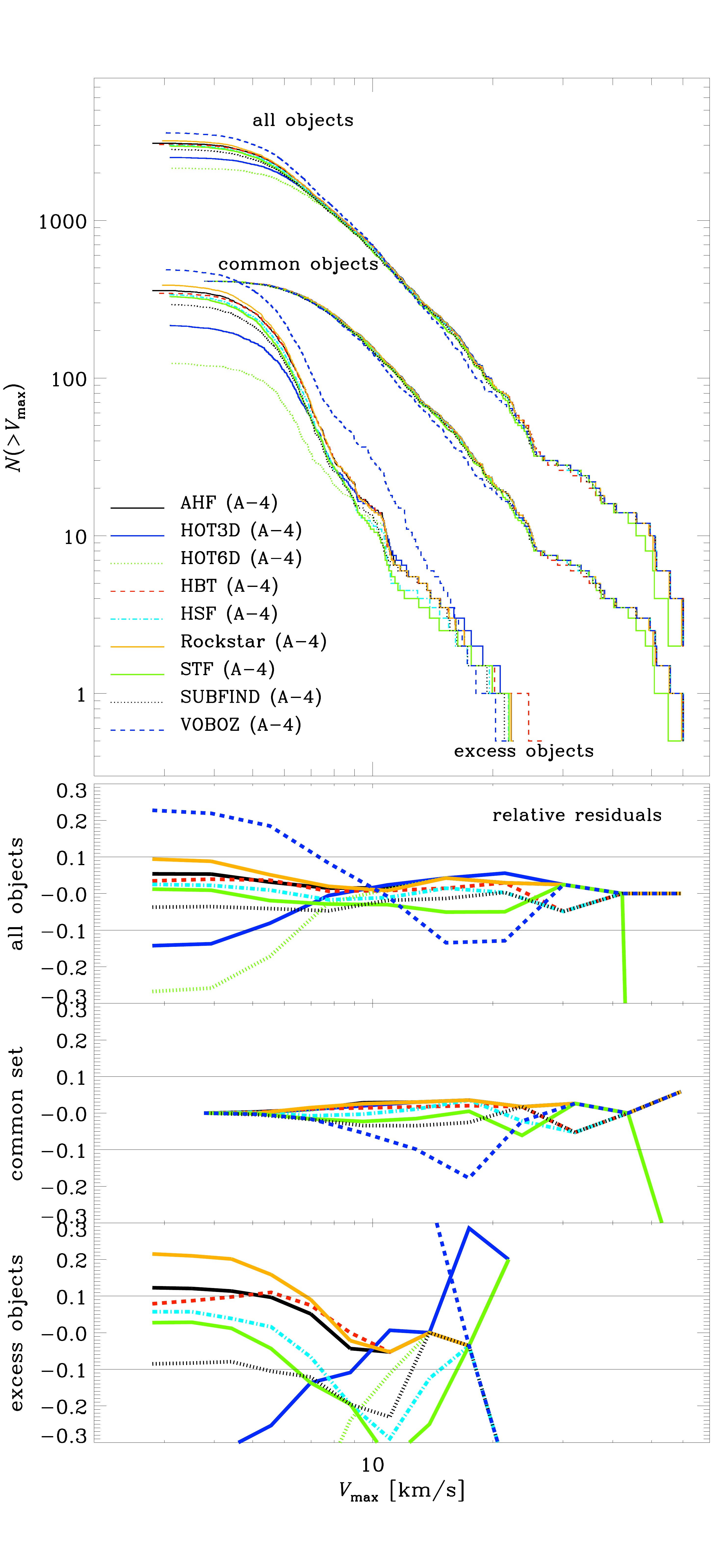}
  \caption{Same as \Fig{fig:submassfuncCroCo} but showing the subhalo  \Vmax\ functions.}
\label{fig:vmaxfuncCroCo}
\end{figure}

There is one element that could make a sizeable contribution to the scatter in the mass and \Vmax\ functions: whether a given halo is detected or not. Remember that the deviations reported in \Tab{tab:summaryrecovery} were based upon a common set of objects, but that each finder certainly found (substantially) more objects than defined by this common set (cf. \Tab{tab:excessobjects}). For every subhalo found by a given reference code in the Aquarius A-4 data we now identify its counterpart in the particle ID lists of all other codes. This was again accomplished by examining the particle lists and maximizing the merit function $C=N_{\rm shared}^2/(N_{1}N_{2})$, where $N_{\rm shared}$ is the number of particles shared by two objects, and $N_{1}$ and $N_{2}$ are the number of particles in each object, respectively (see \App{app:crosscorrelations} for more details). This is the same procedure as applied before, but this time we aim at quantifying how many of the objects found by a given halo finder were found by 
the other finders.

To investigate this, we plot in (the upper panel of) \Fig{fig:superplot-mass-level4-nbins4} the fraction of codes that also found the same object on the $y$-axis against the (binned) number of particles of the object in the reference code.  We can see that there is a group of codes for which all other codes found the same objects (with also approximately the same number of particles). In the lower panel of \Fig{fig:superplot-mass-level4-nbins4} we show the actual number of objects entering into the histograms of the of the upper panel, i.e. the number of haloes found by each finder in the respective number of particles bin.

While there are differences visible in \Fig{fig:superplot-mass-level4-nbins4}, how will they affect, for instance, the (sub-)halo mass or \Vmax\ function -- our usual measure for the scatter? This can be viewed in \Fig{fig:submassfuncCroCo} and \Fig{fig:vmaxfuncCroCo} where we show (in the upper panel) the subhalo mass function and the \Vmax\ function, respectively. The three sets of lines in each panel refer to the original functions based upon all objects found by each finder (shifted upwards by a factor of 2 for clarity, cf.~\Fig{fig:subhaloes}) and the same function using only the set of common objects (middle lines, shifted downwards by a factor of 2 for clarity) as well as the `excess' objects (lower lines, also shifted downwards by a factor of 2) as defined in \Sec{sec:recoveryset}. The three lower panels in each of the figures is the fractional difference of the curve for a given finder to the mean value. Remember that these curves quantify the variation in the number of objects found above a certain mass threshold; they do not directly measure differences in subhalo mass. We see here that restricting ourselves to the common set does not change the observed scatter in the mass function, whereas we find a marginal improvement for \Vmax. Note that the perfect agreement of the functions for the common set at the imposed 20 particle limit is artificial as the number of objects is identical by construction. And even though we clearly observe that the scatter for the excess objects is substantially larger than for the common ones, these results indicate that the overall scatter is not dominated by them. These objects contribute particularly at the low-mass end where their consistent detection becomes difficult, but variations in subhalo properties from code to code are principal source here. We finally note that  the excess subhaloes are primarily of low mass and composed of less than 100 particles, respectively. One last word of caution, the error estimates presented in the previous Sub-Sections~\ref{sec:recoveryposvel} through to \ref{sec:recoveryshapespin} (and summarized in \Tab{tab:summaryrecovery}) are smaller than the scatter seen here as they only took into account the difference in the 3rd and 7th percentile and hence ignoring outliers seen here.

\begin{figure}
  \includegraphics[width=\columnwidth]{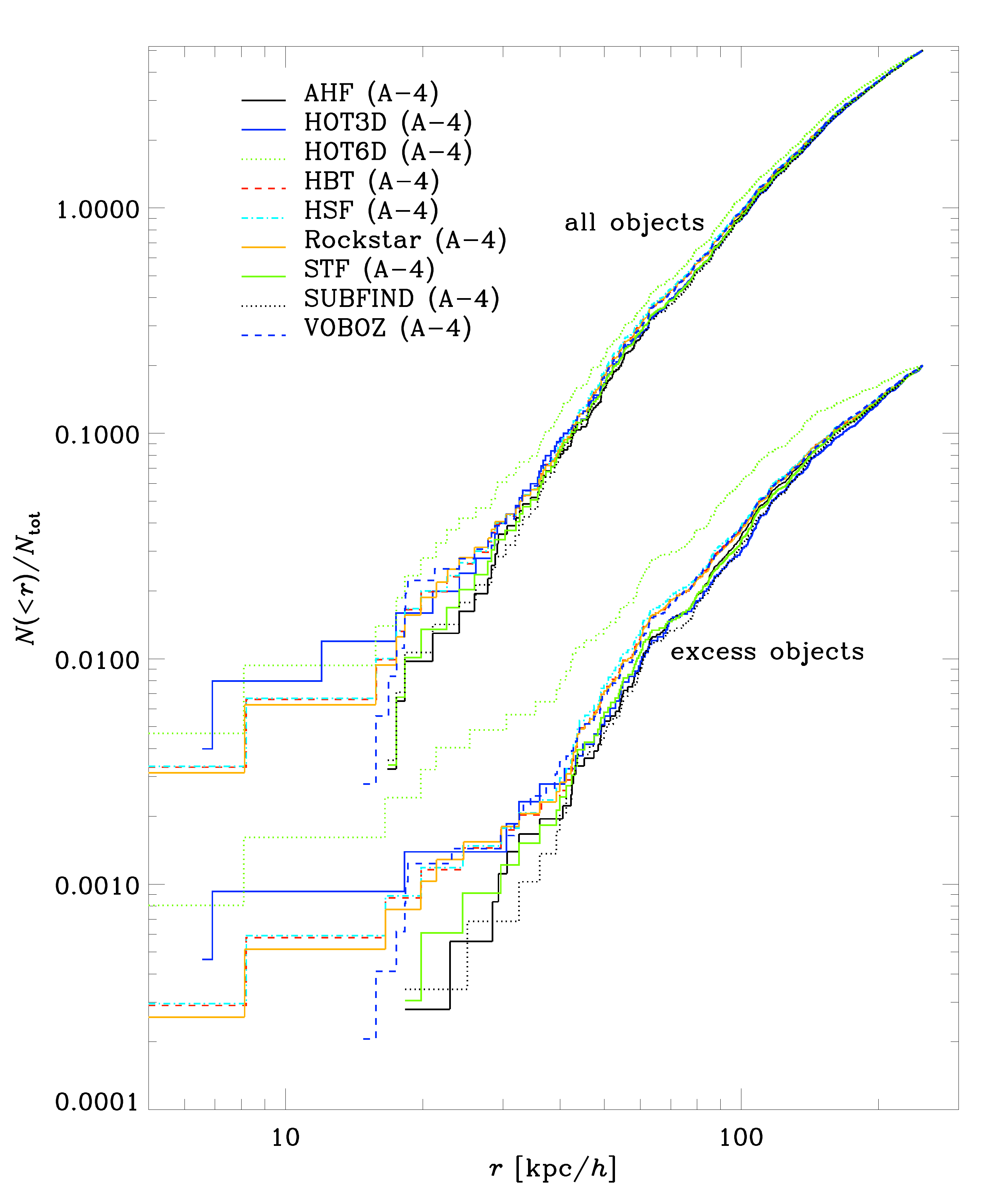}
  \caption{Subhalo radial distance functions for all
  identified objects (upper set of lines) and the `excess' haloes (lower lines). In each case the upper and lower curves have
  been scaled by a factor of 5 to better separate them.  }
\label{fig:subhaloesCroCoNo}
\end{figure}

We have seen before that halo finders perform differently at finding subhaloes close to the centre of their host \citep[][]{Onions12,Knebe11,Muldrew11}. This then raises the question whether or not this is the origin for the existence of the excess objects seen here. We therefore present in \Fig{fig:subhaloesCroCoNo} the cumulative radial distribution of these objects (normalized to the respective total number of excess haloes) always in comparison to the full set of haloes; both curves have been shifted by a factor of 5 for clarity as they would otherwise overlap at he low-$r$ end of the distribution. We can appreciate that excess subhaloes are found at all possible radial distances. And it comes at no surprise that those objects found close to the host centre are forming part of the excess set. We recommend to view the radial distance plot in relation to Fig.7 of \citet{Onions12} that shows the cumulative mass in subhaloes as a function of distance.

\subsubsection{Centre \& Bulk Velocity Determination}\label{sec:precisioncosmologycm}
\begin{figure}
  \includegraphics[width=\columnwidth]{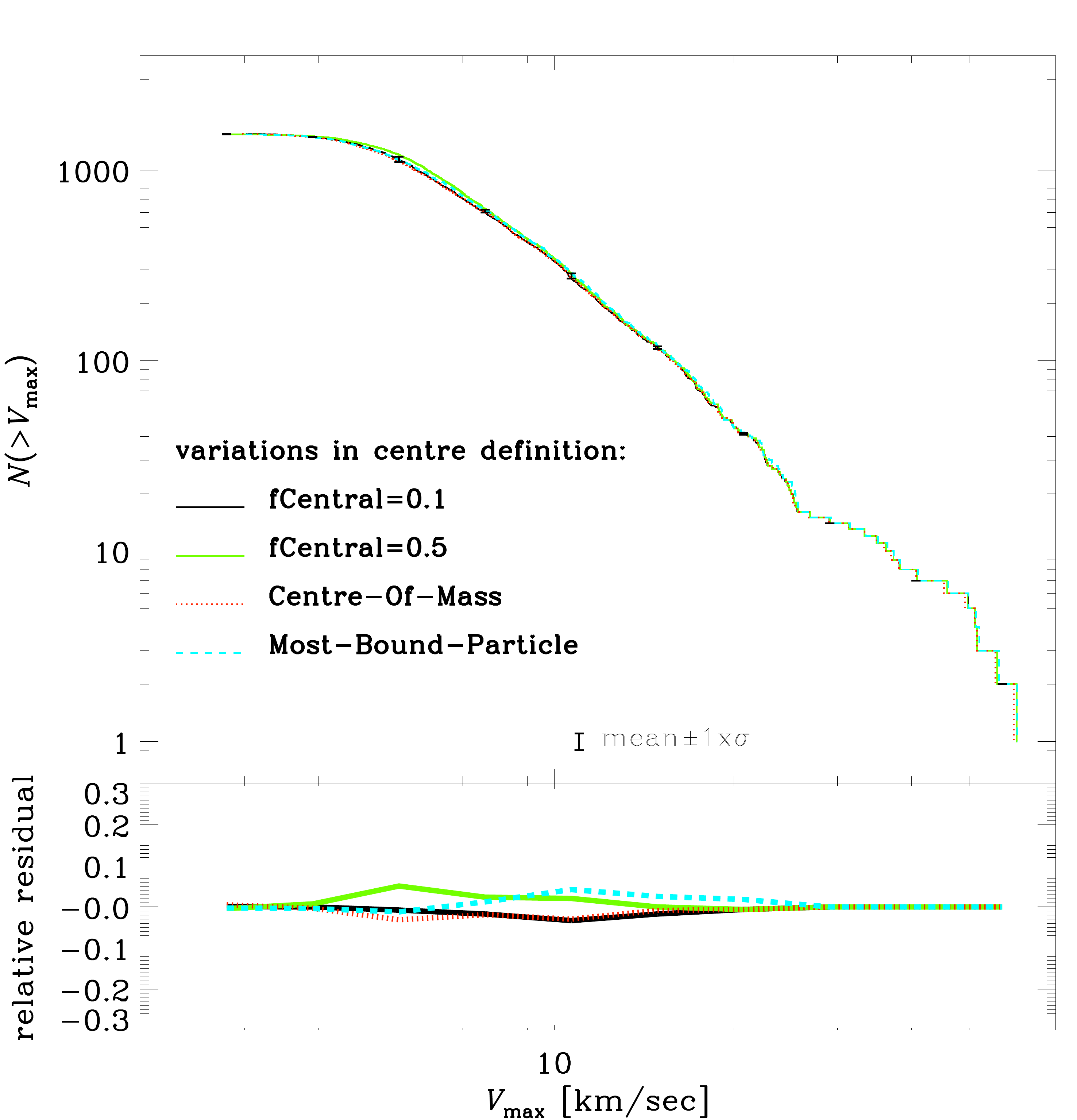}
  \caption{Subhalo \Vmax\ function to various definitions of subhalo centre. Please refer to the main text for more details.}
\label{fig:maxfuncRatio-CentreTests}
\end{figure}

Given that we found a fair agreement for (sub-)halo centres in (the upper panel of) \Fig{fig:errorPosVel}, and that there is little difference between \Fig{fig:fieldhaloes} (no common post-processing) and \Fig{fig:subhaloes} (common post-processing), it seems unlikely that the details of the location of the centre or the calculation of the bulk velocity of an object make a significant contribution to the error budget. But we nevertheless performed the following tests for which the results regarding the \Vmax\ function are presented in \Fig{fig:maxfuncRatio-CentreTests}: using one of the halo finders only we varied the definition of the centre as used with our common post-processing pipeline assessing the effect it has on both the mass and \Vmax. The definitions applied were the overall centre-of-mass, the position of the most bound particle, and an iteratively determined centre using the innermost 10 or 50 per cent of the particles. We found that while the mass is not affected at all (variations below 1 per cent), \Vmax\ can change up to several per cent for certain subhaloes. These results are also in agreement with the findings reported for the mock haloes studied in \citet{Knebe11}. We therefore rather conjecture that either the particle collection method or the particulars of the unbinding procedure may be held responsible for the scatter, both to be studied in the following Sub-Section.

\subsubsection{Particle Collection and Unbinding}\label{sec:precisioncosmologyunbinding}
\begin{figure}
  \includegraphics[width=\columnwidth]{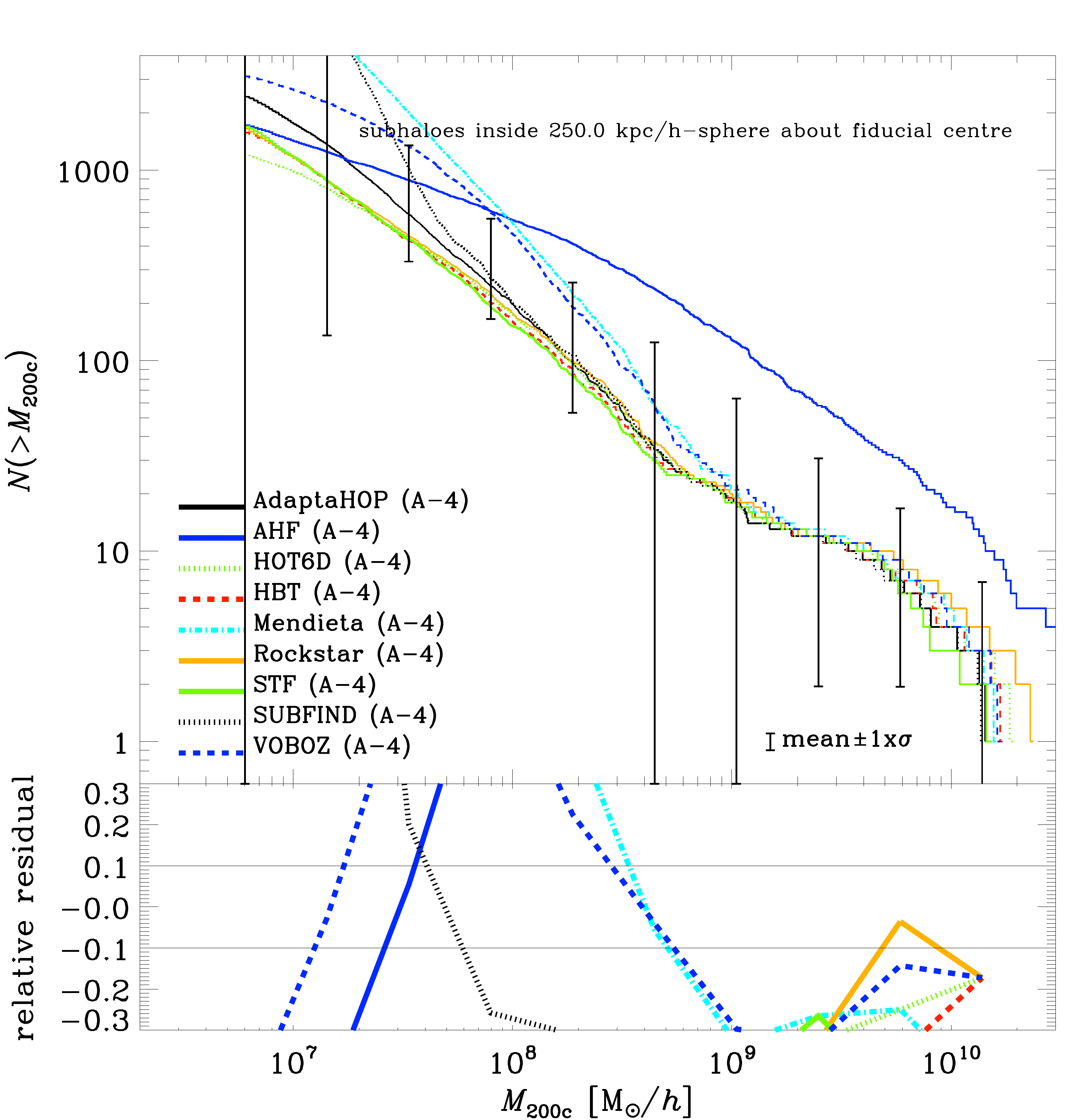}
  \includegraphics[width=\columnwidth]{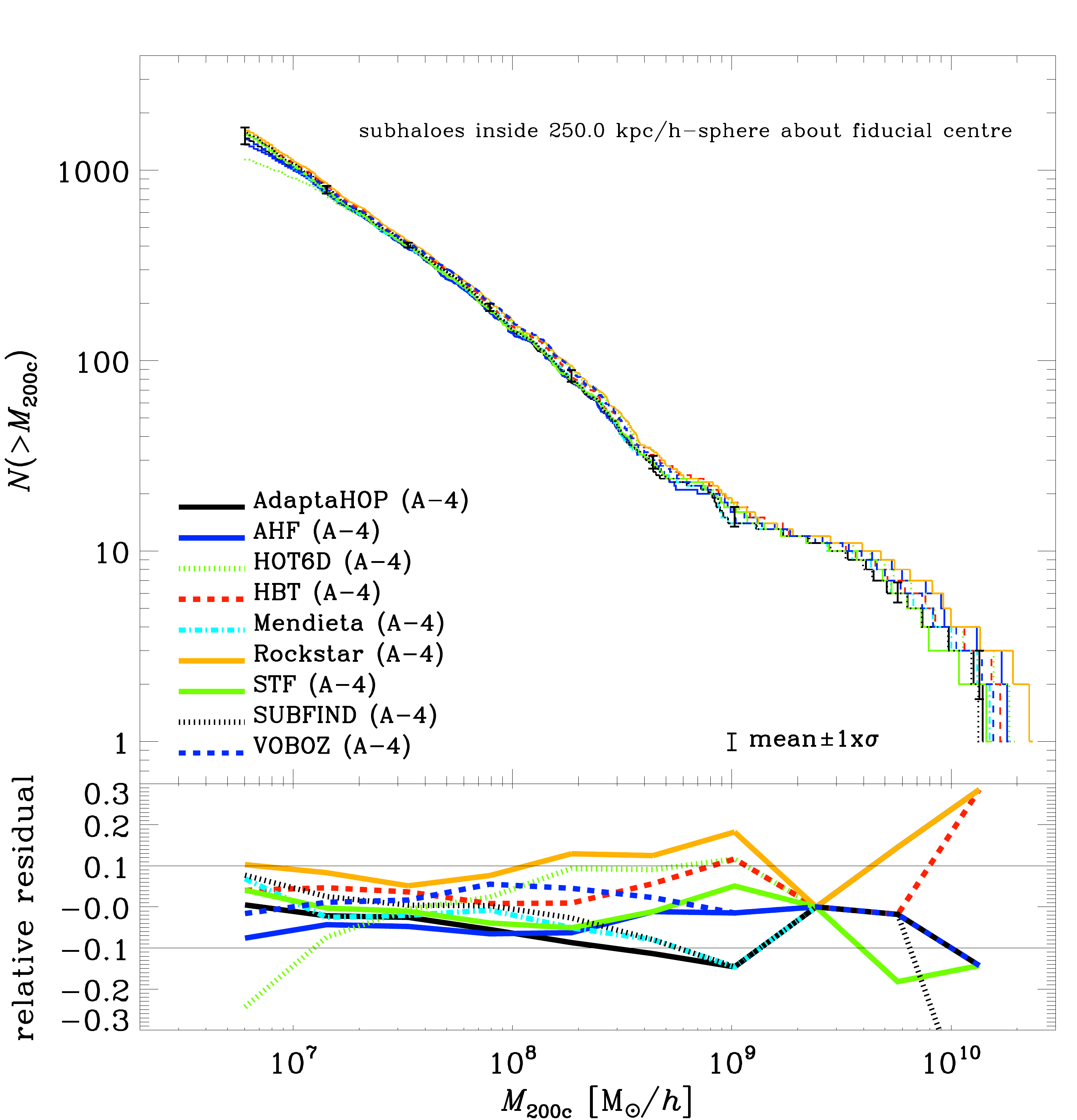}
  \caption{Highlighting the relevance of a common unbinding procedure:
  we show the same subhalo mass function as in \Fig{fig:subhaloes},
  computed from raw particle lists returned by each code
  \emph{without} unbinding (top panel) and with a common
  post-processing routine that also features a common unbinding
  procedure (bottom panel). The pair of solid lines in each of the
  residual plots simply indicates the 10~per cent error bars.}
\label{fig:submassfuncRatio-unbinding}
\end{figure}

\begin{figure*}
  \includegraphics[width=\columnwidth]{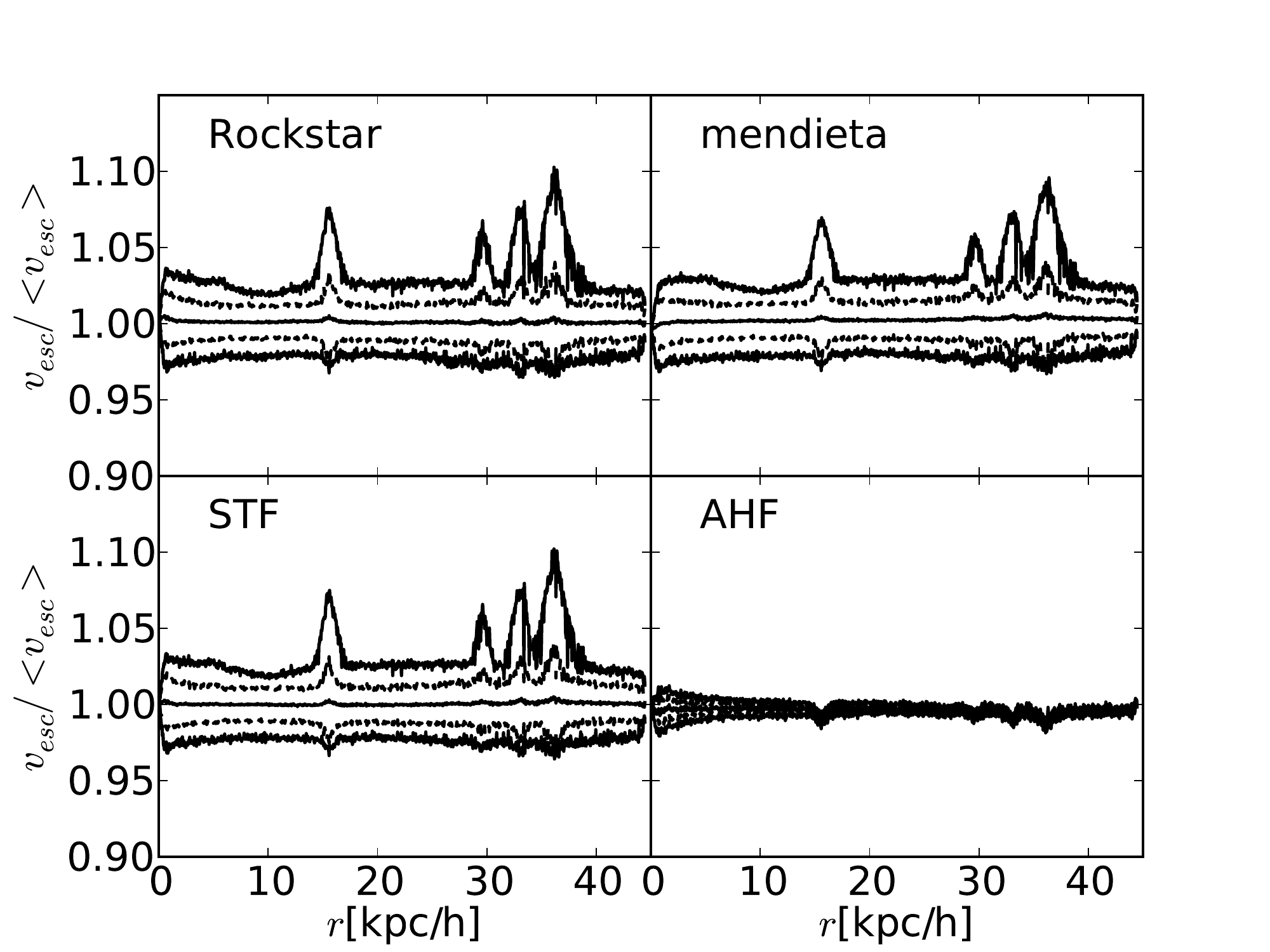}
  \includegraphics[width=\columnwidth]{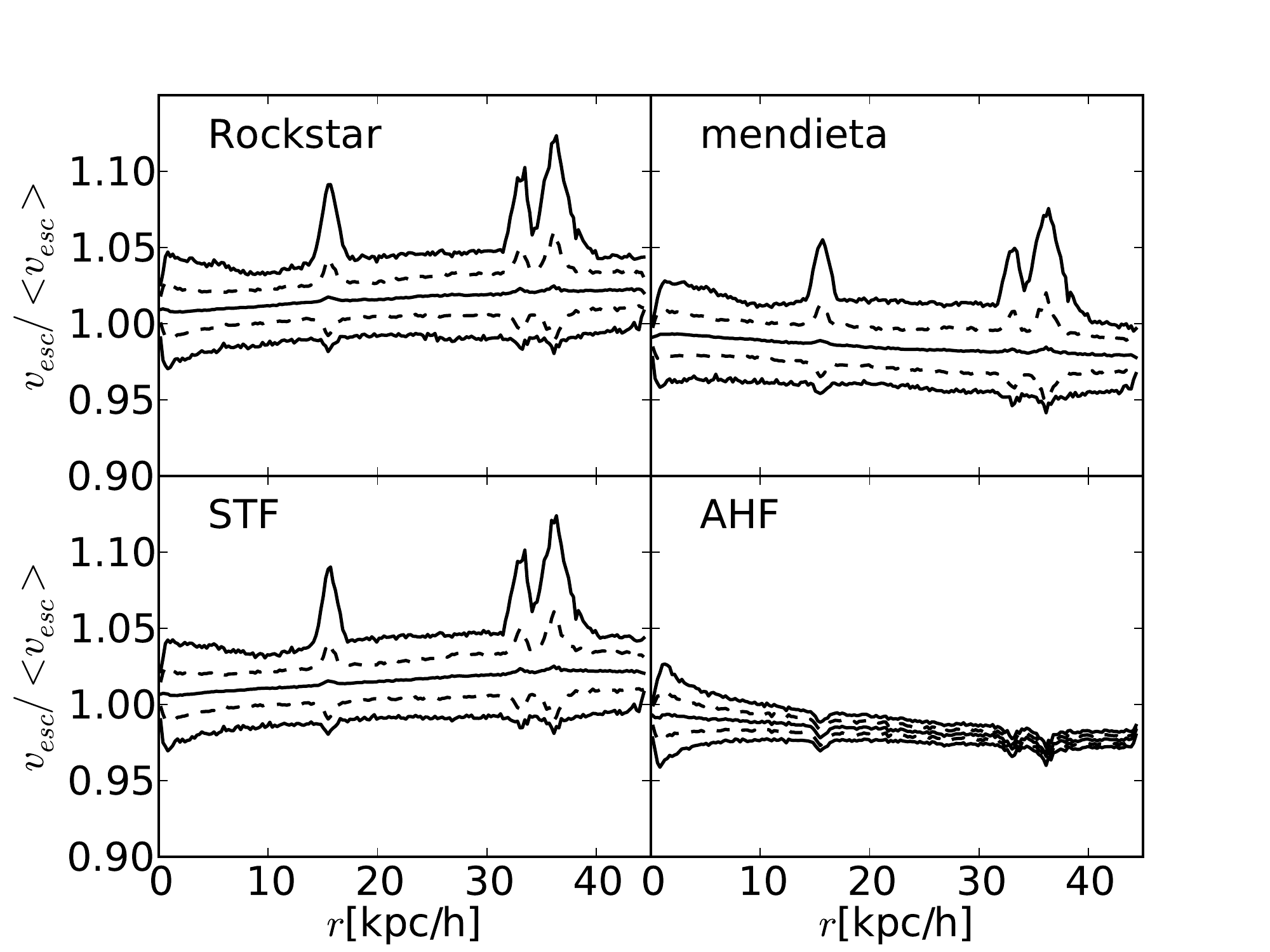}
  \caption{Escape velocity profile for a sample (massive) subhalo as
  determined by a sample of the halo finders prior to their unbinding
  procedure (left panel) and after individual unbinding (right
  panel). The lines indicate the mean (solid), mean $\pm 1\sigma$ (dashed) and min as well as maximum value (solid) in the respective bin for each finder normalised by the binned means of all finders.}
\label{fig:Vesc}
\end{figure*}

Obviously, one cannot come up with either a unique candidate identification or particle collection method, as these clearly \emph{define} the halo finder in question: for instance, phase-space finders base their particle collection on an intrinsically different algorithm than configuration-space finders, and FOF-based methods usually combine these two steps into one single procedure -- and they may also not apply any (additional) edge definition other than the isodensity contour traced by the applied linking length.  Unbinding is also an important -- and delicate -- step, and the particular way of carrying it out may be regarded as an intrinsic piece of some algorithms.

For the Aquarius A-4 data set we now separate the two steps of collecting particles and removing unbound particles from each other and show in \Fig{fig:submassfuncRatio-unbinding} the resulting subhalo mass functions (upper in-set panels) and the resulting variations (lower in-set panels). The top plot shows the subhalo masses as given right after the collection of particles considered potentially belonging to an object by each finder, whereas the bottom plot takes those subhaloes and subjects them to a common unbinding procedure.\footnote{We are applying the unbinding routine of \ahf\ whose mode of operation is described in the Appendix of \citet{Knollmann09}} Please note that not all halo finders participated in this particular test. While there are huge differences in the amount of particles that each algorithm chooses to evaluate for (un)binding, applying a common procedure to all those candidate objects reduces the scatter to the level already seen in \Fig{fig:subhaloes}, where the unbinding was left to the individual halo finders.  On a separate note, this figure also clearly indicates how important unbinding is for configuration-
space finders versus phase-space finders. The mass function for configuration finders such as \ahf, \mendieta, \voboz, and \subfind\ changes significantly after unbinding whereas the distributions for \hotsd, \rockstar, and \stf\ do not.

To quantify the influence of unbinding even more we performed yet another test: we requested participating halo finders to provide the escape velocity profile for a given pre-selected (massive) subhalo originally found across all analyses. The results are presented in the left panel of \Fig{fig:Vesc}, where we plot the calculated escape velocity  $v_{\rm esc}=\sqrt{2|\phi|}$ at each particle position, normalized by the average over the four partaking codes.  First, one can see that different codes do in fact use different methods to calculate the gravitational potential and hence escape velocities. This is particularly the case for \ahf, which uses a spherical approximation when calculating the potential $\phi$, whereas the other three codes base their $\phi$ value on a tree-construction of the particles.  The spikes (dips) in the profile are sub-substructure objects that are not resolved by the (non-realistic) spherical method applied in \ahf. The right panel shows escape velocity profiles \textit{after} 
unbinding. The differences have marginally increased for the tree-based potential calculation, reflecting again the particulars of the individual unbinding procedures, such as what velocity is used to determine whether a particle is (un)bound.  The few percent uncertainty in the escape velocity of the individual particles eventually leads to a $0.5$~per cent error in the different masses after unbinding here. Consequently, unbinding is unlikely to be a primary source of the scatter observed (see also Table \ref{tab:summaryrecovery}).

Before we continue, we should emphasize that the term `unbinding' is a little misleading. First, these structures do not exist in isolation but are treated as such when determining the binding energy. A particle near the edge might be considered bound if an object is treated in isolation but could be stripped by tidal forces. Second, a reference frame must be chosen to determine the kinetic energy of a particle. Some codes may choose to use an object's bulk velocity, others may use the velocity of the density peak, and other may use the velocity of the particle residing at the minimum of the gravitational potential. All of these are valid choices. Third, all so-called unbinding procedures use the instantaneous energy of a particle. Though particles that are false positives with high relative velocities are unlikely to remain with a subhalo, particles that are deemed to be marginally unbound might only be tidally stripped on timescales approaching a dynamical time. Only particles that leave the phase-space 
volume of a subhalo on timescales much shorter than a dynamical time can truly be said to be unbound. However, such a criterion is rarely considered when determining the `bound' mass of a subhalo. 

Despite these complications, we have to acknowledge that even with a common unbinding procedure there still remains a significant degree of scatter. This is not surprising given the large differences in the set of collected particles (i.e. upper panel of \Fig{fig:submassfuncRatio-unbinding}). We conclude that errors in the subhalo mass of order $10$~per cent result from the different initial particle collection schema \emph{intrinsic} to each method. 

\subsection{Potential improvements: facilitating merger trees}\label{sec:precisioncosmologytrees}
As already touched upon in \Sec{sec:methodstracking} it is often the case that we not only have the end-state of a particular simulation but also the evolutionary history at a series of earlier times. We can then make use of this additional information to calculate the temporal evolution of haloes. It has been shown that  software tracking haloes across snapshots can actually improve the results and reliability of halo finders \citep{Springel01subfind,Gill04a,Tormen04,Tweed09,Behroozi12b,Han12,Benson12b}.
This is achieved by explicitly ensuring dynamical consistency of halo properties across timesteps. Note that a lot of these papers primarily deal with subhaloes and following them after infall into the host. But the publicly available tool presented by, for instance, \citet{Behroozi12b} is applicable to both distinct haloes and sub-haloes. It follows a preliminary merger tree but simultaneously integrates the movement of haloes backwards in time: knowing the positions, velocities, and mass 
profiles of haloes at one timestep, one may use the laws of gravity and inertia to predict their properties at adjacent timesteps. This allows the removal of spuriously defined objects or for the insertion of objects not properly identified by the halo finder in the first place. The method employed by \citet{Behroozi12b}  improves the completeness of the halo catalogues in general, particularly at earlier redshifts, even though the results obviously depend on the completeness of the catalogue used as the initial input (generally the $z=0$ snapshot analysis).  For more elaborate details we refer the interested reader to the aforementioned paper.

Such merger trees are now routinely used as inputs to semi-analytic models of galaxy formation  to provide the backbone within which galaxy formation takes place \citep[e.g.][]{Cole00,Benson02,Croton06,DeLucia06,Bower06,Bertone07,Font08}. Therefore, \citet{Benson12b} have studied the convergence in the stellar and total baryonic masses of galaxies, distribution of merger times, stellar mass functions and star formation rates in the \textsc{GALACTICUS} model of galaxy formation \citep{Benson12} as a function of the number of snapshots used to represent dark matter halo merger trees. They found that at least 128 snapshots are required in-between redshifts $z=20$ and $z=0$ to achieve convergence to within 5 per cent for galaxy masses, highlighting again the importance of `sufficient' temporal information for the post-processing of the halo catalogues.

Further, the utilisation of dark matter halo merger trees entails another `feature' for workers in the field of semi-analytical galaxy formation, the so-called `orphan galaxies' \citet{Gao04b}: it can and does happen that a dark matter subhalo dissolves due to tidal forces (and lack of numerical resolution) while orbiting in its host halo \citep[e.g.][for a study of these disrupted subhaloes]{Gill04b}. However, a galaxy having formed prior to this disruption and residing in it should survive longer than this subhalo. Therefore, it became standard practice to keep the galaxy alive even though its subhalo has disappeared, calling it `orphan' galaxy \citep{Springel01subfind,Gao04b,Guo10,Frenk12}. This is another example of the benefits of using merger trees once the simulation of the halo finder does not provide sufficient information anymore. But we also caution the reader that the utilisation of these orphan galaxies is certainly resolution dependent and should be done with care.

But in any case, such merger tree based comparisons and improvements, respectively, are beyond the scope of the present work, but will certainly form part of the next halo finder workshop scheduled for 2014.

\subsection{Summary} \label{sec:precisioncosmologysummary}
Primarily studying subhaloes in the Aquarius simulation, we have been able to associate some of the scatter between the halo finders to inconsistent property definitions and differences in the unbinding routines. However, the most significant part of the scatter seems to stem from the differences in the initial particle collection. Given the present situation, the best halo finders can do for precision cosmology is to provide error bars of 10~per cent on halo mass and \Vmax\ functions of which the prime contribution comes from the initial particle collection and (to a smaller degree) subtleties in the way the unbinding is performed. Note that these two parts are intrinsic to the method of the halo finder. While one might still aim at outsourcing the unbinding, the actual collection of particles \textit{is} the halo finder stripped down to its bare minimum. Making any changes here is equal to simply using another halo finding technique.

We nevertheless need to remind that reader that the situation is a bit different for field haloes and {\em subhaloes}. For the latter, edges and detection thresholds are defined so differently among the finders that a 10~per cent mass difference after common unbinding is not inexplicable.  For the former, the reasons for discrepancies are substantially reduced to issues of definition (and possibly bugs). In that regards, we can only re-iterate that the end-user of any halo catalogues needs to make sure to understand the mode of operation of the halo finder upon which the catalogue is based. But the good news here is that for basically all properties studied here the error decreases with increasing number of particles in the object.

But we also need to bear in mind that there is a subtle difference between disparities in general distribution functions and variations of properties of individual objects.
The aforementioned error of approximately 10~per cent may in part be driven by the fact that halo finders return different numbers of objects above a certain threshold (be that mass, \Vmax, etc.), but we have shown that this effect is virtually negligible at all but the smallest masses (see top panel of Figure~\ref{fig:subhaloesCroCoNo}).
This error is mostly due to the scatter in the mass assigned to each individual halo, but it is exacerbated by the steepness of the cumulative mass function.
Our one-to-one comparison of halo masses (Figure~\ref{fig:recoverymass}) finds that this scatter is substantially smaller, of the order of 3~per cent only.

There are clear indication that a more sophisticated construction of merger trees might reduce incompleteness \citep[][]{Behroozi12b,Han12}, albeit that any possible biases in the initial halo catalogue upon which the revised trees will be based will remain. But the question still is, whether or not (and how) this will affect the end-user of halo finders and the scientific applications across fields that require halo catalogues as an input. We shall discuss these implications in the following Section.

\section{Astrophysical Applications} \label{sec:relationandapplication}

The applications of (sub-)halo catalogues of numerical simulations spread over various fields in astrophysics. These range from facilitating the interpretation of numerical simulations over constituting input to semi-analytic models to comparison and analysis of observational data. In this section, we comment on the different applications and the influence/relevance of the halo finding uncertainties on the results in other fields.

\subsection{Galaxy Formation, Semi-Analytics \& Merger Trees}\label{sec:AppSAM}
We start with one of the biggest topics, i.e. the formation of galaxies within a cosmological context. There are presently two routes to this: one is to directly simulate cosmological volumes including all the relevant baryonic physics \citep[see][for a recent comparison of various baryonic physics and methods]{Scannapieco12}, another is to defer to dark matter only simulations and apply semi-analytical recipes to them \citep{Cole00,Benson02,Croton06,DeLucia06,Bower06,Bertone07,Font08}. And the latter is in fact one of the major areas that requires well understood and carefully constructed halo catalogues. 

\paragraph*{Semi-Analytical Galaxy Formation}
Modern semi-analytical codes \citep{Croton06, Somerville08, Monaco07, Henriques09, Benson12} usually take as input halo merger trees derived from large numerical simulations rather than purely analytical forms such as \citet{Press74} or Extended Press-Schechter \citep{Bond91}. Stable semi-analytic models require stable and physically realistic merger trees: haloes should not dramatically change in mass, size or jump in physical location from one step to the next. As we have seen above all of these changes can result from a poorly constrained halo finder, with halo size being particularly ill-determined. If care is not taken with the choice of the halo centre this can move dramatically from one step to the next, particularly during a large merger event. This dynamical process also leads to large-scale bridging between structures during the initial particle collection phase. This can lead to substantial changes in an objects mass in a very short timescale. Such changes are unphysical. Great 
care also needs to be taken to construct a clean halo catalogue from which to build the trees. The primary concern here is that only gravitationally bound structures are used, otherwise, particularly in regions adjacent to large objects where the background density is already enhanced, spurious groupings can be claimed. Thus we recommend that only halo finding algorithms with a well tested unbinding stage should be used to create catalogues that are intended to form the basis of a semi-analytic model.

Subhalo tracking is also sometimes incorporated into semi-analytic models. We caution that care should be taken with this approach as some algorithms (e.g. \subfind) are very conservative in their allocation of mass to substructures. In this case the subhalo mass can drop dramatically as the structure moves closer to the centre of a host, occasionally vanishing entirely only for it to be partially recovered again afterwards (see, for instance, Figs.~10-12 in \citet{Knebe11} or Figs.~4 and 5 in \citet{Muldrew11}). While all tested finders do a good job of actually locating substructures, even close to the very centre of the host halo, the recovered subhalo mass as a function of the distance to the host halo centre is a function of the particular halo finder you are using \citep{Knebe11,Muldrew11}.

This effect certainly influences the so-called orphan galaxies
(cf.~\Sec{sec:precisioncosmologytrees}), that is galaxies that are no
longer associated with a dark matter subhalo. How these galaxies are
treated varies from model to model and also depends on the resolution of the underlying simulation. The most common approach is to
associate the galaxy with the most bound dark matter particle at the
snapshot just before the subhalo has vanished from the halo
catalogue. Subsequently either this particle is tracked or an orbit is estimated based on the properties of this particle and the galaxy is allowed to survive for a merging time computed using the Chandrasekar formula \cite[e.g.][]{Gao04b,Guo11}. None of the current semi-analytical codes account for the possibility that the subhalo will `reappear' at a later time. However, some codes producing merger trees do attempt to correct for this artificial subhalo removal \citep{Behroozi12b,Han12}. The effect this correction will have on the galaxy population produced by semi-analytic models will certainly be one of the hot topics 
discussed at future workshops. 

A related issue is that of major merger events. As soon as two nearly equal mass haloes approach each other and their radii start to overlap, how will halo finders deal with this? Some codes (e.g. \ahf) have the intrinsic problem that one of them will be tagged `host' and the other `subhalo' at some stage of the merger; this will then evidently lead to a mis-assignment of mass. For a more elaborate study and discussion of this effect we refer the reader to Behroozi et al. (in prep.). 

Another important input to semi-analytical models is the spin
parameter of  dark matter haloes as the size of the galactic disk
residing within them is often derived from this parameter, via
i.e. conservation of angular momentum leading to the formation of a
rotationally supported disk \citep[e.g.][]{Mo98}. Some models assume a
simple relation between a disk size and the spin parameter and use a
fiducial value for the spin parameter, often derived from simulations
\citep{Lu11}. Others use the current angular momentum or spin
parameter of the host halo to determine the angular momentum of the
disk \cite[e.g.][]{Guo11,Benson12} and these codes require this information to be present in the halo merger tree. However, the spin is the least faithfully recovered quantity amongst different halo finders (e.g. \Sec{sec:recovery}) as it is the most sensitive to the unbinding procedure. In fact, \citet{Onions13} showed that the spin parameter can be used as a metric to determine how well a (sub-)halo finder prunes its particle 
collection of high velocity interlopers. The impact of an
unstable or  poorly measured spin parameter is likely small given the
uncertainty in the baryonic physics used in semi-analytical modeling
to convert a halo spin to the spin of a gas and stellar disk. However,
given its use as a metric for assessing the quality of a halo
catalogue and its use as a diagnostic for identifying unrelaxed (read
merging) haloes \citep[e.g.][]{Klypin11}, it warrants further
investigation before its default inclusion in halo merger tree construction. 

One quantity commonly derived with semi-analytical modeling is the luminosity of the galaxy residing within the dark matter (sub-)halo identified by the object finder. And typically observational luminosity functions are better constrained than mass functions that consist of derived quantities. However, comparisons with luminosity functions from numerical simulations are not only challenging because of the differences between various (sub-)halo finders: currently, the variety of subgrid models in hydrodynamical simulations of galaxy formation introduces significantly larger uncertainties \citep{Scannapieco:2011p2069}, and the same is true for semi-analytical modeling \citep[e.g.][]{Snaith11}. This shows that with the current state of (sub-)halo finders luminosity functions from numerical simulations can be used to investigate the physical effects of different subgrid models without concern with respect to any specific halo finder used.

Finally,  aside from variations introduced by using different halo finders, \cite{2003ApJ...598...36A} investigate the effects of non-Gaussian initial conditions on (sub-)structure in CDM. They found that the spin parameter distribution depends on the amount of non-Gaussianity in the initial conditions, though at 100 particles their minimum halo mass is significantly below what is required for reliable spin measurements \citep[][]{Bett07,Onions13}. 

\paragraph*{Galaxy Formation}
Simulating the evolution of the \textit{visible} Universe is much more complex than just following its dark components, because it requires understanding the many astrophysical processes which drive the evolution of the baryonic component under the gravitational influence of the dark matter. But nevertheless, this is the same problem semi-analytical modelers are facing, with the only difference here that the dynamics of the baryonic/collisional component is followed by means of integrating the corresponding equations of hydrodynamics. This entails that one specific model for sub-grid physics like star formation and stellar feedback requires one full simulation to be run. The difference to semi-analytics is on the one hand the presence of gas and stars in the simulation and on the other hand the intrinsic handling of merger trees. While the latter alleviates part of the aforementioned problems with the construction of merger trees, one nevertheless needs to follow the formation and evolution of galaxies 
throughout the simulation: some of the hottest topics in the field
nowadays,  have to do with the morphological evolution of galaxies and
the properties of the stellar populations building up the galaxies. In
these scenarios, the role of mergers \citep{Toomre77,White78} and its
connection to the cosmic web \citep{Aragon-Calvo07,Hahn07,Falck12,Libeskind12}
seems to be crucial. These are topics where hydrodynamic simulations
produce the most important advances in order to compute the required
merger rates, fraction of masses in mergers, and stellar populations
brought by merger. It is therefore fundamental to have a well defined
way to identify host haloes and subhaloes -- and to construct reliable
merger trees for them -- as well as cosmic web classifiers
\citep{Aragon-Calvo07,Hahn07,Hoffman12}, something which is beyond the
aim of this work though. The identification of objects (haloes and
galaxies) in such hydrodynamical simulations poses further challenges to halo finding techniques.
Though we find excellent agreement between different finders regarding the dark matter and stellar mass associated with a galaxy, we find significant differences in the gas content, highlighting the need for a well-defined treatment of gas \citep[cf.~\Sec{sec:recoverygalaxies} and][]{Knebe13}.

\subsection{Large-Scale Structure}\label{sec:AppLSS}

Precision cosmology is often synonymous with large-scale structure (LSS). The growth rate of LSS is directly sensitive to the expansion rate of the Universe, and hence is an excellent probe of cosmological parameters. The question is whether the required precision in theoretically derived quantities is attainable. For instance, the theoretical dark matter halo mass function must be known to an accuracy of $\lesssim1$~per cent to constrain the time evolution of dark energy models with surveys such as DES \citep{Wu10}. \cite{Reed12} showed that this type of accuracy is achievable if difficult for pure dark matter only simulations, but they only considered the FOF halo finder. However, the mass function depends on how haloes are identified
\cite[e.g.][]{Lacey94,Cole96,Tinker08,Lukic09,More11,Watson13}.
Moreover, there is no guarantee that differences between halo finders remain fixed at higher redshifts -- in our comparisons we only considered redshift $z=0$. As a consequence, deviations 
from a universal behaviour will probably depend on applied halo finder \cite[e.g][]{Warren06,Tinker08,Courtin11,Watson13}.

Present and future large-scale structure surveys of the Universe (just to name a few, BOSS, PAU, WiggleZ, eBOSS, BigBOSS, DESpec, PanSTARRS, DES, HSC, Euclid, WFIRST, etc.) will aim to constrain the cosmological model and the true nature of dark energy with unprecedented accuracy. This very high required accuracy for surveys is primarily driven by the number density of clusters. Cluster mass dark matter haloes represent rare objects that lie on the exponential tail of the halo mass function. As they arise from extremely rare density fluctuations, not only are they probes of cosmological parameters such as $\Omega_m$, they are sensitive to any non-Gaussianities present in the primordial density field \cite[e.g.][]{Matarrese00,Pillepich10,Marian11}. Accurately recovering cluster properties may present a special challenge to certain object finders since many clusters are unrelaxed or in the process of merging. For instance, it has been shown that FOF groups will `merge' whereas the equivalent haloes are found by other algorithms such as SO \cite[e.g.][]{Klypin11,Behroozi12}. Care must also be taken when unbinding candidate clusters since merging objects will produce large velocity offsets, which subsequently will result in particles being considered unbound to either of the two merging halo cores while still possibly being bound to the merging system as a whole. 

In order for the aforementioned surveys to be used for precision constraints on the cosmological parameters the 2-point galaxy correlation function (or alternatively the power spectrum) needs to be determined to unprecedented accuracy \citep[e.g.][]{Smith13}. This also entails an unparalleled understanding of the galaxy bias \citep[e.g.][]{Nuza13}, non-linear effects \citep[e.g.][]{Chuang12} and any non-Gaussianities present in the primordial density field \cite[e.g.][]{Matarrese08,Jeong09,Pillepich10}. Using the correlation function for these studies hinges on accurately determining the spatial distribution of haloes {\em and subhaloes} and understanding how galaxies are hosted in these dark matter potential wells. The spatial distribution of subhaloes of different masses is particularly sensitive to the subhalo finder in question and will leave its imprint in the clustering properties \citep{Zentner05b,Tinker10,Watson11}. For instance, an analysis of the MultiDark simulation\footnote{\url{http://www.
multidark.org}} with \rockstar, \ahf, and \bdm\ shows that missing subhaloes close to the centre of the host will lead to a dramatic decrease of the two-point correlation function on small scales (Knebe et al., in prep.). Unfortunately this phenomenon has not arisen in any of the previous comparison projects as it requires a comparison of (sub-)halo finders in large-scale structure simulations with sufficient resolution to resolve subhaloes. Therefore, it is important to know the limitations of the halo finder applied to such large-scale simulations.


LSS measurements and statistics also offer many avenues for investigating deviations from the standard $\Lambda$CDM cosmology. From the theoretical point of view, studying LSS in simulations with differing dark matter models, dark energy types or modified gravity theories (cf. also \Sec{sec:AppModifiedGravitySimulations} below), we conclude from the results presented here and in previous comparisons that it is important to always use the same halo finder. Care should be taken when choosing that finder as some may behave poorly in certain situations. For instance, if a particular cosmological simulation results in a very high merger rate and many filamentary structures, one might expect the mass function produced by FOF to be artificially biased to high masses due to the presence of filaments connecting haloes (see related discussion on halo mergers in \citealp{Klypin11,Behroozi12}). Furthermore, one should in general be careful when studying questions such as the universality of the mass 
function or trying to reproduce simulation results with Excursion Set Theory \citep[][]{Zentner07}. In both cases the halo finder should be specified and the main parameters given to the end-user who develops the theory. 

From the observational point of view, all of these issues boil down to the fact that end-users should have a clear understanding of the main characteristics of the halo finder used in the \nbody\ simulation -- and its limitations. The onus is on developers of halo finders to make the object selection criteria clear (and to clearly define the code's deficiencies) preferably even allowing observational astronomers to apply the halo finder to their data. For that reason, for instance, \cite{Tinker08} have argued in favour of using SO halo catalogues (or other like algorithms which use density to define halo edges) over FOF halo catalogues, since defining haloes using isodensity contours offers a more direct comparison to X-ray observations of clusters. However, since this this is not always possible and nevertheless also involves systematics, it is important to know that (halo finder) errors will propagate to uncertainties in estimates of cosmological parameters, for example the dark energy equation of 
state. We close by remarking that it is not only the halo finding
community that is under pressure to supply high-precision results:
\citet{Smith13} have recently shown that rigorous convergence testing
of $N$-body codes themselves is also needed to meet the future challenges of precision cosmology.

\subsection{Near-Field Cosmology}\label{sec:AppNearCosmology}

Ever since the overmerging problem in simulations of cosmic structure
formation had been overcome \citep{Klypin99}, the observed/simulated
Milky Way satellite mass function has been used as a test for our
standard cosmology, $\Lambda$CDM. It failed and immediately led to the
so-called missing satellite problem \citep{Moore99,Klypin99s}, i.e. an
over-prediction of subhaloes as compared to observations. Further, the
possibility of numerically modeling and studying the dynamics of halo
substructure has spawned a new `industry' of (computational)
Near-Field Cosmology, a term coined by \citet{Freeman02}. Since then a multitude of articles emerged all dealing with the analysis of subhaloes and their orbits within dark matter host haloes \citep[e.g.][for the first such studies]{Stoehr02,DeLucia04,Diemand04b,Gill04b,Gao04,Kravtsov04a}.

We have seen that the theoretical subhalo mass function suffers from variations of $\sim10$~per cent arising from using different subhalo finders (cf.~\Fig{fig:subhaloes}). Observationally its uncertainties are currently dominated by incompleteness effects at the low mass end \citep[e.g.][]{Simon07, Walsh09}. In any case, these uncertainties are unlikely to account for the disparity first pointed out by \citet{Klypin99s} and \citet{Moore99} and still not fully explained. The disagreement between the theoretical and observed satellite mass function has driven a large number of studies attempting to reduce this disparity using baryonic physics \cite[e.g.][]{Nickerson11}, e.g. invoking stellar feedback \cite[e.g.][]{MacLow1999}, reionization \cite[e.g.][]{Bullock00,Benson02}, and ram pressure stripping \cite[e.g][]{Mayer06}, or by invoking different dark matter models
\cite[e.g.][]{Colin00,Bode01,Knebe02,Knebe08wdm,Maccio10wdm,Lovell12}
which have stronger effects on the abundance of 
subhaloes than the scatter introduced by using different halo finders.

However, the cumulative mass (or luminosity) function is not the sole place where tensions exist between theory and observations. The properties of the most massive subhaloes extracted from numerical simulations of Milky Way like haloes are at odds with those observed \cite[e.g.][]{Boylan-Kolchin11,DiCintio11,Boylan-Kolchin12,Vera-Ciro12,DiCintio12}. Constraining the masses of individual Milky Way satellites is observationally challenging due to the large uncertainties in the 3D velocity dispersion and the low number of member stars for the ultra-faint dwarf galaxies (errors of order 50~per cent; \citealp{Simon07}). Still, upcoming surveys like GAIA\footnote{\url{http://gaia.esa.int}} are expected to improve both the completeness and the accuracy of the mass determination of Milky Way satellites. Using Jeans modeling with various anisotropy parameters \cite{Wolf10} have shown that the mass within the 3D deprojected half light radius $M_{1/2}$ is an accurate mass estimator for dispersion supported systems. However, 
they still quote uncertainties of 10-20~per cent, well above the scatter in individual subhaloes from numerical simulations seen in this study and hence also here code scatter is not the dominant problem.

The orbits of satellites and their alignment with the surrounding environment also offers an avenue for (computational) Near-Field Cosmology \cite[e.g.][]{Zentner05,Libeskind05,Libeskind10,Knebe10a,Deason11,2011MNRAS.413.3013L,2011MNRAS.417L..56K,Libeskind12}. However, knowing the accurate 6D position and velocity of a (Milky Way) satellite is crucial for any kind of orbit determination \citep{Lux10} with the exception of known tidal streams \cite[e.g.][]{2010ApJ...714..229L}. The positions and bulk velocities of subhaloes are well constrained, with accuracies of $\lesssim1$~per cent. This is significantly better than the current observational constraints for both the satellites within the Milky Way and Andromeda \cite[e.g.][]{Lux10,Watkins13} which are needed to test the significance of the recently proposed thin plane of co-rotating subhaloes in Andromeda \citep{Ibata13}  and the previously reported disk of satellites in the MW  \citep{Kroupa05,Metz08,Metz09}.  Additionally, determining the centre of a galaxy and how that corresponds to the host halo's centre is not so clear cut observationally. 
Different observational methods will give different results, with
these differences being more pronounced for irregular galaxies, such
as the LMC and SMC. For galaxies close to the detection limit
false/non-detection of member stars add to the uncertainties in the
position of the centre. For example the centering of the ultra faint
Milky Way dwarf galaxies \cite{2007ApJ...670..313S} that contain $\sim100$ stars is accurate to a few arc seconds. This is better for the smaller dwarf galaxies that contain more stars, e.g. ultra compact dwarfs.

In any case, given the observational challenge to determine positions, velocities and mass (even for future missions such as GAIA) as well as the complexity of the (baryonic) physics involved in properly modeling satellite galaxies the differences found across subhalo finders are the smallest source of uncertainty.

\subsection{Streams}\label{sec:AppStreams}
Haloes contain not only bound subhaloes but a wealth of substructure such as tidal debris from disrupted subhaloes. Streams and other unbound structures may constitute an important component of the make-up of a halo \cite[e.g.][]{Carollo:2007p1011,Helmi:2008p942}. Observationally the past years have seen the discovery of tremendous amounts of such structures, primarily in the stellar component: the Orphan stream  \citep{2007ApJ...658..337B,2008MNRAS.389.1391S,2010ApJ...711...32N},  the Monoceros stream
\citep{2002ApJ...569..245N,Ibata03,Yanny03},
the Sagittarius stream \citep{2001ApJ...547L.133I}, moving groups \citep[e.g. Hercules Corona Borealis,][]{2010MNRAS.405.1796H}, the Hercules-Aquila cloud \citep{2007ApJ...657L..89B}, the Cetus polar stream \citep{2009ApJ...700L..61N}, the Virgo stellar stream
\citep{2002ApJ...569..245N,2007ApJ...660.1264M},
the Virgo overdensity \citep{
2007ApJ...660.1264M,2008ApJ...673..864J}, and the Pisces overdensity
\citep{2007AJ....134.2236S,2009MNRAS.398.1757W}.
All this observational work raises the question of whether or not we can also find these structures in cosmological simulations.

Unfortunately, typical (sub-)halo finders cannot be used to detect streams in galaxy formation simulations as most are configuration/density based finders. Such codes effectively collect particles by searching for clustering in configuration-space and remove false positives using an unbinding procedure, converting them to pseudo phase-space finders. As streams will not appear as an overdensity in configuration-space, these finders simply cannot identify them, despite their pseudo phase-space nature. 

\par
In order to detect streams, previous methods have focused on tracking the particles of an accreted halo in time. However, requiring temporal information severely limits the search for streams and such techniques cannot be directly applied to observational data sets, which only provide an instantaneous snapshot of `particle' (star) positions. Moreover, particles originating from the same progenitor need not be dynamically related to one another, especially if the progenitor has completed many orbits in an evolving potential. Tracking necessitates the application of a dynamically motivated criterion on the particles as opposed to an energy based one.

\par
However, there are several promising avenues for detecting streams without requiring temporal information \citep[][]{Sharma06,Diemand08,Zemp09,Ascasibar10,Elahi11}, and we recently studied the performance of several substructure -- where substructure refers to both subhaloes and streams -- detectors in a separate paper (Elahi et al., submitted). By including velocity-space information in the initial particle collection, an object finder could in principle detect streams. The complication lies in the fact that due to their unbound nature, one cannot use unbinding to remove false positives. Naturally, due to the more complex phase-space volumes occupied by these structures, these methods may be worse than currently used particle tagging methods at identify streams in cosmological simulations \citep[e.g.][]{Warnick08,2010MNRAS.406..744C,2011ApJ...733L...7H,Rashkov12}. The advantage of these methods is that they do not require numerous snapshots, and could potentially be transferred to stream detection in real 
data sets, e.g. from SDSS\footnote{\url{http://www.sdss3.org/dr8}} or the upcoming GAIA mission. The comparison of various snapshot-based stream detectors to a full tracking code revealed that  basic properties of the total substructure distribution (mass, velocity dispersion, position) are recovered with a scatter of $\sim$20 per cent; and tidal debris with purities of $\sim$50 per cent -- where purity is defined as the fraction of particles in debris originating from a distinct progenitor halo (Elahi et al., submitted).

The wealth of extra information provided by identifying the
tidal features associated with a subhalo and tidal debris has a
number of applications. For instance, the orientation and shape of
a subhaloÕs tidal features could be incorporated into semi-analytic
models of galaxy formation to determine the morphology of the
satellite galaxy. The velocity distribution of tidal debris may have
significant ramifications for direct dark matter detectors \citep[e.g.][and see \Sec{sec:AppDarkMatterDetection}]{Fairbairn09,Kuhlen10,Kuhlen12}
We conclude that tidal debris fields and streams are an extremely
important field of research in the near-future when observations will
provide a means to fully facilitate them. Substructure finding is
moving in the right direction with the first indications of
successfully utilizing stream properties as a possible discriminator
between cosmological models, casting light on the nature of dark matter. And for the first comparison of codes capable of detecting tidal debris fields we refer the reader to \citet{Elahi13} -- a comparison project emerging from the 'Subhaloes going Notts' workshop.



\subsection{Gravitational Lensing}\label{sec:AppLensing}
Gravitational lensing is the astrophysical phenomenon whereby the
propagation of light is affected by the distribution of mass in the
universe. It therefore provides a unique and direct probe of the
matter distribution in and about cosmic structures such as dark matter
haloes. Gravitational lensing actually comes in three flavours,
i.e. strong, weak, and micro-lensing. Strong lensing leads to multiple
images of a background sources whereas for weak lensing the field of
the deflector is only strong enough to produce generic distortions
detectable in a statistical sense. For historical reasons strong
lensing events leading to very small angular separations between the
multiple images (as produced by stars, for instance) is referred to as
micro-lensing and is not of immediate relevance for halo finding. The
former two nevertheless are and hence shall be discussed further here.

\paragraph*{Strong Lensing}
In the strong lensing regime one can use the particulars of the
multiply-imaged background sources to infer the mass distribution of
the lens. But this (iterative) process requires reliable models for
these mass distributions which come primarily from simulations of
cosmic structure formation. And these in turn made use of one or other
halo finder to find the dark matter haloes in the first place. But it is apparent from our own and other studies that halo density profiles and the concentration-mass relation are  subject to biases introduced by the applied halo finder  \cite[e.g.][]{Cole96,Lukic09,More11,Falck12,Bhattacharya11}. Further, more sophisticated models have to drop the assumption of spherical symmetry and make use of the triaxial shape of the dark matter \citep[e.g.][]{Jing02,Oguri05,Gavazzi05,Sereno12,Limousin13} distribution. While there is no doubt that simulated dark matter (sub-)haloes have triaxial shapes
\citep[e.g.][]{Warren92,Allgood06,Knebe08,Vera-Ciro11}
we have just seen that the scatter in the actual value of, for instance, the halo sphericity is at best as small as 5 per cent (cf. \Fig{fig:recoveryshapespin} and \Tab{tab:summaryrecovery}). And baryonic processes -- routinely simulated these days too (cf. \Sec{sec:methodsbaryons}) -- and the precise way how to measure shapes \citep[cf.][]{Zemp11} will certainly also influence the applicability of halo catalogues to strong lensing studies.

While major improvements in the observations and numerical simulations
have not yet significantly alleviated the aforementioned satellite crisis (cf. \Sec{sec:AppNearCosmology}), new techniques -- involving gravitational lensing -- have been proposed for the indirect and direct detection of subhaloes, primarily studying flux ratio anomalies introduced by the presence of substructure 
\citep[e.g.][]{Mao98,Metcalf01,Dalal02,Koopmans05}.
However, the modeling of these substructures for gravitational lensing heavily depends on whether the simulation includes baryonic effects or not \citep{Maccio06,Xu09} and certainly on the capabilities of the applied halo finder: will the finder be able to properly find all substructures? And this not only refers to bound subhaloes but also more diffuse streams that still might have a surviving core capable of strong lensing. Maybe the reported \textit{under-}prediction of subhaloes in galaxy clusters in \LCDM\ simulations \citep{Maccio06,Xu09}
 is still related to halo finder issues?

\paragraph*{Weak Lensing}
Weak gravitational lensing has become one of the key probes of the
cosmological model, dark energy, and dark matter, providing insight
into both the cosmic expansion history and large scale structure
growth history
\citep[][]{Kaiser93,Wilson96a,Bartelmann01a,Schneider06}.
Early work on measuring halo mass distributions using weak galaxy-galaxy
lensing was performed by \cite{Kaiser93}, \citet{Wilson96a,Wilson96b},
\citet{Schneider97a}, and \citet{Schneider97b}. Following these,
\citet{Natarajan00} proposed a technique for using weak gravitational
lensing to measure the ellipticity of haloes. Recently a lot of effort
has gone into the application of measuring the shear of the matter
distribution by means of statistical distortions of the background
images of galaxies
\citep[][]{Mellier99,Refregier03,Schneider06,Munshi08}. Examining the substructure content of galaxy clusters \citet{Natarajan07} find good agreement 
between the distribution of substructure properties retrieved using
their weak lensing analysis and those obtained from the Millennium simulation. And weak lensing is also being used to infer halo shapes both observationally 
\citep[][]{Natarajan00,Hoekstra04,Mandelbaum06,vanUitert12}
as well as in simulations \citep{Bett12}.

All this indicates that the same limitations coming from halo finders
that apply to strong lensing will also affect the applications of weak
lensing. In addition, the large-scale distribution of haloes will be relevant for the determination of cosmological parameters and hence problems in that area (cf. \Sec{sec:AppLSS}) naturally enter here, too.

\subsection{Dark Matter Detection}\label{sec:AppDarkMatterDetection}
Dark matter detection relies on two distinct avenues. Indirect searches are attempting to observe the possible secondary particles that originate from either the decay or the self-annihilation of dark matter particles. Direct detection experiments rely on identifying the nuclear recoil signature of a dark matter particle colliding with a target atom in the detector volume. Each of these methods are sensitive to the dark matter substructures present in dark matter haloes, though in different fashions. As the emission from dark matter decay or self-annihilation scale with the density and the square of the density respectively, subhaloes can enhance the signal relative to that predicted for a smooth dark matter halo
\citep[e.g.][]{Stoehr03,Diemand06,Elahi09b,Maciejewski11,Blanchet12,Gao12}.
But the measured contribution of resolved substructure is affected by both the numerical resolution of the underlying simulation and the capability to correctly identify substructure. Direct detection is extremely sensitive to the local velocity distribution of dark 
matter, thus both the presence of bound subhaloes and unbound tidal streams can significantly distort the signals observed by these detectors. Consequently, for a theoretical modeling of the expected signal, an accurate determination of the full substructure distribution function, from bound subhaloes to tidal debris, and how substructure alters the density profile and velocity distribution of a halo is very important.

While we have seen here and in previous works \citep{Onions12,Onions13,Knebe13} how different finders perform with respect to the identification of (sub-)haloes, a comparison of the radial profile of the respective (sub-)haloes would be required to shed light into the subject of dark matter detection: while different finders might in fact find the same objects with comparable masses, are the associated particles distributed in the same way? While one naively might answer `yes' to this question, it should not be taken for granted. It all comes down again to the particle collection method and unbinding procedure. For instance, different ways of calculating the reference escape velocity as a function radius during the unbinding might lead to preferential removal/keeping of particles in certain regions like the centre or the outskirts and vice versa. 

However, we are not arguing against the established notion that dark
matter haloes follow a universal density profile \citep[first reported
by][]{Navarro95,Navarro96,Navarro97}. But even this universality has
been questioned with the first indications of scatter in the profiles
between haloes by \citet{Jing00} and \citet{Bullock01}. Presently the
question of the precise value of the logarithmic central slope is
still debated -- a question related to the `cusp-core crisis of
\LCDM', i.e. the discrepancy between cuspy profiles as predicted by
simulations and cored profiles as derived observationally
\citep[see][for a recent review]{deBlok10}. All we can do at the
moment is to alert the reader to the fact that any possible dependence
of the radial halo profile (and in particular the much sought-after central slope) on the applied halo finder has not been tested yet. As putative as this might be, it cannot be ruled out at this stage.

Further, the inclusion of baryons in simulations as well as during the halo finding process will certainly alter the (central) density profile
\citep[e.g.][]{Blumenthal86,Tissera98,Romano-Diaz10,DiCintio11,Zemp12}.
But the uncertainties in the modeling of baryonic processes will likely leave us with a larger scatter than finder-to-finder variations and hence should be of greater concern in the end.

\subsection{Modified Gravity Simulations}\label{sec:AppModifiedGravitySimulations}
Though possibly less developed than \LCDM\ simulations, there now exist several suites of \nbody\ simulations that solve the gravitational evolution of particles according to a particular model that modifies general relativity (GR) as an alternative to dark energy \cite[e.g.][]{Schmidt09,Zhao11a,Zhao11b,Li12} or assume some other form of modifications to gravity and/or the expansion of the Universe
\citep[e.g.][]{Knebe04b,Llinares08,Li11,Carlesi12,Hellwing12,Baldi12}.
Currently, these studies primarily focus on the (sub-)halo mass function, particularly the high mass end, using the frequency of massive bound objects as discriminators for these models \citep[e.g.][]{LoVerde11,Hoyle11,Carlesi11,Baldi12b}.

However, along with the production simulation code the halo finding
algorithm also requires adjustment in compliance with the adopted non-standard model: the calculation of many halo properties (for example virial radius, rotation curve and thus \Vmax, spin, and concentration) assume general relativity (GR) and so must be modified for modified gravity (MG) simulations according to the specific model. This is obviously best left to the users and not the code providers to derive, but it is important that enough GR-independent (i.e. dynamical) parameters are included in halo catalogs to allow for calibration. This also means that results for MG simulations depend strongly on the unbinding procedure, which itself depends on environment. For example, in most MG models gravity is very different for field haloes and haloes in low-density environments, so the unbinding procedures in the above codes would be invalid, although, as noted in Section \ref{sec:recoveryfieldhaloes}, the unbinding procedure has little impact on the {\em 
mass} of field haloes. In clusters, GR is recovered in most models, so in principle the unbinding would be the same. Regardless of the details, the unbinding procedure in MG models must be addressed as we saw in Section \ref{sec:precisioncosmologyunbinding} the absence of unbinding leads to large scatter in the subhalo mass functions and without it, a number of physical parameters, such as spin, are poorly recovered. 

\subsection{Summary}\label{sec:AppSummary}
Fortunately, we find that most applications of (sub-)halo catalogues are not affected by the uncertainties discussed here.
However, future goals of improved precision will demand higher accuracy in the determination of (sub-)halo properties (see discussion in \Sec{sec:precisioncosmology}).
It should be emphasized that the potential errors introduced by halo finders, as well as the interplay with observable quantities, come with the proviso that the baryonic physics and the biases and changes it introduces are well understood.
This is by no means the case, but given that baryons are unlikely to drastically alter the performance of current halo finders or alter their systematic biases, we argue that testing and improving these object finders using pure dark matter simulations is sufficient for the time being.

\section{Summary \& Conclusions}\label{sec:conclusion}
With the ever increasing size and complexity of fully self-consistent
simulations of cosmic structure formation, the demands upon object
finders for these simulations has simultaneously grown. These codes
not only need to locate haloes residing in the cosmic web, they are
also often required to (correctly) identify substructure living in the
inhomogeneous background of their host haloes. The last decade, and in
particular the last couple of years, has seen an immense boost in the
number of techniques and codes specifically developed for these
tasks. But while a lot of effort has been put into validating the
results coming from different simulation codes \citep[e.g.][]{Frenk99,
Knebe00, OShea05, Agertz07, Heitmann08cccp, Tasker08}, until recently
it was unclear how different structure finders compare. To this extent
we initiated the Halo Finder Comparison Project that gathered together
all experts in the field and has so far led to a series of comparison
papers \citep[][Elahi et al. submitted]{Knebe11,Onions12,Onions13,Knebe13}, emerging out of two 
workshops.\footnote{``Haloes going MAD''
(\url{http://popia.ft.uam.es/HaloesGoingMAD}) and ``Subhaloes going Notts''
(\url{http://popia.ft.uam.es/SubhaloesGoingNotts})}

In this overview article, we summarise the results of both workshops (partially published in the two previous comparison papers, \citet{Knebe11} and \citet{Onions12}) and take the analysis one step further. We aimed at familiarizing the
reader with the general concepts commonly applied for halo finding
clearly separating methods from definitions: the method is intrinsic
to the code whereas the definition of a given halo property should be
independent of the applied finder. While both -- method and definition
-- should be physically motivated, the latter certainly is less
technical and should be made abundantly clear to any end-user of halo
finders. For instance, the same matter distribution describing a dark
matter halo could be assigned different masses if the definition for
the edge (and hence mass) is not identical. Or put differently, even
if two finders consider the same particles belonging to an object,
they may still write different masses to their halo catalogue
depending on the assumed values for $\Delta_{\rm ref}$ and/or
$\rho_{\rm ref}$ of \Eq{eq:virialradius}; or they may even apply
another mass/edge definition.

Any further differences between finders now originate from various
error sources, primarily the following two:

\begin{itemize}
 \item codes recover different values for the same property (even when using the same definition), and
 \item codes find different (numbers of) objects.
\end{itemize}

\noindent
While all previous comparison projects mainly dealt with distribution
functions of properties, we were so far unable to disentangle the
relative strength of these two sources. Here (i.e. in
Secs.~\ref{sec:recoveryset} through~\ref{sec:recoveryshapespin}) we
now focused on the former by restricting the analysis to only those
objects that had been found by all finders. We further utilized a
common post-processing pipeline that only deals with particle ID lists
as returned by each code and calculates all halo properties in the
same manner to avoid contamination from different definitions. Our
results are best summarized in \Tab{tab:summaryrecovery} indicating
that for the most basic properties (i.e. position, velocity, 
\Vmax, and possibly mass) and structures that are found by all participating finders the agreement across codes is at the 1~per cent level, sufficient for
the so-called era of `Precision Cosmology' \citep{Smoot03,Primack05,Coles05,Primack07}. Note, that these errors have been derived as lower limits for subhalo catalogues based upon
a common subset of objects and post-processing pipeline avoiding scatter from various distributions. On the other hand, finding distinct haloes rather than substructure is less challenging and the errors can be expected to be of this order.

More involved halo properties such as spin parameter suffer from a larger scatter \citep[see also][]{Onions13}. However, we also need to acknowledge
that the general (sub-)halo mass and \Vmax-functions including all objects identified by each finder suffer from a larger
scatter than given in \Tab{tab:summaryrecovery}. This is accounted for (though not explained) by different
numbers of objects: while our common set consists of some 800 haloes,
\Tab{tab:excessobjects} clearly shows that there exist of the same
order of objects \textit{not} found by all the other codes. These
`excess' objects then lead to an upwards boost in the (cumulative)
distribution functions. We have shown (\Fig{fig:superplot-mass-level4-nbins4}) that the majority of these missing objects are small, containing less than a few hundred particles in general. Further, they occur at all radii and are not predominantly near the centre of the halo. We therefore suggest that if a well matched, high purity catalogue is required for the scientific study in question that a higher particle limit than the usual 20 is required and that adopting 300 particles, the limit usually suggested for obtaining stable derived halo properties such as the spin parameter would be a good idea.

We further investigated the possible origin of the differences in halo
properties when using different finders by trying to decode the
influence of varying methodologies in \Sec{sec:precisioncosmology}. We
found that both the collection of particles and the particulars of the
unbinding procedure have an impact. In particular, some codes return
unbound particles to the pool of all particles to be considered bound
to any other object whereas other codes completely remove unbound
particles from the set. However, the characteristics of how to obtain
the potential entering the formula for the escape velocity during the
unbinding appears to have only marginal effects.

But this also brings us back to (some of) the points raised in \Sec{sec:methodsmass}
and discussed during the last ``Subhaloes going Notts'' workshop: what is the proper
definition for the mass of a halo? Practically all codes prune their initial particle
collection by some sort of unbinding procedure. But for subhaloes, for instance,
`boundness' may not be that well-defined; and remember, all codes extract the
subhalo particles and remove unbound particles as if the object were in isolation.
But what is the right way to treat subhaloes then and remove particles that do not 
belong to it? One of the bullet points in \Sec{sec:methodsmass} was to define
objects in general (and not only subhaloes) dynamically, i.e. only those particles
that stay with the halo over (at least) a dynamical time should be considered `bound'.
These thoughts naturally lead to potential refinements of halo finding techniques.

Possible improvements have also been briefly touched upon in this article pointing
towards halo tracking methods: most workers in the field and end-users
of halo catalogues, respectively, are not only interested in single
temporal snapshots of a simulation but also like to trace objects
backwards (or forward) in time. And it has recently been shown that
this approach can be used to actually `correct' halo catalogues
\citep{Behroozi12b}. Further, one of the halo finders presented here
is in fact based upon this approach \citep[HBT,][]{Han12}.
We reiterate that basically all the results presented here and elsewhere
are based upon a single snapshot analysed at redshift $z=0$. We leave comparisons at
higher redshift where, for instance, mergers play a major role, to a future study/workshop.

We closed the paper with a discussion of the relevance of halo finding
for various astrophysical applications such as galaxy formation
(either semi-analytical modeling or direct simulation), the
interpretation of large-scale structure surveys, near-field cosmology,
gravitational lensing, dark matter detection, (stellar) streams, and
modified gravity models. While the requirements are quite diverse we
nevertheless conjecture that intrinsic uncertainties in the respective
application might be larger than variations introduced by using
different halo finders. For example, the adoption of different
sub-grid physics to model galaxy formation will certainly lead to more
pronounced variations in the final results than changing the halo
finder for a given model. Nevertheless, we are not claiming that the
observed finder-to-finder scatter should be neglected: only with
credible and reliable halo catalogues can we adequately (and
scientifically) address the open questions. And there appears to
remain some work to be done to fully align the outcomes of the different halo finders.

We conclude that while the agreement across different halo finding
techniques is converging towards the requirements for precision
cosmology, there is still room for improvement. It remains unclear
where part of the observed scatter stems from and why all finders
do not find the same objects. We aim to address these issues at the next
halo finder comparison workshop, due to take place in 2014.

\section*{Acknowledgments}
This paper was initiated at the Subhaloes going Notts workshop in Dovedale, UK, which was funded by the European Commissions Framework Programme 7, through the Marie Curie Initial Training Network Cosmo- Comp (PITN-GA-2009-238356). It also uses data and results from the first halo finder comparison workshop ``Haloes going MAD'' in Madrid, Spain, which was funded by the ASTROSIM network of the European Science Foundation (Science Meeting 2910).

AK is supported by the {\it Spanish Ministerio de Ciencia e Innovaci\'on} (MICINN) in Spain through the Ram\'{o}n y Cajal programme as well as the grants AYA 2009-13875-C03-02, AYA2009-12792-C03-03, CSD2009-00064, CAM S2009/ESP-1496 (from the ASTROMADRID network) and the {\it Ministerio de Econom'a y Competitividad} (MINECO) through grant AYA2012-31101. He further thanks Calexico for tres avisos.

HL acknowledges a fellowship from the European Commissions Framework Programme 7, through the Marie Curie Initial Training Network CosmoComp (PITN-GA-2009-238356).

YA receives financial support from project AYA2010-21887-C04-03 from the  MICINN (Spain), as well as the Ram\'{o}n y Cajal programme (RyC-2011-09461), now managed by the MINECO (fiercely cutting back on the Spanish scientific infrastructure).

PB received support from NASA HST Theory grant HST-AR-12159.01-A and was additionally supported by the U.S. Department of Energy under contract number DE-AC02-76SF00515.

JC is supported by a contract from MICINN/MINECO (Spain) through the grant AYA2009-12792-C03-03 from the PNAyA.

RDT is supported by the MICINN (Spain) through the grant AYA2009-12792-C03-03 from the PNAyA, as well as by the regional Madrid V PRICIT program through the ASTROMADRID network (CAM S2009/ESP-1496) and the ''Supercomputaci\'on y e-Ciencia'' Consolider-Ingenio CSD2007-0050 project. She also thanks the computer resources provided by BSC/RES (Spain) and the Centro de Computaci\'on Cientif\'ica (UAM, Spain).

BF acknowledges support from the Gordon and Betty Moore Foundation and the STFC grants ST/H002774/1 and ST/K0090X/1.

JH acknowledges a fellowship from the European Commissions Framework Programme 7, through the Marie Curie Initial Training Network CosmoComp (PITN-
GA-2009-238356), too.

SIM acknowledges the support of the STFC Studentship Enhancement Programme (STEP).

SP also acknowledges a fellowship from the European Commission's Framework Programme 7, through the Marie Curie Initial Training Network CosmoComp (PITN-GA-2009-238356) as well as from the PRIN-INAF09 project ``Towards an Italian Network for Computational Cosmology''.  SP and VQ thank partial financial support from Spanish MINECO (grant AYA2010-21322-C03-01).

MZ is supported by a 985 grant from Peking University and the International Young Scientist grant 11250110052 by the National Science Foundation of China.

We thank the anonymous referee for constructive comments that helped to improve the paper.

This research has made use of NASA's Astrophysics Data System (ADS) and the arXiv preprint server.

\bibliographystyle{mn2e}
\bibliography{archive}

\appendix

\section{Code performance} \label{app:codeperformance}
\begin{table*}
  \caption{Timing results (in seconds) for analysing both Aquarius A-5 and A-4 at redshift $z=0$ on a dedicated compute server (all using one thread and the same compiler.}
\label{tab:timing}
\begin{center}
\begin{tabular}{lcccccccc}
\hline
Data	& \adaptahop	& \ahf	& \grasshopper	& \hbt	& \rockstar	& \stf		& \subfind		& \voboz \\
\hline
A-5   & 170 		& 42		& 108		& ---		& 24			& 52		& 209		& 3533 \\
A-4	& 1966		& 359	& 997		& 2940	& 227		& 503	& 2238		& 7469\\
\end{tabular}
\end{center}
\end{table*}

One question that gets repeatedly asked is the relative speed of the
different codes. Table~\ref{tab:timing} gives the time taken in
seconds to analyse the Aquarius A-5 and A-4 datasets at redshift $z=0$ on the same
dedicated compute server. All codes used one thread and the same compiler.
These numbers should be taken with a large
pinch of salt. For instance, \hbt, a halo tracking finder, needs to
analyse multiple outputs to produce results (even though only the time for the $z=0$ analysis is reported here) and \voboz\ timings are
dependent on the size of the surrounding region chosen. This included the entire
simulation in the A-5 case, but subregions were analyzed in other cases. \mendieta\ also
returned timing information, but only after the workshop (i.e. not run on the same machine as the other codes) stating that A-5 (A-4) took 328 (25844) seconds on a Xeon E5520 cpu using only one core. Aquarius A-4 contains roughly 8 times more particles than
Aquarius A-5, and all the codes returning data appear to scale
relatively well. The take home message is simple: for a wide variety
of theoretical approaches it is possible to extract a good subhalo
catalogue from the Aquarius A-5 dataset in a few minutes, whereas for
the A-4 dataset this process takes of order an hour even on a single
processor. 

\section{Cross-Correlations} \label{app:crosscorrelations}
\subsection{Procedure to obtain the common set}
In order to identify the common set of objects used in \Sec{sec:recovery} haloes were required to be within $250$ kpc/h of the fiducial host center $x=57060.4$\hkpc, $y=52618.6$\hkpc/h, $z=48704.8$\hkpc\ and to have more than 20 particles. 	For the purposes of the analysis in this paper, the counterpart to halo $A_1$ in catalog 1 from all haloes $B$ in catalog 2 is computed by finding that object $B_i$ in catalog 2 which maximizes a merit function $M_i$
\begin{equation}
	M_i=\frac{N_{A_1 B_i}^2}{N_{A_1}N_{B_i}}
\end{equation}
where $N_{A_1B_i}$ is the number of shared particles between halo $A_1$ and halo $B_i$, $N_{A_1}$ the number of particles in $A_1$ and $N_{B_i}$ the number of particles in $B_i$.

For each of the nine considered finders for this exercise a list containing the matches to each of the remaining eight finders has been generated. We then cross-references these lists restricting the analysis only to the set of objects found by every halo finder, i.e. a table has been created listing only those objects for each finder that form part of the common set. 
	
\subsection{Alternative Merit Functions}
The merit function deployed in this paper is based upon the \texttt{MergerTree} tool from the \ahf\ halo finder distribution \citep{Knollmann09} and has been previously used successfully \citep{Klimentowski10,Libeskind10}. To verify the suitability of this choice we present a brief comparison of alternative merit functions. These are detailed in Table \ref{counterpartalgos} and their success at maximizing the set of of counterparts is summarized in Table \ref{excess}. Note that for this exercise we worked with the Aquarius A-5 data set and hence the lower numbers than those given in \Tab{tab:excessobjects}. We can see that our standard merit function $M_i$ and the somewhat simpler $M^a_i$ have exactly the same performance; indeed their global set of common objects identified is identical. This is due to the fact that the host halo is not included in the cross-correlation: the normalisation to $N_{A_1}N_{B_i}$ in $M_i$ is being used to avoid matching subhaloes to their host in case of inclusive particle ID lists. Other, more complicated metrics only degrade performance. We note that the common set of objects identified by $M^q_i$ and $M^r_i$ are proper subsets of the global set of common objects identified by $M_i$. And please note that the rather low number for the common set is determined by \hottd\ that only found a total of 58 subhaloes \cite[[cf. Table 1 in][]{Onions12}.

\begin{center}
\begin{table*}
\hfill{}
\caption{Counterpart merit functions. Note that some are combination of others and hence we show in bold-face those entering the actual comparison presented in \Tab{excess}.}
\begin{center}
\begin{tabular}{lll}
\hline
merit function & formula & description\\
\hline
\hline
${\bf M_i}$ & $\frac{N_{A_1 B_i}^2}{N_{A_1}N_{B_i}}$ & normalized shared particles\\
${\bf M^a_i}$ & $N_{A_1 B_i}$ & shared particles \\
${\bf M^c_i}$ & $r=\sqrt{(X_{A_1}-X_{B_i})^2+(Y_{A_1}-Y_{B_i})^2+(Z_{A_1}-Z_{B_i})^2}$ & $r$ is radial distance from $A_1$ to $B_i$\\
$M^b_i(P)$ & $P_{A_1 B_i}=|P_{A_1}-P_{B_i}|$ & property P difference\\
$M^v_i(P)$ & $\bar P_{A_1 B_i}=\frac{P_{A_1}-P_{B_i}}{2}$ & property P average \\
$M^w_i(P)$ & $\sigma_{P_{A_1 Bi}}=\frac{(P_{A_1}-\bar P_{A_1 Bi})^2+(P_{B_i}-\bar P_{A_1 Bi})^2}{2}$ & property P standard deviation \\
$M^h_i$ & $\frac{P_{A_1}+P_{B_i}}{2}$ & average property P value \\
$M^o_i(P)$ & $\frac{{M^h_i}^2}{P_{A_1B_i}}$  & normalized property P value average \\
$M^p_i(P)$ & $\log{\frac{1}{M^b_i(P)}}$& downweighted property P comparison \\
${\bf M^q_i}$ & $\log{\frac{1}{{M^c_i}^2}}$& downweighted radius $r^2$ comparison \\
${\bf M^d_i}$ & $M_i+ M^c_i$ & -\\
${M^e_i}$ & $M_i+ M^b_i(v_{\textrm{max}})$ & -\\
${\bf M^r_i}$ & $M^q_i+M^p_i(v_{\textrm{max}})$ & - \\
${\bf M^s_i}$ & $M^q_i+\sum_{p\in\{v_{\textrm{max},\textrm{mass}}\}} M^p_i(p)$ & - \\
${\bf M^t_i}$ & $\sum_{p\in\{v_{\textrm{max},\textrm{mass},n_{\textrm{vpart}}}\}} M^p_i(p)$ & - \\
${\bf M^u_i}$ & $\sum_{p\in\{v_{\textrm{max},\textrm{mass},n_{\textrm{vpart},b,c}}\}} M^p_i(p)$ & - \\
${\bf M^g_i}$ & $ M_i+ M^c_i + \sum_{p\in\{v_{\textrm{max},b,c}\}} M^b_i(p)$ & - \\
${\bf M^f_i}$ & $ M_i+ M^c_i+ M^b_i(v_{\textrm{max}})$ & -\\
\hline
\end{tabular}
\end{center}
\hfill{}
\label{counterpartalgos}
\end{table*}
\end{center}

\begin{center}
\begin{table*}
\hfill{}
\caption{Number of haloes in the common set and found in excess of it for the Aquarius A-5 data set.}
\begin{center}
\begin{tabular}{ccccccccccc}
\hline
metric & common objects & \multicolumn{9}{c}{excess objects}\\
 & 
 &
\ahf &
\hottd &
\hotsd &
\hbt &
\hsf &
\rockstar &
\stf &
\subfind &
\voboz \\
\hline
\hline
$M_i$ & 39 &
191 &
19  &
97  &
189 &
192 &
233 &
106 &
175 &
218 \\
$M^a_i$ & 39 &
191 &
19  &
97  &
189 &
192 &
233 &
106 &
175 &
218 \\
$M^q_i$ & 18 &
     212 &
      40 &
     118 &
     210 &
     213 &
     254 &
     187 &
     196 &
     239 \\
$M^r_i$& 18 &
     212 &
      40 &
     118 &
     210 &
     213 &
     254 &
     187 &
     196 &
     239 \\
$M^s_i$ & 19 &
     211 &
      39 &
     117 &
     209 &
     212 &
     253 &
     186 &
     195 &
     238 \\
$M^t_i$ & 18 &
     212 &
      40 &
     118 &
     210 &
     213 &
     254 &
     187 &
     196 &
     239 \\
$M^u_i$ & 18 &
     212 &
      40 &
     118 &
     210 &
     213 &
     254 &
     187 &
     196 &
     239 \\
$M_i^d$ & 1 &
     229 &
      57 &
     135 &
     227 &
     230 &
     271 &
     204 &
     213 &
     256 \\
$M_i^g$ & 1&
     229 &
      57 &
     135 &
     227 &
     230 &
     271 &
     204 &
     213 &
     256 \\
$M_i^f$ & 1&
     229 &
      57 &
     135 &
     227 &
     230 &
     271 &
     204 &
     213 &
     256 \\
$M_i^c$ & 1&
     229 &
      57 &
     135 &
     227 &
     230 &
     271 &
     204 &
     213 &
     256 \\
\hline
\end{tabular}
\end{center}
\hfill{}
\label{excess}
\end{table*}
\end{center}

\section{Code Descriptions} \label{app:codedescriptions}
While the information presented here can be found in various other publications, we nevertheless considered it helpful to compile it in one single place here. We are giving here a very brief and hopefully concise description of all those halo finders that participated in any of the comparison projects ``Haloes going MAD" \citep{Knebe11}, ``Subhaloes going Notts'' \citep{Onions12}, and ``Galaxies going MAD'' \citep{Knebe13}. And to facilitate the understanding of the mode of operation of each of the codes, we present a brief summary in \Tab{tab:codes}, too.

\begin{table}
  \caption{Brief summary of the codes listing the dimensionality of the (primary) metric to find objects, the assumed geometry, whether the code features an unbinding procedure and can sufficiently handle subhalo detection.}
\label{tab:codes}
\begin{center}
\begin{tabular}{lllllllll}
\hline
code             		& metric		& geometry		& unbinding	& subhaloes 	\\
\hline
 \sdfof			& 6D			& arbitrary			& no			& yes		\\
 \adaptahop 		& 3D			& spherical		& no			& yes		\\
 \ahf               		& 3D			& spherical		& yes		& yes		\\
 \asohf 			& 3D			& spherical		& yes		& yes		\\
 \bdm 			& 3D			& spherical		& yes		& yes		\\
 \fof				& 3D			& arbitrary			& no			& limited		\\
 \grasshopper		& 3D			& spherical		& yes		& yes		\\
 \hbt				& 3D+time	& arbitrary			& yes		& yes		\\
 \hot 				& 3D/6D		& arbitrary			& no			& yes		\\
 \hsf 				& 6D			& arbitrary			& yes		& yes		\\
 \jumpd			& 3D			& spherical		& no			& yes		\\
 \lanl 			& 3D			& spherical		& no			& no			\\
 \mendieta		& 3D			& arbitrary			& yes		& yes		\\
 \ntropyfof			& 3D			& arbitrary			& no			& no			\\
 \origami 			& 3D			& arbitrary			& yes		& no			\\
 \pfof 			& 3D			& arbitrary			& no			& no			\\
 \pso 			& 3D			& spherical		& no			& no			\\
 \rockstar    		& 6D			& arbitrary			& yes 		& yes 		\\
 \skid 			& 3D			& spherical		& yes		& yes		\\
 \stf				& 6D			& arbitrary			& yes		& yes		\\
 \subfind			& 3D			& arbitrary			& yes		& yes		\\
 \voboz			& 3D			& arbitrary			& yes		& yes		\\
\hline
\end{tabular}
\end{center}
\end{table}

\subsection{\sdfof} \label{sec:6dfof}
\sdfof\ is a simple extension of the well known FOF method
which also includes a proximity condition in velocity space.  Since
the centres of all resolved haloes and subhaloes reach a similar peak
phase space density they can all be found at once with \sdfof.
The algorithm was first presented in \cite{Diemand06}.  The
\sdfof\ algorithm links two particles if the following
condition
\begin{equation}
\frac{({\bf x}_1 - {\bf x}_2)^2}{\Delta x^2} +
\frac{({\bf v}_1 - {\bf v}_2)^2}{\Delta v^2} < 1
\end{equation}
is fulfilled.  There are three free parameters: $\Delta x$, the
linking length in position space, $\Delta v$, the linking length in
velocity space, and $N_\mathrm{min}$, the minimum number of particles
in a linked group so that it will be accepted.  For $\Delta v
\rightarrow \infty$ it reduces to the standard FOF scheme.  The
\sdfof\ algorithm is used for finding the phase space
coordinates of the high phase space density cores of haloes on all
levels of the hierarchy and is fully integrated in parallel within the
MPI and OpenMP parallelised code \textsc{PKDGRAV} \citep{Stadel01}.
For this work we used the following values: $\Delta x$ = 70 kpc (physical), $\Delta v$ = 250 km s$^{-1}$ and $N_\mathrm{min}$ = 15.

The centre position and velocity of a halo are then determined from
the linked particles of that halo.  For the centre position of a halo,
one can choose between the following three types: 1) the
centre-of-mass of its linked particles, 2) the position of the
particle with the largest absolute value of the potential among its
linked particles or 3) the position of the particle which has the
largest local mass density among its linked particles.  For the
analysis presented here, we chose type 3) as our halo centre position
definition.  The centre velocity of a halo is calculated as the
centre-of-mass velocity of its linked particles.  Since in
\sdfof\ only the particles with a high phase space density in
the very centre of each halo (or subhalo) are linked together, it
explains the somewhat different halo velocities (compared to the other
halo finders) and slightly offset centres in cases only a few
particles were linked.

Other properties of interest (e.g. mass, size or maximum of the
circular velocity curve) and the hierarchy level of the individual
haloes are then determined by a separate profiling routine in a post
processing step.  For example, a characteristic size and mass scale
definition (e.g. $r_\mathrm{200c}$ and $M_\mathrm{200c}$) for field
haloes based on traditional spherical overdensity criteria can be
specified by the user.  For subhaloes, a truncation scale can be
estimated as the location where the mass density profile reaches a
user specified slope.  During the profiling step no unbinding
procedure is performed.  Hence, the profiling step does not base its
(sub-)halo properties upon particle lists but rather on spherical
density profiles.  Therefore, \sdfof\ directly returned halo
properties instead of the (requested) particle ID lists.

\subsection{\adaptahop} \label{sec:adaptahop}
The code \adaptahop\ is described in Appendix A of
\citet{Aubert04}. The first step is to compute an SPH density for each
particle from the 20 closest neighbours. Isolated haloes are then
described as groups of particles above a density threshold $\rho_t$,
where this parameter is set to 80, which closely matches results of a
FOF group finder with parameter $b=0.2$. To identify subhaloes within
those groups, local density maxima and saddle points are
detected. Then, by increasing the density threshold, it is a simple
matter to decompose haloes into nodes that are either density maxima,
or groups of particles whose density is between two values of saddle
points. A node structure tree is then created to detail the whole
structure of the halo itself. Each leaf of this tree is a local
density maximum and can be interpreted as a subhalo. However, further
post-processing is needed to define the halo structure tree,
describing the host halo itself, its subhaloes and subhaloes within
subhaloes. This part of the code is detailed in \citet{Tweed09}; the
halo structure tree is constructed so that the halo itself contains
the most massive local maximum (Most massive Sub maxima Method: MSM).
This method gives the best result for isolated snapshots, as used in
this paper.

In more detail, \adaptahop\ needs a set of seven
parameters. The first parameter is the number of neighbours $n_{nei}$
used with a $k$D-tree scheme in order to estimate the SPH
density. Among these $n_{nei}$ neighbours, the $n_{hop}$ closest are
used to sweep through the density field and detect both density maxima
and saddle points. As previously mentioned, the parameter $\rho_t$
sets the halo boundary. The decomposition of the halo itself into
leaves that are to be redefined as subhaloes has to fulfil certain
criteria set by the remaining four parameters. The most relevant is
the statistical significance threshold, set via the parameter $fudge$,
defined via $(\langle\rho\rangle - \rho_t)/\rho_t > fudge/\sqrt{N}$,
where $N$ is the number of particles in the leaves. The minimal mass
of a halo is limited by the parameter $n_{members}$, the minimum
number of particles in a halo.  Any potential subhalo has also to
respect two conditions with respect to the density profile and the
minimal radius, through the parameters $\alpha$ and
$f_{\epsilon}$. These two values ensure that a subhalo has a maximal
density $\rho_{max}$ such as $\rho_{max} > \alpha\langle\rho\rangle$
and a radius greater than $f_{\epsilon}$ times the mean interparticle
distance. We used the following set of parameters
($n_{nei}=n_{hop}=20$, $\rho_t=80$, $fudge=4$, $\alpha=1$,
$f_{\epsilon}=0.05$, $n_{members}=20$).  It is important to understand
that all nodes are treated as leaves and must comply with
aforementioned criteria before being further decomposed into separate
structures. As for defining haloes and subhaloes themselves, this is
done by grouping linked lists of particles corresponding to different
nodes and leaves from the node structure tree. Further, the halo and
subhalo centres are defined as the position of the particle with the
highest density and the velocity is the centre-of-mass velocity. The halo edge corresponds to the $\rho_t$ density
threshold, whereas the saddle points define the subhalo edge.

Please note that \adaptahop\ is a mere topological code that
does \textit{not} feature an unbinding procedure. For substructures
(whose boundaries are chosen from the saddle point value) this may
impact on the estimate of the mass as well as lead to contamination by
host particles.

\subsection{\ahf} \label{sec:ahf}
The MPI+OpenMP parallelised halo finder
\ahf\footnote{\ahf\ is freely available from
  \texttt{http://www.popia.ft.uam.es/AHF}} \citep[\textsc{AMIGA}
Halo Finder,][]{Knollmann09}, is an improvement of the \mhf\
halo finder \citep{Gill04a}, which employs a recursively refined grid
to locate local overdensities in the density field. The identified
density peaks are then treated as centres of prospective haloes. The
resulting grid hierarchy is further utilized to generate a halo tree
readily containing the information which halo is a (prospective) host
and subhalo, respectively. We therefore like to stress that our halo
finding algorithm is fully recursive, automatically identifying
haloes, sub-haloes, sub-subhaloes, etc. Halo properties are calculated
based on the list of particles asserted to be gravitationally bound to
the respective density peak. To generate this list of particles we
employ an iterative procedure starting from an initial guess of
particles. This initial guess is based again upon the adaptive grid
hierarchy: for field haloes we start with considering all particles
out to the iso-density contour encompassing the overdensity defined by
the virial criterion based upon the spherical top-hat collapse model;
for subhaloes we gather particles up to the grid level shared with
another prospective (sub-)halo in the halo tree which corresponds to
the upturn point of the density profile due to the embedding within a
(background) host. This tentative particle list is then used in an
iterative procedure to remove unbound particles, not changing the halo centre though: In each step of the
iteration, all particles with a velocity exceeding the local escape
velocity, as given by the potential based on the particle list at the
start of the iteration, are removed.  The process is repeated until no
particles are removed anymore. At the end of this procedure we are
left with bona fide haloes defined by their bound particles and we can
calculate their integral and profiled quantities.

The only parameter to be tuned is the refinement criterion used to
generate the grid hierarchy (usually set to 3-4 particles per cell) that serves as the basis for the halo tree
and also sets the accuracy with which the centres are being
determined. The virial overdensity criterion applied to find the
(field) halo edges is determined from the cosmological model of the
data though it can readily be tailored to specific needs; for the
analysis presented here we used $200\times\rho_{\rm crit}$. For more
details on the mode of operation and actual functionality we refer the
reader to the two code description papers by \citet{Gill04a} and
\citet{Knollmann09}, respectively.

\subsection{\asohf} \label{sec:asohf}
The \asohf\ finder \citep{Planelles10} is based on the
spherical overdensity (SO) approach. Although it was originally
created to be coupled to an Eulerian cosmological code, in its actual
version, it is a stand-alone halo finder capable of analysing the
outputs from cosmological simulations including different components
(i.e., dark matter, gas, and stars). The algorithm takes advantage of
an AMR scheme to create a hierarchy of nested grids placed at
different levels of refinement. All the grids at a certain level,
named patches, share the same numerical resolution. The higher the
level of refinement the better the numerical resolution, as the size
of the numerical cells gets smaller. The refining criteria are open
and can be chosen depending on the application. For a general purpose,
\asohf\ refines when the number of particles per cell exceeds a
user defined parameter. Once the refinement levels are set up, the
algorithm applies the SO method independently at each of those levels.

The parameters needed by the code are the following: 
i) the cosmological parameters when analysing cosmological simulations (given by the simulation itself), 
ii) the size of the coarse cells (determined by the ratio between the box size and the number of cells in 
the coarse level of refinement), the maximum number of refinement levels ($N_{\rm levels}$), and the 
maximum number of patches ($N_{\rm patch}$) for all levels in order to build up the AMR hierarchy of nested grids
(these parameters, which only represent maximum values in order to avoid
memory problems, are usually taken as $N_{\rm levels}=10-20$ and $N_{\rm patch}=10^5-10^6$),
iii) the number of particles per cell in order to choose the cells to
be refined (usually set to 3-4 particles per cell), and 
iv) the minimum number of particles in a halo (usually set to 10-20).

After this first step, the code naturally produces a tentative list of
haloes of different sizes and masses. Moreover, a complete description
of the substructure (haloes within haloes) is obtained by applying the
same procedure on the different levels of refinement.  A second step,
not using the cells but the particles within each halo, makes a more
accurate study of each of the previously identified haloes.  These
prospective haloes (subhaloes) may include particles which are not
physically bound. In order to remove unbound particles, the local
escape velocity is obtained at the position of each particle. To
compute this velocity we integrate Poisson equation assuming spherical
symmetry.  If the velocity of a particle is higher than the escape
velocity, the particle is assumed to be unbound and is therefore
removed from the halo (subhalo) being considered. Following this
procedure, unbound particles are removed iteratively along a list of
radially ordered particles until no more of them need to be
removed. In the case that the number of remaining particles is less
than a given threshold the halo is dropped from the list.

After this cleaning procedure, all the relevant quantities for the
haloes (subhaloes) as well as their evolutionary merger trees are
computed.  The lists of (bound) particles are used to calculate
canonical properties of haloes (subhaloes) like the position of the
halo centre, which is given by the centre of mass of all the bound
particles, the bulk velocity, and the size of the haloes, given by the distance of the
farthest bound particle to the centre.

The ability of the \asohf\ method to find haloes and their
substructures is limited by the requirement that appropriate
refinements of the computational grid exist with enough resolution to
spot the structure being considered. In comparison to algorithms based
on linking strategies, \asohf\ does not require a linking
length to be defined, although at a given level of refinement the size
of the cell can be considered as the linking length of this particular
resolution.

The version of the code used in this comparison is serial, although
there is already a first parallel version based on OpenMP.

\subsection{\bdm} \label{sec:bdm}
The Bound Density Maxima (\bdm) halo finder originally
described in \citet{Klypin97} uses a spherical 3D overdensity
algorithm to identify haloes and subhaloes. It starts by finding the
local density at each individual particle position. This density is
defined using a top-hat filter with a constant number of particles
$N_{\rm filter}$, which typically is $N_{\rm filter}=20$. The code
finds all maxima of density, and for each maximum it finds a sphere
containing a given overdensity mass $M_\Delta=(4\pi/3)\Delta\rho_{\rm
  cr}R^3_\Delta$, where $\rho_{\rm cr}$ is the critical density and
$\Delta$ is the specified overdensity.

For the identification of distinct haloes, the code uses the density
maxima as halo centres; amongst overlapping sphere the code finds the
one that has the deepest gravitational potential. Haloes are ranked by
their (preliminary) size and their final radius and mass are derived by
a procedure that guarantees smooth transition of properties of small
haloes when they fall into a larger host) halo becoming subhaloes:
this procedure either assigns $R_\Delta$ or $R_{\rm dist}$ as the
radius for a currently infalling halo as its radius depending on the
environmental conditions, where $R_{\rm dist}$ measures the distance
of the infalling halo to the surface of the soon-to-be host halo.

The identification of subhaloes is a more complicated procedure:
centres of subhaloes are certainly density maxima, but not all density
maxima are centres of subhaloes. \bdm\ eliminates all density
maxima from the list of subhalo candidates which have less than
$N_{\rm filter}$ self-bound particles. For the remaining set of
prospective subhaloes the radii are determined as the minimum of the
following three distances: (a) the distance to the nearest barrier
point (i.e. centres of previously defined (sub-)haloes), (b) the
distance to its most remote bound particle, and (c) the truncation
radius (i.e.  the radius at which the average density of bound
particles has an inflection point). This evaluation involves an
iterative procedure for removing unbound particles and starts with the
largest density maximum.

The unbinding procedure requires the evaluation of the gravitational
potential which is found by first finding the mass in spherical shells
and then by integration of the mass profile. The binning is done in
$\log$ radius with a very small bin size of $\Delta\log(R) =0.005$.

The bulk velocity of either a distinct halo or a subhalo is defined as
the average velocity of the 30 most bound particles of that halo or by
all particles, if the number of particles is less than 30. The number
30 is a compromise between the desire to use only the central
(sub)halo region for the bulk velocity and the noise level.

The code uses a domain decomposition for MPI parallelization and
OpenMP for the parallelization inside each domain.

\subsection{\fof}\label{sec:fof}
In order to analyse large cosmological simulations with up to $2048^3$
particles we have developed a new MPI version of the hierarchical
Friends-Of-Friends algorithm with low memory requests. It allows us to
construct very fast clusters of particles at any overdensity
(represented by the linking length) and to deduce the
progenitor-descendant-relationship for clusters in any two different
time steps. The particles in a simulation can consist of different
species (dark matter, gas, stars) of different mass. We consider them
as an undirected graph with positive weights, namely the lengths of
the segments of this graph. For simplicity we assume that all weights
are different. Then one can show that a unique minimum spanning tree
(MST) of the point distribution exists, namely the shortest graph
which connects all points. If subgraphs cover the graph then the MST
of the graph belongs to the union of MSTs of the subgraphs. Thus
subgraphs can be constructed in parallel. Moreover, the geometrical
features of the clusters, namely the fact that they occupy mainly
almost non-overlapping volumes, allow the construction of fast
parallel algorithms.  If the MST has been constructed all possible
clusters at all linking lengths can be easily determined. To represent
the output data we apply topological sorting to the set of clusters
which results in a cluster ordered sequence. Every cluster at any
linking length is a segment of this sequence. It contains the
distances between adjacent clusters. Note, that for the given MST
there exist many cluster ordered sequences which differ in the order
of the clusters but yield the same set of clusters at a desired
linking length.  If the set of particle-clusters has been constructed
further properties (centre of mass, velocity, shape, angular momentum,
orientation etc.) can be directly calculated. Since this concept is by
construction aspherical a circular velocity (as used to characterise
objects found with spherical overdensity algorithms) cannot be
determined here. The progenitor-descendant-relationship is calculated
for the complete set of particles by comparison of the cluster-ordered
sequences at two different output times.

The hierarchical \fof\ algorithm identifies objects at
different overdensities depending on the chosen linking length
\citep{More11}. In order to avoid artificial misidentifications of
subhaloes on high overdensities one can add an additional
criterion. Here we have chosen the requirement that the spin parameter
of the subhalo should be smaller than one. All subhaloes have been
identified at 512 times the virial overdensity. Thus only the highest
density peak has been taken into account for the mass determination
and the size of the object, which are therefore underestimated. The
velocity of the density peak is estimated correctly but without
removing unbound particles.

\subsection{\grasshopper} \label{sec:grasshopper}
\grasshopper\ (Stadel, in prep.) is based on a reworking of the \skid\ group finder \citep[][see \Sec{sec:skid} below]{Stadel01}. It finds density peaks and subsequently determines all associated bound particles thereby identifying haloes. Particles are slowly slid along the local density gradient until they pool at a maximum, each pool corresponding to each initial group. This first phase of \grasshopper\ can be computationally very expensive for large simulations, but is also quite robust.
Each pool is then unbound by iteratively evaluating the binding energy of every particle in their original positions and then removing the most non-bound particle until only bound particles remain. This removes all particles that are not part of substructure either because they are part of larger scale structure or because they are part of the background. The halo's position and velocity is given by the position and velocity of its centre-of-mass. For more details please refer to the \skid\ description below in \Sec{sec:skid}.

\subsection{\hbt}
\hbt\ \citep{Han12} is a tracing algorithm\footnote{It should be noted that \hbt\ had access to the full snapshot data for Aquarius-A.} working in the time domain of each subhaloesÕ evolution. Haloes are identified with a Friends-of-Friends algorithm, with the standard linking length of 0.2 times the average inter-particle separation, and halo merger trees are constructed. HBT then traverses the halo merger trees from the earliest to the latest time and identifies a self-bound remnant for every halo at every snapshot after infall. We apply an iterative unbinding procedure to derive self-bound remnants. Specifically, at each unbinding step, for each particle that is not yet removed, we calculate its potential energy using all the remaining particles with a tree code, and its kinetic energy with respect to the average velocity of a minimum potential core, including contribution from the Hubble flow with respect to the centre of mass of the core. The core consists of approximately 25 per cent of the remaining particles with the lowest potential energy, and it is used to define the halo's centre(-of-mass) and bulk velocity. Particles with positive total energy are then removed and the iteration continues. We stop the iteration until the relative change in the remaining mass between two iterations is smaller than 0.5 percent, or when the remaining mass falls below a lower mass limit of 20 particles.  To ensure that subhaloes are robustly traced over long periods, unbound particles from a subhalo at reshift $z_1$ are allowed to rebind to its descendent at a lower redshift $z_2$, as long as the descendent mass at $z_2$ is above 25 per cent the progenitor mass at $z_1$. We also record the merging hierarchy of progenitor haloes, to efficiently allow satellite-satellite mergers or satellite accretion.

\subsection{\hot} \label{sec:hot}
This algorithm, still under development, computes the Hierarchical Overdensity Tree (\hot) of a point distribution in an arbitrary multidimensional space.
\hot\ is introduced as an alternative to the minimal spanning tree for spaces where a metric is not well defined, like the phase space of particle positions and velocities. Rather than assuming an Euclidean metric, distance estimates are based on the Field Estimator for Arbitrary Spaces \citep[FiEstAS, ][]{Ascasibar05, Ascasibar10}, where the data space is tessellated one dimension at a time, until it is divided into a set of hypercubical cells containing exactly one particle.
In the \textsc{HOT+FiEstAS} scheme, objects correspond to the peaks of the density field, and their boundaries are set by the isodensity contours at the saddle points. At each saddle point, the object containing less particles is attached to the most massive one, which may then be incorporated into even more massive objects in the hierarchy. This idea can be implemented by computing the MST of the data distribution, defining the distance between two neighbouring particles as the minimum density along an edge connecting them (i.e. the smallest of the two densities, or the density of the saddle point when it exists).  Once the distances are defined, \texttt{HOT+FiEstAS} computes the MST of the data distribution by means of Kruskal's algorithm \citep{Kruskal56}. The output of the algorithm consists of the tree structure, given by the parent of each data point in \hot, and a catalogue containing an estimate of the centroid (given by the density-weighted centre of mass) as well as the number of particles in the object (both including and excluding substructures).  In order to discard spurious density fluctuations, a minimum number of points and density contrast are required for an object to be output to the catalogue.  Currently, these parameters are set to $N>20$ particles and a contrast threshold $\rho_{\rm peak}/\rho_{\rm background}>5$.  Although these values seem to yield reasonable results, more experimentation is clearly needed.

Exactly the same algorithm has been applied to the particle positions only (\hottd) and the full set of phase-space coordinates (\hotsd).
In order to optimize the method for the specific problem of halo finding, a post-processing routine, akin to a `hard' expectation maximization, has been developed, where \Rmax\ and \Vmax\ are computed for every object in the catalogue, and objects with more than 10 particles within \Rmax\ are labelled as (sub)halo candidates.
Then, particles are assigned to the candidate that contributes most to the phase-space density at their location, approximating each candidate by a \citet{Hernquist90} sphere.
The final catalog consists of all objects that contain more than five particles within \Rmax\ and an associated density above 100 times the critical value.

\subsection{\hsf} \label{sec:hsf}
The Hierarchical Structure Finder \citep[\hsf,][]{Maciejewski09}
identifies objects as connected self-bound particle sets above some
density threshold.  This method consists of two steps. Each particle
is first linked to a local DM phase-space density maximum by following
the gradient of a particle-based estimate of the underlying DM
phase-space density field. The particle set attached to a given
maximum defines a candidate structure. In a second step, particles
which are gravitationally unbound to the structure are discarded until
a fully self-bound final object is obtained.

In the initial step the phase-space density and phase-space gradients
are estimated by using a six-dimensional SPH smoothing kernel with a
local adaptive metric as implemented in the \textsc{EnBiD} code
\citep{Sharma06}. For the SPH kernel we use $N_{\rm sph}$ between $20$
and $64$ neighbours whereas for the gradient estimate we use $N_{\rm
  ngb}=20$ neighbours.

Once phase-space densities have been calculated, we sort the particles
according to their density in descending order. Then we start to grow
structures from high to low phase-space densities. While walking down
in density we mark for each particle the two closest (according to the
local phase-space metric) neighbours with higher phase-space density,
if such particles exist. In this way we grow disjoint structures until
we encounter a saddle point, which can be identified by observing the
two marked particles and seeing if they belong to different
structures. A saddle point occurs at the border of two
structures. According to each structure mass, all the particles below
this saddle point can be attached to only one of the structures if it
is significantly more massive than the other one, or redistributed
between both structures if they have comparable masses.  This is
controlled by a simple but robust cut or grow criterion depending on a
{\em connectivity parameter} $\alpha$ which is ranging from $0.2$ up
to $1.0$. In addition, we test on each saddle point if structures are
statistically significant when compared to Poisson noise (controlled by
a $\beta$ parameter). At the end of this process, we obtain a
hierarchical tree of structures.

In the last step we check each structure against an unbinding
criterion. Once we have marked its more massive partner for each
structure, we sort them recursively such that the larger partners
(parents) are always after the smaller ones (children). Then we
unbind structure after structure from children to parents and add
unbound particles to the larger partner. If the structure has less
than $N_{\rm cut}=20$ particles after the unbinding process, then we
mark it as not bound and attach all its particles to its more
massive partner. The most bound particle of each halo/subhalo defines its position centre.

Although \hsf\ can be used on the entire volume, to speed up the
process of identification of the structures in the cosmological
simulation volume we first apply the FOF method to disjoint the
particles into smaller FOF groups.

\subsection{\jumpd} \label{sec:jumpd}
\jumpd\ is a galaxy finder and \textit{not} a sub-\textit{halo} finder and hence is treated differently than the other finders in this work. It aims at finding and measuring central and satellite galaxies within given host haloes, i.e. \textit{baryonic} substructure objects within a sphere of given radius $R_{\rm lim}$ about the centre of the host. To this extent the stellar and gas mass profiles are searched for jumps (and hence the name) in the three-dimensional cumulative mass profiles  from the host halo centre out to the limiting radius $R_{\rm lim}$ (i.e. usually the host halo's virial radius). The jump detection criterion is based on the detection of changes in the first and second derivatives of the respective mass profiles in the $r, \theta$ and $\phi$ variables at the substructure locations corresponding to the humps they cause. For the stellar object, the jump in the stellar mass profile is used as a first satellite detection (i.e., location and velocity),  that is later on refined by searching 
for maxima in 6-dimensional phase-space within 
 an allowance region about that first center, returning  the object stellar sizes $r_{\rm star}$ as well. The jumps  in the gas profile are then matched to the stellar objects and gas particles inside a spherical region defined by the radial extend of the gas jump ($r_{\rm gas}$) are then  associated to  the stellar object. Note that for the detection of the jumps only cold gas is considered.

Please note that the approach of \jumpd\ is substantially different to halo finders in general. The code \textit{only} locates a baryonic object (`galaxy') without considering the dark matter. To this extent \jumpd\ cannot be subjected to the common post-processing pipeline when it comes to subhaloes as that pipeline heavily relies on the embedding of satellite galaxies within dark matter subhaloes.

\subsection{\lanl} \label{sec:lanl}
The \lanl\ halo finder is developed to provide on-the-fly halo
analysis for simulations utilizing hundreds of billions of particles,
and is integrated into the \texttt{HACC} code \citep{Habib09,Habib12},
although it can also be used as a stand-alone halo finder. Its core is
a fast $k$D-tree FOF halo finder which uses 3D (block), structured
decomposition to minimize surface to volume ratio of the domain
assigned to each process. As it is aimed at large-scale structure
simulations (100+ Mpc/$h$ on the side), where the size of any single
halo is much smaller than the size of the whole box, it uses the
concept of `ghost zones' such that each process gets all the
particles inside its domain as well as those particles which are
around the domain within a given distance (the overload size, a code
parameter chosen to be larger then the size of the biggest halo we
expect in the simulation). After each process runs its serial version
of a FOF finder, MPI based `halo stitching' is performed to ensure that
every halo is accounted for, and accounted for only once.

If desired, spherical `SO' halo properties can be found using the
FOF haloes as a proxy. Those SO haloes are centred at the particle
with the lowest gravitational potential, while the edge is at
$R_{\Delta}$ -- the radius enclosing an overdensity of $\Delta$. It is
well known that percolation based FOF haloes suffer from the
over-bridging problem; therefore, if we want to ensure completeness of
our SO sample we should run FOF with a smaller linking length than
usual in order to capture all density peaks, but still avoid
over-bridging at the scale of interest (which depends on our choice of
$\Delta$). Overlapping SO haloes are permitted, but the centre of one
halo may not reside inside another SO halo (that would be considered
as a substructure, rather than a `main' halo). The physical code
parameters are the linking length for the FOF haloes, and overdensity
parameter $\Delta$ for SO haloes, which have been chosen as 0.2 and 200 for the present study, respectively. Technical parameters are the overload
size and the minimum number of particles in a halo.

The \lanl\ halo finder is included in the standard
distributions of \texttt{PARAVIEW}\footnote{http://www.paraview.org/}
package, enabling researchers to combine analysis and visualization of
their simulations.

\subsection{\mendieta}
The \mendieta\ subhaloes finder is based on the Friends-of-Friends (FoF) algorithm which is successively applied using a
shorter linking length with the aim of finding the substructures. The \mendieta\ algorithm involves $1 + N_s$ steps, the
first of which consists in the identification of dark matter haloes by applying the FoF algorithm with a linking length
$L_{FoF}$, whereas the remaining $N_s$ steps are used to find the corresponding substructures. The two free parameters
are the initial linking length $L_{FoF}$ and the number of steps $N_s$. In each of $N_s$ steps the linking length is
reduced by a factor $1/(1 + N_s)$. The standard choice for $L_{FoF}$ parameter is 0.2 times the mean inter-particle
separation, whereas for the $N_s$ parameter a value of 9 is the recommended option. It is worth mentioning that the
value of $N_s$ fix the hierarchy of substructures that can be found.

In order to explain the algorithm, let A be the set of (sub)haloes identified in the step $i - 1$. The first part of the
$i^{th}$ step consists in applying a FoF identification with a linking length equal to $i/(1 + N_s)$. As result, in
general a (sub)halo of A is decomposed in two or more fragments: a  main substructure, a set of smaller substructures
and a set of unlinked particles. Once the identification has been carried out, an unbinding procedure is applied over
each (sub)halo of A. With this purposes,  all unlinked particles are associated to the main substructure. For each small
substructure, all unbounded particles are identified as those with positive total energy (potential and kinetic energy).
The  potential energy of one particle is computed by considering the gravitational interaction of the particle itself
with the rest of the substructure particles added to its own potential energy. The kinetic energy is calculated with
respect to the velocity of the centre of mass of the substructure summed to the Hubble flow. The latter operation is
performed assuming that the centre of the (sub)halo is located at the position of the most bounded particle (that one
with more negative potential energy). These particles are removed from their host substructure and assigned to the main
one. The last part of this step consists in applying the same unbinding algorithm to the main substructure. In this
case, all unbounded particles are marked as free particles and linked with no substructure.

The standard output of the total procedure consist in a dark matter haloes catalogue (i.e. the standard FoF haloes) and
a subhaloes catalogue each of which is associated with the corresponding host halo. \mendieta\ is described more fully in
\citet{Sgro10}.

\subsection{\ntropyfof}\label{sec:ntropy-fofsv}
The Ntropy parallel programming framework is derived from $N$-body codes
to help address a broad range of astrophysical problems\footnote{http://www.phys.washington.edu/users/gardnerj/ntropy}.  This
includes an implementation of a simple but efficient FOF halo finder,
\ntropyfof, which is more fully described in
\citet{Gardner07a} and \citet{Gardner07b}. Ntropy provides a
`distributed shared memory' (DSM) implementation of a $k$D-tree,
where the application developer can reference tree nodes as if they
exist in a global address space, even though they are physically
distributed across many compute nodes.  Ntropy uses the $k$D-tree data
structures to speed up the FOF distance searches.  It also employs an
implementation of the \citet{Shiloach82} parallel connectivity
algorithm to link together the haloes that span separate processor
domains.  The advantage of this method is that no single computer node
requires knowledge of all of the groups in the simulation volume,
meaning that \ntropyfof\ is scalable to petascale platforms
and can handle large data input. This algorithm was used in the mock halo
test cases to stitch together particle groups found across many
threads into the one main FOF halo. As FOF is a deterministic
algorithm, \ntropyfof\ takes a single physical linking
length to group particles into FOF haloes without performing any
particle unbinding or subhalo identification.  The halo centres for
the analysis presented here use centre-of-mass estimates based on the
FOF particle list, obtained by using a linking length of 0.2.  Ntropy achieves parallelisation by calling
`machine dependent library' (MDL) that consists of high-level
operations such as `acquire\_treenode' or `acquire\_particle.'  This
library is rewritten for a variety of models (MPI, POSIX Threads, Cray
SHMEM, etc.), allowing the framework to extract the best performance
from any parallel architecture on which it is run.

\subsection{\origami} \label{sec:origami}
\origami\ \citep[Order-ReversIng Gravity, Apprehended Mangling
Indices,][]{Falck12} uses a natural, parameter-free definition of the
boundary between haloes and the non-halo environment around them: halo
particles are particles that have experienced shell-crossing.  This
dynamical definition does not make use of the density field, in which
the boundary can be quite ambiguous.  In one dimension, shell
crossings can be detected by looking for pairs of particles whose
positions are out-of-order compared with their initial positions. In
3D, then, a halo particle is defined as a particle that has undergone
shell crossings along 3 orthogonal axes. Similarly, this would be 2
axes for a filament, 1 for a wall, and 0 for a void. There is a huge
number of possible sets of orthogonal axes in the initial grid to use
to test for shell-crossing, but we only used four simple ones, which
typically suffice to catch all the shell-crossings.  We used the
Cartesian $x$, $y$, and $z$ axes, as well as the three sets of axes
consisting of one Cartesian axis and two ($45\degr$) diagonal axes in
the plane perpendicular to it.

Once halo particles have been tagged, there are many possible ways of
grouping them into haloes.  For this paper, we grouped them on a
Voronoi tessellation of final-conditions particle positions.  This
gives a natural density estimate \citep[e.g.][\texttt{VTFE}, Voronoi
Tessellation Field Estimator]{Schaap00} and set of neighbours for each
particle.  Haloes are sets of halo particles connected to each other
on the Voronoi tessellation.  To prevent haloes from being unduly
linked, we additionally require that a halo contain at most one halo
`core', defined as a set of particles connected on the tessellation
that all exceed a \texttt{VTFE} density threshold. This density
threshold is the only parameter in our algorithm, since the initial
tagging of halo particles is parameter-free; for this study, we set it
to 200 times the mean density.  We partition connected groups of halo
particles with multiple cores into haloes as follows: each core
iteratively collects particles in concentric rings of Voronoi
neighbours until all halo particles are associated.  The tagging
procedure establishes halo boundaries, so no unbinding procedure is
necessary.  Also, we note that currently, the algorithm does not
identify subhaloes. We remove haloes with fewer than 20 particles from
the \origami\ halo catalogue, and the halo centre reported is
the position of the halo's highest-density particle.

\subsection{\pfof}\label{sec:pfof}
Parallel FOF (\pfof) is a MPI-based parallel
Friends-of-Friends halo finder which is used within the DEUS
Consortium \footnote{\texttt{www.deus-consortium.org}} at LUTH
(Laboratory Universe and Theories). It has been parallelized by Roy
and was used for several studies involving large $N$-body simulations
such as \cite{Courtin11, Rasera10}. The principle is the following:
first, particles are distributed in cubic subvolumes of the simulation
and each processor deals with one `cube', and runs
Friends-of-Friends locally. Then, if a structure is located close to
the edge of a cube, \pfof\ checks if there are particles
belonging to the same halo in the neighbouring cube. This process is
done iteratively until all haloes extending across multiple cubes have
been merged. Finally, particles are sorted on a per halo basis, and
the code writes two kinds of output: particles sorted per region,
particles sorted per halo. This makes any post-processing
straightforward because each halo or region can be analysed
individually on a single CPU server.  \pfof\ was successfully used on 32768 cores for more than one hundred snapshots with 8192$^3$ particles each \citep[DEUS Full Universe Run][]{Alimi13}. In this article, the serial version was used for
mock haloes and small cosmological simulations, and the parallel
version for larger runs. The linking length was set to $b=0.2$
\citep[however see][for a discussion on the halo
definition]{Courtin11}, and the minimum halo mass to 100
particles. And the halo centres reported here are the centre-of-mass
of the respective particle distribution.

\subsection{\pso} \label{sec:pso}
The parallel spherical overdensity (\pso) halo finder is a
fast, highly scalable MPI-parallelized tool directly integrated into
the \textsc{FLASH} simulation code that is designed to provide
on-the-fly halo finding for use in subgrid modeling, merger tree
analysis, and adaptive refinement schemes \citep{Sutter10}. The
\pso\ algorithm identifies haloes by growing SO spheres. There
are four adjustable parameters, controlling the desired overdensity
criteria for centre detection and halo size, the minimum allowed halo
size, and the resolution of the halo radii relative to the grid
resolution. For the cases here the overdensity thresholds were both chosen
to be $200\times \rho_{\rm crit}$,  the minimum halo size the
the spacing of the grid, and the radius resolution half of the grid spacing with the grid resolution chosen to be of order the force resolution of the simulation.

The algorithm discovers halo centres by mapping dark
matter particles onto the simulation mesh and selecting cell centres
where the cell density is greater than the given overdensity
criterion. The algorithm then determines the halo edge using the SO
radius by collecting particles using the \textsc{FLASH} AMR tree
hierarchy.  The algorithm determines the halo centre, bulk velocity,
mass, and velocity dispersion from all enclosed particles without additional
post-processing. \pso\ is provided as both an API for use
in-code and as a stand-alone halo finder.

\subsection{\rockstar} \label{sec:rockstar}
\rockstar\footnote{\rockstar\ is freely available from \url{http://code.google.com/p/rockstar}} is a  phase-space based halo finder designed to
maximize halo consistency across timesteps; as such, it is especially
useful for studying merger trees and halo evolution \citep{Behroozi12}.  \rockstar\ first selects particle groups with a
3D Friends-of-Friends variant with a very large linking length
($b=0.28$).  For each main FOF group, \rockstar\ builds a
hierarchy of FOF subgroups in phase space by progressively and
adaptively reducing the linking length, so that a tunable fraction
(70~per cent, for this analysis) of particles are captured at each subgroup
as compared to the immediate parent group.  For each subgroup, the
phase-space metric is renormalized by the standard deviations of
particle position and velocity.  That is, for two particles $p_1$ and
$p_2$ in a given subgroup, the distance metric is defined as:
\begin{equation}
d(p_1, p_2) = \left(\frac{({\bf x}_1-{\bf x}_2)^2}{\sigma_x^2} +
  \frac{({\bf v}_1-{\bf v}_2)^2}{\sigma_v^2}\right)^{1/2},
\end{equation}
where $\sigma_x$ and $\sigma_v$ are the particle position and velocity
dispersions for the given subgroup. This metric ensures an adaptive
selection of overdensities at each successive level of the FOF
hierarchy.

When this is complete, \rockstar\ converts FOF subgroups into
haloes beginning at the deepest level of the hierarchy.  For a subgroup
without any further sublevels, all the particles are assigned to a
single seed halo.  If the parent group has no other subgroups, then
all the particles in the parent group are assigned to the same seed
halo as the subgroup.  However, if the parent group has multiple
subgroups, then particles are assigned to the subgroups' seed haloes
based on their phase-space proximity.  In this case, the phase-space
metric is set by halo properties, so that the distance between a halo
$h$ and a particle $p$ is defined as:
\begin{equation}
d(h, p) = \left(\frac{({\bf x}_h-{\bf x}_p)^2}{r_{vir}^2} +
  \frac{({\bf v}_h-{\bf v}_p)^2}{\sigma_v^2}\right)^{1/2},
\end{equation}
where $r_{vir}$ is the current virial radius of the seed halo and
$\sigma_v$ is the current particle velocity dispersion.  This process
is repeated at all levels of the hierarchy until all particles in the
base FOF group have been assigned to haloes.  Unbinding is performed
using the full particle potentials (calculated using a modified Barnes
\& Hut method, \citet{Barnes86}); halo centres are defined by
averaging particle positions at the FOF hierarchy level which yields
the minimum estimated Poisson error---which in practice amounts to
averaging positions in a small region close to the phase-space density
peak.  For further details about the unbinding process and for details
about accurate calculation of halo properties, please see Behroozi et
al. in prep.

\rockstar\ is a massively parallel code (hybrid OpenMP/MPI
style); it can already run on up to $10^5$ CPUs and on the very
largest simulations ($> 10^{10}$ particles).  Additionally, it is very
efficient, requiring only 56 bytes of memory per particle and 4-8
(total) CPU hours per billion particles in a simulation snapshot.  
  
\subsection{\skid} \label{sec:skid}
\skid\ (Spline Kernel Interpolative Denmax)\footnote{The OpenMP
  parallelized version of \skid\ can be freely downloaded from
  \texttt{https://hpcforge.org/projects/skid}}, first mentioned in
\citet{Governato97} and extensively described in \citet{Stadel01},
finds density peaks within $N$-body simulations and subsequently
determines all associated bound particles thereby identifying
haloes. It is important to stress that \skid\ will only find
the smallest scale haloes within a hierarchy of haloes as is generally
seen in cosmological structure formation simulations.  Unlike original
DENMAX \citep{Bertschinger91, Gelb92} which used a fixed grid based
density estimator, \skid\ uses SPH (i.e., smoothed particle
hydrodynamics) kernel averaged densities which are much better suited
to the Lagrangian nature of $N$-body simulations and allow the method
to locally adapt to the large dynamic range found in cosmological
simulations.

Particles are slowly slid (each step moving the particles by a
distance of order the softening length in the simulation) along the
local density gradient until they pool at a maximum, each pool
corresponding to each initial group. This first phase of \skid\
can be computationally very expensive for large simulations, but is
also quite robust. Each pool is then `unbound' by iteratively evaluating the binding
energy of every particle in their original positions and then removing
the most non-bound particle until only bound particles remain.  This
removes all particles that are not part of substructure either because
they are part of larger scale structure or because they are part of
the background.

The halo's position is given by the coordinates of the density maximum and its velocity by the center of mass velocity of the particles belonging to it.

\skid\ can also identify structure composed of gas and stars in
hydrodynamical simulations using the dark matter only for its
gravitational binding effect. The ``Haloes going MAD'' meeting has
motivated development of an improved version of the algorithm (now called \grasshopper, see \Sec{sec:grasshopper}) capable
of also running on parallel computers.

\subsection{\stf}
The STructure Finder \citep[\stf\ a.k.a VELOCIraptor,][]{Elahi11} is a hybrid OpenMP+MPI code that identifies objects by utilizing the fact that dynamically distinct substructures in a halo will have a {\em local} velocity distribution that differs significantly from the mean, {\em i.e.} smooth background of the halo. This method consists of two main steps, identifying particles that appear dynamically distinct and linking this outlier population using a Friends-of-Friends-like approach. Specifically, outlier particles are identified by estimating the mean velocity distribution function using a coarse grain approach and comparing the predicted distribution to that of a particle's local velocity distribution, which is calculated using a near-neighbour kernel technique. Specifically, the local velocity density is estimated using 32 nearest neighbours in velocity space drawn from a larger sample of 256 nearest neighbours in physical space. By taking the ratio of the local velocity distribution density relative to the expected mean velocity distribution density at a particle's phase-space position, the contrast of particles resident in substructure relative to those in the background are greatly enhanced. The scatter in this estimator is determined by examining the distribution of this ratio, $\mathcal{L}$, which is characterised by a Gaussian distribution corresponding to the virialized background, and numerous secondary peaks located at large values of the ratio arising from particles resident in substructure. The variance about the central main peak is estimated and only outlier particles, which have ratios lying several $n_{\mathcal{L}}\sigma$ away from the main peak, are searched. In this way, we quantify how dynamically different a particle is and the likelihood that a particle is resident in substructure. 
The criteria used to link particles together are
\begin{gather}
    \frac{({\bf x}_i-{\bf x}_j)^2}{\ell_x^2}<1,\notag\\
    1/V_r\leq  v_i/v_j\leq V_r,\notag\\
    \cos\theta_{\rm op}\leq \frac{{\bf v}_i\cdot{\bf v}_j}{v_i v_j}, \label{eqn:stffof}
\end{gather}
where the $\ell_x$ is the physical linking length, $V_r$ is a velocity ratio and $\cos\theta_{\rm op}$ is a velocity opening angle. Typical values are $n_{\mathcal{L}}=2.5$, $\ell_x=0.20$ times the inter-particle spacing, the linking length used to find haloes, $V_r\sim2$ and $\theta_{\rm op}\sim20^\circ$. These parameters would generally link entire halo if this method was not limit to the outlier population. Since this approach is capable of not only finding subhaloes, but tidal features surrounding subhaloes as well as tidal debris from completely disrupted subhaloes, for this study we also ensure that a group is self-bound. We do this by calculating the potential of the particles in the (sub)halo while ignoring the background using a tree code. Particles with positive energy relative to the halo's bulk velocity are considered gravitationally unbound and are discarded until a fully self-bound (sub)halo is obtained or the (sub)halo consists of fewer than 20 particles, at which point the group is removed entirely. The properties are then calculated about the (sub)halo's centre, which is determined by calculating the centre-of-mass using an iterative approach to determine the inner most 10\% of the particles.
\footnote{Those interested in obtaining a copy of the code should contact the author at \href{mailto:pelahi@physics.usyd.edu.au}{pelahi@physics.usyd.edu.au}.  Current acceptable input formats for simulation files are \textsc{Tipsy} and \textsc{GADGET-2}.}

\subsection{\subfind}\label{sec:subfind}
\subfind\ \citep{Springel01subfind} identifies gravitationally
bound, locally overdense regions within an input parent halo,
traditionally provided by a FOF group finder, although other group
finders could be used in principle as well. The densities are
estimated based on the initial set of all particles via adaptive
kernel interpolation based on a number $N_{\rm dens}$ of smoothing
neighbours. For each particle, the nearest $N_{\rm ngb}$ neighbours
are then considered for identifying local overdensities through a
topological approach that searches for saddle points in the isodensity
contours within the global field of the halo. This is done in a
top-down fashion, starting from the particle with the highest
associated density and adding particles with progressively lower
densities in turn. If a particle has only denser neighbours in a
single structure it is added to this region. If it is isolated it
grows a new density peak, and if it has denser neighbours from two
different structures, an isodensity contour that traverses a saddle
point is identified. In the latter case, the two involved structures
are joined and registered as candidate subhaloes if they contain at
least $N_{\rm ngb}$ particles. These candidates, selected according to
the spatial distribution of particles only, are later processed for
gravitational self-boundness. Particles with positive total energy
are iteratively dismissed until only bound particles remain. The
gravitational potential is computed with a tree algorithm, such that
large haloes can be processed efficiently. If the remaining bound
number of particles is at least $N_{\rm ngb}$, the candidate is
ultimately recorded as a subhalo. The set of initial substructure
candidates forms a nested hierarchy that is processed from inside out,
allowing the detection of substructures within substructures. However,
a given particle may only become a member of one substructure,
i.e. \subfind\ decomposes the initial group into a set of
disjoint self-bound structures. Particles not bound to any genuine
substructure are assigned to the `background halo'. This component
is also checked for self-boundness, so that some particles that are
not bound to any of the structures may remain. For all substructures
as well as the main halo, the particle with the minimum gravitational
potential is adopted as (sub)halo centre. For the main halo,
\subfind\ additionally calculates a SO virial mass around this
centre, taking into account all particles in the simulation (i.e. not
just those in the FOF group that is analyzed). The values adopted for the studies presented
here were $N_{\rm dens}=N_{\rm ngb}=32$ (and a linking length of 0.17 times the interparticle separation) for the comparison of field haloes and  $N_{\rm dens}=N_{\rm ngb}=20$  (and a linking length of 0.2) for the subhalo analysis. There exist both serial
and MPI-parallelized versions of \subfind, which implement the
same underlying algorithms. For more details we refer the reader to
the paper by \citet{Springel01subfind}.

\subsection{\voboz} \label{sec:voboz}
Conceptually, a \voboz\ \citep[VOronoi BOund
  Zones,][]{Neyrinck05} halo or subhalo is a density peak surrounded
by gravitationally bound particles that are down steepest-density
gradients from the peak.  A statistical significance is measured for
each (sub)halo, based on the probability that Poisson noise would
produce it.

The only physical parameter in \voboz\ is the density threshold
characterizing the edge of (parent) haloes (set to 200 times the mean
density here), which typically only affects their measured masses.  To
return a definite halo catalog, we also impose a
statistical-significance threshold (set to 4-$\sigma$ here), although
depending on the goal of a study, this may not be necessary.

Density peaks are found using a Voronoi tessellation (parallelizable
by splitting up the volume), which gives an adaptive, parameter-free
estimate of each particle's density and set of neighbours
\citep[e.g.][]{Schaap00}.
Each particle is joined to the peak particle
(whose position is returned as the halo centre) that lies up the
steepest density gradient from that particle.  A halo associated with
a high density peak will also contain smaller density peaks.  The
significance of a halo is judged according to the ratio of its central
density to a saddle point joining the halo to a halo with a higher
central density, comparing to a Poisson point process.  Pre-unbinding
(sub)halo boundaries are defined along these density ridges.

Unbinding evaporates many spurious haloes, and often brings other halo
boundaries inward a bit, reducing the dependence on the outer density
contrast.  Particles not gravitationally bound to each halo are
removed iteratively, by comparing their potential energies (measured
as sums over all other particles) to kinetic energies with respect to
the velocity centroid of the halo's core (i.e.\ the particles that
directly jump up density gradients to the peak).  The unbinding is
parallelized using OpenMP. 

\bsp

\label{lastpage}

\end{document}